\documentclass[pdftex,twocolumn,epjc3_preprint,runningheads]{svjour3}

\usepackage[T1]{fontenc}
\usepackage{lmodern}
\usepackage{calc}
\usepackage{graphicx}
\usepackage{booktabs}
\usepackage{textcomp}
\usepackage{xspace}
\usepackage{relsize}
\usepackage{amssymb}
\usepackage{amsmath}
\usepackage{listings}
\usepackage{microtype}
\usepackage{multirow}
\usepackage{tabularx}
\usepackage{array}
\usepackage{placeins}
\usepackage{cuted}
\usepackage{soul} 
\usepackage{fixltx2e}
\usepackage[numbers,sort&compress]{natbib}
\usepackage[labelfont=bf,font=small]{caption}
\usepackage[skip=-2pt]{subcaption}
\usepackage[colorlinks,citecolor=blue,urlcolor=blue,linkcolor=blue,breaklinks=breakall]{hyperref}
\usepackage{breakurl}
\usepackage[dvipsnames]{xcolor}
\usepackage[clockwise,figuresright]{rotating}
\usepackage{siunitx}
\usepackage{tikz}

\usepackage{etoolbox}
\AfterEndEnvironment{strip}{\leavevmode}

\allowdisplaybreaks

\newcolumntype{L}{>{\raggedright\let\newline\\\arraybackslash\hspace{0pt}}X}
\newcolumntype{R}{>{\raggedleft\let\newline\\\arraybackslash\hspace{0pt}}X}
\newcolumntype{C}{>{\centering\let\newline\\\arraybackslash\hspace{0pt}}X}

\newcommand{\imperial}{Department of Physics, Imperial College London, Blackett Laboratory, Prince Consort Road, London SW7 2AZ, UK}
\newcommand{\nordita}{NORDITA, Roslagstullsbacken 23, SE-10691 Stockholm, Sweden}
\newcommand{\oslo}{Department of Physics, University of Oslo, N-0316 Oslo, Norway}
\newcommand{\adelaide}{Department of Physics, University of Adelaide, Adelaide, SA 5005, Australia}
\newcommand{\glasgow}{SUPA, School of Physics and Astronomy, University of Glasgow, Glasgow, G12 8QQ, UK}
\newcommand{\monash}{School of Physics and Astronomy, Monash University, Melbourne, VIC 3800, Australia}
\newcommand{\coepp}{Australian Research Council Centre of Excellence for Particle Physics at the Tera-scale}
\newcommand{\okc}{Oskar Klein Centre for Cosmoparticle Physics, AlbaNova University Centre, SE-10691 Stockholm, Sweden}
\newcommand{\su}{Department of Physics, Stockholm University, SE-10691 Stockholm, Sweden}
\newcommand{\mcgill}{Department of Physics, McGill University, 3600 rue University, Montr\'eal, Qu\'ebec H3A 2T8, Canada}
\newcommand{\ucla}{Physics and Astronomy Department, University of California, Los Angeles, CA 90095, USA}
\newcommand{\annecy}{LAPTh, Universit\'e de Savoie, CNRS, 9 chemin de Bellevue B.P.110, F-74941 Annecy-le-Vieux, France}
\newcommand{\harvard}{Department of Physics, Harvard University, Cambridge, MA 02138, USA}
\newcommand{\grappa}{GRAPPA, Institute of Physics, University of Amsterdam, Science Park 904, 1098 XH Amsterdam, Netherlands}
\newcommand{\sydney}{Centre for Translational Data Science, Faculty of Engineering and Information Technologies, School of Physics, The University of Sydney, NSW 2006, Australia}

\newcommand{\cernth}{Theoretical Physics Department, CERN, CH-1211 Geneva 23, Switzerland}
\newcommand{\lyon}{Univ Lyon, Univ Lyon 1, ENS de Lyon, CNRS, Centre de Recherche Astrophysique de Lyon UMR5574, F-69230 Saint-Genis-Laval, France}
\newcommand{\iuf}{Institut Universitaire de France, 103 boulevard Saint-Michel, 75005 Paris, France}
\newcommand{\zurich}{Physik-Institut, Universit\"at Z\"urich, Winterthurerstrasse 190, 8057 Z\"urich, Switzerland}
\newcommand{\krakow}{H.~Niewodnicza\'nski Institute of Nuclear Physics, Polish Academy of Sciences, 31-342  Krak\'ow, Poland}

\newcommand{\valencia}{Instituto de F\'isica Corpuscular, IFIC-UV/CSIC, Valencia, Spain}

\newcommand{\gambitacknospmare}{We warmly thank the Casa Matem\'aticas Oaxaca, affiliated with the Banff International Research Station, for hospitality whilst part of this work was completed, and the staff at Cyfronet, for their always helpful supercomputing support.  \GB has been supported by STFC (UK; ST/K00414X/1, ST/P000762/1), the Royal Society (UK; UF110191), Glasgow University (UK; Leadership Fellowship), the Research Council of Norway (FRIPRO 230546/F20), NOTUR (Norway; NN9284K), the Knut and Alice Wallenberg Foundation (Sweden; Wallenberg Academy Fellowship), the Swedish Research Council (621-2014-5772), the Australian Research Council (CE110001004, FT130100018, FT140100244, FT160100274), The University of Sydney (Australia; IRCA-G162448), PLGrid Infrastructure (Poland), Red Espa\~nola de Supercomputaci\'on (Spain; FI-2016-1-0021), Polish National Science Center (Sonata UMO-2015/17/D/ST2/03532), the Swiss National Science Foundation (PP00P2-144674), the European Commission Horizon 2020 Marie Sk\l{}odowska-Curie actions (H2020-MSCA-RISE-2015-691164), the ERA-CAN+ Twinning Program (EU \& Canada), the Netherlands Organisation for Scientific Research (NWO-Vidi 680-47-532), the National Science Foundation (USA; DGE-1339067), the FRQNT (Qu\'ebec) and NSERC/The Canadian Tri-Agencies Research Councils (BPDF-424460-2012).}

\makeatletter

\newcommand{\preprintnumber}[1]{\gdef\@preprintnumber{\begin{flushright}{#1}\end{flushright}}}

\g@addto@macro\bfseries{\boldmath}
\makeatother

\newcommand{\subparagraph}{} 
\usepackage{titlesec}
\titleformat*{\paragraph}{\bfseries}

\journalname{Eur. Phys. J. C}
\smartqed
\let\underscore\_
\renewcommand{\_}{\discretionary{\underscore}{}{\underscore}}

\makeatletter
\let\orgdescriptionlabel\descriptionlabel
\renewcommand*{\descriptionlabel}[1]{%
  \let\orglabel\label
  \let\label\@gobble
  \phantomsection
  \protected@edef\@currentlabel{#1}%
  \let\label\orglabel
  \orgdescriptionlabel{#1}%
}
\makeatother

\newcommand\postnewlinemarker{\hbox{\ensuremath{\hookrightarrow}}}
\newcommand\cpp[1]{{\lstinline!#1!}}  

\newcommand\yaml[1]{{\lstset{style=yaml}\lstinline!#1!\lstset{style=cpp}}}

\newcommand\term[1]{{\lstset{style=terminal}\lstinline!#1!\lstset{style=cpp}}}
\newcommand\fortran[1]{{\lstset{style=fortran}\lstinline!#1!\lstset{style=cpp}}}
\newcommand\py[1]{{\lstset{style=python}\lstinline!#1!\lstset{style=cpp}}}
\newcommand\customtilde{{\raisebox{0.2ex}{\scalebox{0.6}{\boldmath$\sim$}}}}
\newcommand\mathematica[1]{{\lstset{style=Mathematica}\lstinline!#1!\lstset{style=cpp}}}

\lstnewenvironment{lstlistingyaml}{\lstset{style=yaml}}{\lstset{style=cpp}}
\lstnewenvironment{lstlistingterm}{\lstset{style=terminal}}{\lstset{style=cpp}}
\lstnewenvironment{lstlistingfortran}{\lstset{style=fortran}}{\lstset{style=cpp}}
\lstnewenvironment{lstcpp}{\lstset{style=cpp}}{\lstset{style=cpp}}
\lstnewenvironment{lstcppalt}{\lstset{style=cppalt}}{\lstset{style=cpp}}
\lstnewenvironment{lstcppnum}{\lstset{style=cppnum}}{\lstset{style=cpp}}
\lstnewenvironment{lstyaml}{\lstset{style=yaml}}{\lstset{style=cpp}}
\lstnewenvironment{lstterm}{\lstset{style=terminal}}{\lstset{style=cpp}}
\lstnewenvironment{lsttermalt}{\lstset{style=terminalalt}}{\lstset{style=cpp}}
\lstnewenvironment{lsttext}{\lstset{style=text}}{\lstset{style=cpp}}
\lstnewenvironment{lstfortran}{\lstset{style=fortran}}{\lstset{style=cpp}}
\lstnewenvironment{lstpy}{\lstset{style=python}}{\lstset{style=cpp}}
\lstnewenvironment{lstmathematica}{\lstset{style=mathematica}}{\lstset{style=cpp}}

\newcommand{\tmpname}{}
\newcommand{\tmplistingname}{}
\makeatletter
\newif\ifATOlabelname
\lst@Key{labelname}{Listing}{\def\ATOlabelname{#1}\global\ATOlabelnametrue}
\makeatother
\lstnewenvironment{lstcpplabel}[1][]{
  \lstset{style=cpp,#1} 
  \ifATOlabelname
    \renewcommand{\tmpname}{\lstlistingname}
    \renewcommand{\tmplistingname}{\lstlistlistingname}
    \renewcommand{\lstlistingname}{\ATOlabelname}
    \renewcommand{\lstlistlistingname}{List of \lstlistingname s}
  \fi
}{
  \renewcommand{\lstlistingname}{\tmpname}
  \renewcommand{\lstlistlistingname}{\tmplistingname}
  \lstset{style=cpp}
}
\definecolor{solarized@base03}{HTML}{002B36}
\definecolor{solarized@base02}{HTML}{073642}
\definecolor{solarized@base01}{HTML}{586e75}
\definecolor{solarized@base00}{HTML}{657b83}
\definecolor{solarized@base0}{HTML}{839496}
\definecolor{solarized@base1}{HTML}{93a1a1}
\definecolor{solarized@base2}{HTML}{EEE8D5}
\definecolor{solarized@base3}{HTML}{FDF6E3}
\definecolor{solarized@yellow}{HTML}{B58900}
\definecolor{solarized@orange}{HTML}{CB4B16}
\definecolor{solarized@red}{HTML}{DC322F}
\definecolor{solarized@magenta}{HTML}{D33682}
\definecolor{solarized@violet}{HTML}{6C71C4}
\definecolor{solarized@blue}{HTML}{268BD2}
\definecolor{solarized@cyan}{HTML}{2AA198}
\definecolor{solarized@green}{HTML}{859900}
\definecolor{darkred}{HTML}{550003}
\definecolor{darkgreen}{HTML}{00AA00}

\newcommand\YAMLstringstyle{\footnotesize\color{solarized@green}\mdseries}
\newcommand\YAMLkeystyle{\footnotesize\color{solarized@blue}\ttfamily}
\newcommand\YAMLvaluestyle{\footnotesize\color{blue}\mdseries}
\newcommand\ProcessThreeDashes{\llap{\color{cyan}\mdseries-{-}-}}

\newcommand\CPPcommentstyle{\color{solarized@violet}\footnotesize\ttfamily}
\newcommand\CPPdirectivestyle{\color{solarized@magenta}\footnotesize\ttfamily}
\newcommand\termplainstyle{\footnotesize\ttfamily}

\newcommand\processLongMacroDelimiter
{%
\CPPdirectivestyle%
\#define%
}

\lstdefinestyle{cpp}
{
  language=C++,
  basicstyle=\footnotesize\ttfamily,
  basewidth={0.53em,0.44em},
  numbers=none,
  tabsize=2,
  breaklines=true,
  escapeinside={@}{@},
  showstringspaces=false,
  numberstyle=\tiny\color{solarized@base01},
  keywordstyle=\color{solarized@orange},
  stringstyle=\color{solarized@red}\ttfamily,
  identifierstyle=\color{solarized@blue},
  commentstyle=\CPPcommentstyle,
  directivestyle=\CPPdirectivestyle,
  emphstyle=\color{solarized@green},
  frame=single,
  rulecolor=\color{solarized@base2},
  rulesepcolor=\color{solarized@base2},
  literate={~} {\customtilde}1,
  moredelim=*[directive]\ \ \#,
  moredelim=*[directive]\ \ \ \ \#
}

\lstdefinestyle{cppalt}
{
  language=C++,
  basicstyle=\footnotesize\ttfamily,
  basewidth={0.53em,0.44em},
  numbers=none,
  tabsize=2,
  breaklines=true,
  escapeinside={*@}{@*},
  showstringspaces=false,
  numberstyle=\tiny\color{solarized@base01},
  keywordstyle=\color{solarized@orange},
  stringstyle=\color{solarized@red}\ttfamily,
  identifierstyle=\color{solarized@blue},
  commentstyle=\CPPcommentstyle,
  directivestyle=\CPPdirectivestyle,
  emphstyle=\color{solarized@green},
  frame=single,
  rulecolor=\color{solarized@base2},
  rulesepcolor=\color{solarized@base2},
  literate={~}{\customtilde}1,
  moredelim=**[is][\processLongMacroDelimiter]{BeginLongMacro}{EndLongMacro} 
}

\lstdefinestyle{cppnum}
{
  language=C++,
  basicstyle=\footnotesize\ttfamily,
  basewidth={0.53em,0.44em}, 
  numbers=none,
  tabsize=2,
  breaklines=true,
  escapeinside={@}{@},
  numberstyle=\tiny\color{solarized@base01},
  showstringspaces=false,
  numberstyle=\tiny\color{solarized@base01},
  keywordstyle=\color{solarized@orange},
  stringstyle=\color{solarized@red}\ttfamily,
  identifierstyle=\color{solarized@blue},
  commentstyle=\CPPcommentstyle,
  directivestyle=\CPPdirectivestyle,
  emphstyle=\color{solarized@green},
  frame=single,
  rulecolor=\color{solarized@base2},
  rulesepcolor=\color{solarized@base2},
  literate={~} {\customtilde}1,
  moredelim=*[directive]\ \ \#,
  moredelim=*[directive]\ \ \ \ \#
}

\lstdefinestyle{python}
{
  language=Python,
  basicstyle=\footnotesize\ttfamily,
  basewidth={0.53em,0.44em},
  numbers=none,
  tabsize=2,
  breaklines=true,
  escapeinside={@}{@},
  showstringspaces=false,
  numberstyle=\tiny\color{solarized@base01},
  keywordstyle=\color{blue},
  stringstyle=\color{orange}\ttfamily,
  identifierstyle=\color{darkred},
  commentstyle=\color{purple},
  emphstyle=\color{green},
  frame=single,
  rulecolor=\color{solarized@base2},
  rulesepcolor=\color{solarized@base2},
  literate = {~}{\customtilde}1
             {\ as\ }{{\color{blue}\ as\ \color{black}}}3
}

\lstdefinestyle{fortran}
{
  language=Fortran,
  basicstyle=\footnotesize\ttfamily,
  basewidth={0.53em,0.44em},
  numbers=none,
  tabsize=2,
  breaklines=true,
  escapeinside={@}{@},
  showstringspaces=false,
  numberstyle=\tiny\color{solarized@base01},
  keywordstyle=\color{blue},
  stringstyle=\color{orange}\ttfamily,
  identifierstyle=\color{Periwinkle},
  commentstyle=\color{purple},
  emphstyle=\color{green},
  morekeywords={and, or, true, false},
  frame=single,
  rulecolor=\color{solarized@base2},
  rulesepcolor=\color{solarized@base2},
  literate={~}{\customtilde}1
}

\lstdefinestyle{terminal}
{
  language=bash,
  basicstyle=\termplainstyle,
  numbers=none,
  tabsize=2,
  breaklines=true,
  escapeinside={@}{@},
  frame=single,
  showstringspaces=false,
  numberstyle=\tiny\color{solarized@base01},
  keywordstyle=\color{solarized@orange},
  stringstyle=\color{solarized@red}\ttfamily,
  identifierstyle=\color{black},
  commentstyle=\color{solarized@violet},
  emphstyle=\color{solarized@green},
  frame=single,
  rulecolor=\color{solarized@base2},
  rulesepcolor=\color{solarized@base2},
  morekeywords={gambit, cmake, make, mkdir},
  deletekeywords={test},
  literate = {\ gambit}{{\ }{\color{black}}gambit}7
             {/gambit}{{/}{\color{black}}gambit}6
             {gambit/}{{\color{black}}gambit{/}}6
             {/include}{{/}{\color{black}}include}8
             {cmake/}{{\color{black}}cmake/}6
             {.cmake}{{.}{\color{black}}cmake}6
             {~}{\customtilde}1
}

\lstdefinestyle{terminalalt}
{
  language=bash,
  basicstyle=\footnotesize\ttfamily,
  numbers=none,
  tabsize=2,
  breaklines=true,
  escapeinside={*@}{@*},
  frame=single,
  showstringspaces=false,
  numberstyle=\tiny\color{solarized@base01},
  keywordstyle=\color{solarized@orange},
  stringstyle=\color{solarized@red}\ttfamily,
  identifierstyle=\color{black},
  commentstyle=\color{solarized@violet},
  emphstyle=\color{solarized@green},
  frame=single,
  rulecolor=\color{solarized@base2},
  rulesepcolor=\color{solarized@base2},
  morekeywords={gambit, cmake, make, mkdir},
  deletekeywords={test},
  literate = {\ gambit}{{\ }{\color{black}}gambit}7
             {/gambit}{{/}{\color{black}}gambit}6
             {gambit/}{{\color{black}}gambit{/}}6
             {/include}{{/}{\color{black}}include}8
             {cmake/}{{\color{black}}cmake/}6
             {.cmake}{{.}{\color{black}}cmake}6
             {~}{\customtilde}1
}

\lstdefinestyle{text}
{
  language={},
  basicstyle=\footnotesize\ttfamily,
  identifierstyle=\color{black},
  numbers=none,
  tabsize=2,
  breaklines=true,
  escapeinside={*@}{@*},
  showstringspaces=false,
  frame=single,
  rulecolor=\color{solarized@base2},
  rulesepcolor=\color{solarized@base2},
  literate={~}{\customtilde}1
}

\lstdefinestyle{yaml}
{
  language=bash,
  escapeinside={@}{@},
  keywords={true,false,null},
  otherkeywords={},
  keywordstyle=\color{solarized@base0}\bfseries,
  basicstyle=\footnotesize\color{black}\ttfamily,
  identifierstyle=\YAMLkeystyle,
  sensitive=false,
  commentstyle=\color{solarized@orange}\ttfamily,
  morecomment=[l]{\#},
  morecomment=[s]{/*}{*/},
  stringstyle=\YAMLstringstyle\ttfamily,
  moredelim=**[s][\YAMLkeystyle]{,}{:},   
  moredelim=**[l][\YAMLvaluestyle]{:},    
  morestring=[b]',
  morestring=[b]",
  literate =    {---}{{\ProcessThreeDashes}}3
                {>}{{\textcolor{solarized@red}\textgreater}}1
                {|}{{\textcolor{solarized@red}\textbar}}1
                {\ -\ }{{\mdseries\color{black}\ -\ \negmedspace}}3
                {\}}{{{\color{black} \}}}}1
                {\{}{{{\color{black} \{}}}1
                {[}{{{\color{black} [}}}1
                {]}{{{\color{black} ]}}}1
                {~}{\customtilde}1,
  breakindent=0pt,
  breakatwhitespace,
  columns=fullflexible
}

\lstdefinestyle{mathematica}
{
  language={Mathematica},
  basicstyle=\footnotesize\ttfamily,
  basewidth={0.53em,0.44em},
  numbers=none,
  tabsize=2,
  breaklines=true,
  escapeinside={@}{@},
  numberstyle=\tiny\color{black},
  showstringspaces=false,
  numberstyle=\tiny\color{solarized@base01},
  keywordstyle=\color{solarized@orange},
  stringstyle=\color{solarized@red}\ttfamily,
  identifierstyle=\color{solarized@orange}\ttfamily,
  commentstyle=\color{solarized@gray}\ttfamily,
  directivestyle=\color{solarized@orange}\ttfamily,
  emphstyle=\color{solarized@green},
  frame=single,
  rulecolor=\color{solarized@base2},
  rulesepcolor=\color{solarized@base2},
  literate={~} {\customtilde}1,
  moredelim=*[directive]\ \ \#,
  moredelim=*[directive]\ \ \ \ \#,
  mathescape=true
}

\newcommand{\doublecross}[2]{\hyperref[#2]{\textbf{#1}}}
\newcommand{\doublecrosssf}[2]{\hyperref[#2]{\textbf{\textsf{#1}}}}

\newcommand{\gsfitem}[1]{\item[\textbf{\textsf{#1}}\label{#1}]}

\newcommand{\startglossary}{\section{Glossary}\label{glossary}Here we explain some terms that have specific technical definitions in \GB.\begin{description}}
\newcommand{\finishglossary}{\end{description}}

\newcommand{\bcode}{\begin{lstlisting}}
\newcommand{\ecode}{\end{lstlisting}}

\DeclareMathOperator\erf{erf}

\newcommand{\eV}{\ensuremath{\text{e}\mspace{-0.8mu}\text{V}}\xspace}

\newcommand{\GeV}{\text{G\eV}\xspace}

\newcommand{\MSbar}{$\MSBar$\xspace}
\newcommand{\MSBar}{\overline{MS}}
\newcommand{\CL}{\text{CL}\xspace}

\newcommand{\CLsb}{\ensuremath{\CL_{s+b}}\xspace}

\newcommand{\gambit}{\textsf{GAMBIT}\xspace}

\newcommand{\darkbit}{\textsf{DarkBit}\xspace}
\newcommand{\colliderbit}{\textsf{ColliderBit}\xspace}
\newcommand{\flavbit}{\textsf{FlavBit}\xspace}
\newcommand{\specbit}{\textsf{SpecBit}\xspace}
\newcommand{\decaybit}{\textsf{DecayBit}\xspace}
\newcommand{\precisionbit}{\textsf{PrecisionBit}\xspace}
\newcommand{\scannerbit}{\textsf{ScannerBit}\xspace}

\newcommand{\GB}{\gambit}

\newcommand{\omp}{\textsf{OpenMP}\xspace}

\newcommand{\buckfast}{\textsf{BuckFast}\xspace}

\newcommand{\pythiaeight}{\textsf{Pythia\,8}\xspace}

\newcommand{\higgsbounds}{\textsf{HiggsBounds}\xspace}

\newcommand{\higgssignals}{\textsf{HiggsSignals}\xspace}
\newcommand{\ds}{\textsf{DarkSUSY}\xspace}
\newcommand{\darksusy}{\ds}
\newcommand{\wimpsim}{\textsf{WimpSim}\xspace}

\newcommand{\feynhiggs}{\textsf{FeynHiggs}\xspace}
\newcommand{\FH}{\feynhiggs}

\newcommand\FS{\FlexibleSUSY}
\newcommand\flexiblesusy{\FlexibleSUSY}
\newcommand\FlexibleSUSY{\textsf{FlexibleSUSY}\xspace}

\newcommand\SOFTSUSY{\textsf{SOFTSUSY}\xspace}

\newcommand\HDECAY{\textsf{HDECAY}\xspace}

\newcommand\SUSYHIT{\textsf{SUSY-HIT}\xspace}

\newcommand\susyhit{\SUSYHIT}
\newcommand\gmtwocalc{\textsf{GM2Calc}\xspace}
\newcommand\SARAH{\textsf{SARAH}\xspace}

\newcommand\superiso{\textsf{SuperIso}\xspace}

\newcommand\FeynHiggs{\textsf{FeynHiggs}\xspace}

\newcommand\nulike{\textsf{nulike}\xspace}
\newcommand\gamLike{\textsf{gamLike}\xspace}
\newcommand\gamlike{\gamLike}

\newcommand\MultiNest{\textsf{MultiNest}\xspace}
\newcommand\multinest{\MultiNest}

\newcommand\diver{\textsf{Diver}\xspace}
\newcommand\ddcalc{\textsf{DDCalc}\xspace}

\newcommand\beq{\begin{equation}}
\newcommand\eeq{\end{equation}}

\renewcommand{\url}[1]{\href{#1}{#1}}

\def\at{\alpha_t}
\def\ab{\alpha_b}
\def\as{\alpha_s}
\def\atau{\alpha_{\tau}}

\def\oatab{\mathcal{O}(\at\ab)}
\def\oatas{\mathcal{O}(\at\as)}
\def\oabas{\mathcal{O}(\ab\as)}

\def\oatq{\mathcal{O}(\at^2)}
\def\oabq{\mathcal{O}(\ab^2)}
\def\oatauq{\mathcal{O}(\atau^2)}

\begin{document}

\preprintnumber{CERN-TH-2017-168, CoEPP-MN-17-9, NORDITA 2017-080}

\title{Global fits of GUT-scale SUSY models with GAMBIT}

\author
{
The GAMBIT Collaboration:
Peter Athron\thanksref{inst:a,inst:b,e1} \and
Csaba Bal\'azs\thanksref{inst:a,inst:b} \and
Torsten Bringmann\thanksref{inst:c} \and
Andy Buckley\thanksref{inst:d} \and
Marcin Chrz\k{a}szcz\thanksref{inst:e,inst:f} \and
Jan Conrad\thanksref{inst:g,inst:h} \and
Jonathan M.~Cornell\thanksref{inst:i} \and
Lars A.~Dal\thanksref{inst:c} \and
Joakim Edsj\"o\thanksref{inst:g,inst:h} \and
Ben Farmer\thanksref{inst:g,inst:h,e2} \and
Paul Jackson\thanksref{inst:k,inst:b} \and
Abram Krislock\thanksref{inst:c} \and
Anders Kvellestad\thanksref{inst:m,e3} \and
Farvah Mahmoudi\thanksref{inst:n,inst:o,e6} \and
Gregory D.\ Martinez\thanksref{inst:p} \and
Antje Putze\thanksref{inst:r} \and
Are Raklev\thanksref{inst:c} \and
Christopher Rogan\thanksref{inst:s} \and
Roberto Ruiz de Austri\thanksref{inst:v} \and
Aldo Saavedra\thanksref{inst:t,inst:b} \and
Christopher Savage\thanksref{inst:m} \and
Pat Scott\thanksref{inst:q,e4} \and
Nicola Serra\thanksref{inst:e} \and
Christoph Weniger\thanksref{inst:u} \and
Martin White\thanksref{inst:k,inst:b,e5}
}

\institute{%
  \monash\label{inst:a} \and
  \coepp\label{inst:b} \and
  \oslo\label{inst:c} \and
  \glasgow\label{inst:d} \and
  \zurich\label{inst:e} \and
  \krakow\label{inst:f} \and
  \okc\label{inst:g} \and
  \su\label{inst:h} \and
  \mcgill\label{inst:i} \and
  \adelaide\label{inst:k} \and
  \nordita\label{inst:m} \and
  \lyon\label{inst:n} \and
  \cernth\label{inst:o} \and
  \ucla\label{inst:p} \and
  \annecy\label{inst:r} \and
  \harvard\label{inst:s} \and
  \valencia\label{inst:v} \and
  \sydney\label{inst:t} \and
  \imperial\label{inst:q} \and
  \grappa\label{inst:u}
}

\thankstext{e1}{peter.athron@coepp.org.au}
\thankstext{e2}{benjamin.farmer@fysik.su.se}
\thankstext{e3}{anders.kvellestad@nordita.org}
\thankstext{e4}{p.scott@imperial.ac.uk}
\thankstext{e5}{martin.white@adelaide.edu.au}
\thankstext[*]{e6}{Also \iuf.}

\titlerunning{GUT-scale SUSY with GAMBIT}
\authorrunning{The GAMBIT Collaboration}

\date{Received: date / Accepted: date}

\maketitle

\begin{abstract}
We present the most comprehensive global fits to date of three supersymmetric models motivated by grand unification: the Constrained Minimal Supersymmetric Standard Model (CMSSM), and its Non-Universal Higgs Mass generalisations NUHM1 and NUHM2.  We include likelihoods from a number of direct and indirect dark matter searches, a large collection of electroweak precision and flavour observables, direct searches for supersymmetry at LEP and Runs I and II of the LHC, and constraints from Higgs observables.  Our analysis improves on existing results not only in terms of the number of included observables, but also in the level of detail with which we treat them, our sampling techniques for scanning the parameter space, and our treatment of nuisance parameters. We show that stau co-annihilation is now ruled out in the CMSSM at more than 95\% confidence.  Stop co-annihilation turns out to be one of the most promising mechanisms for achieving an appropriate relic density of dark matter in all three models, whilst avoiding all other constraints.  We find high-likelihood regions of parameter space featuring light stops and charginos, making them potentially detectable in the near future at the LHC.  We also show that tonne-scale direct detection will play a largely complementary role, probing large parts of the remaining viable parameter space, including essentially all models with multi-TeV neutralinos.

\end{abstract}

\tableofcontents

\section{Introduction}
\label{intro}

Although the Standard Model (SM) of particle physics has long provided a spectacularly successful
description of physics at and below the electroweak scale, it remains incomplete.
Explaining dark matter (DM), the asymmetry between matter and antimatter, the hierarchy
between the Planck and electroweak scales, the origin of the
fundamental forces and charges, or anomalies in low-energy precision and flavour
measurements, requires extending the SM by adding one or more new particles.

The Minimal Supersymmetric extension of the SM (MSSM)
offers solutions to many of these shortcomings, with substantial implications for dark
matter~\cite{Profumo:2016zxo, Roy:2016zst, arXiv:1602.08103,
arXiv:1602.01030, arXiv:1602.00590, arXiv:1601.04718, IC79_SUSY,
arXiv:1511.05964, arXiv:1511.05386, arXiv:1510.07616,
arXiv:1510.07501, arXiv:1510.06295, arXiv:1510.05378,
arXiv:1510.04291, arXiv:1510.03498, arXiv:1510.03460,
arXiv:1510.02470, arXiv:1510.02473, arXiv:1509.09159,
arXiv:1509.05076, arXiv:1508.04383, arXiv:1508.04373,
arXiv:1507.06164, arXiv:1412.4789, arXiv:1507.05584, arXiv:1507.04644,
arXiv:1505.04595, Crivellin:2015oha, arXiv:1504.05554,
arXiv:1504.05091, arXiv:1504.00915, arXiv:1504.00504,
arXiv:1503.07142, arXiv:1503.03478, arXiv:1503.00599,
arXiv:1502.06000, arXiv:1502.05703, arXiv:1502.05672,
arXiv:1502.05406, arXiv:1412.8698}, the cosmic matter-antimatter
asymmetry~\cite{arXiv:1512.09172, arXiv:1508.04144, arXiv:1508.00011},
Higgs physics~\cite{Athron:2016fuq,
arXiv:1608.02573, Bahl:2016brp, arXiv:1608.00638, arXiv:1601.01890,
arXiv:1512.00437, arXiv:1511.08461, arXiv:1511.07853,
arXiv:1511.06002, arXiv:1507.04469, arXiv:1506.08462,
arXiv:1505.01059, arXiv:1504.06932, arXiv:1504.06625,
arXiv:1504.05200, arXiv:1504.04308, arXiv:1502.05653}, the unification
of gauge forces~\cite{arXiv:1702.05431, arXiv:1611.08341,
arXiv:1610.10084, Ellis:2016tjc, arXiv:1512.09148,
arXiv:1511.06205, arXiv:1509.08838, arXiv:1508.04176,
arXiv:1506.05962, arXiv:1506.05850, arXiv:1505.04950,
arXiv:1504.00904, arXiv:1504.00505, arXiv:1501.05307,
arXiv:1501.02906, arXiv:1412.5766}, the stability of the electroweak
vacuum~\cite{arXiv:1606.08356, Bagnaschi:2015pwa,
Blinov:2013fta,Carena:2012mw}, cosmological
inflation~\cite{Ferrara:2016vzg, arXiv:1503.08867, arXiv:1407.4110,
arXiv:1405.4125, arXiv:1312.3623, arXiv:1309.7788, arXiv:1305.1066,
arXiv:1304.5202, arXiv:1303.5351, arXiv:1205.2815}, precision measurements \cite{Kobakhidze:2016mdx, gm2calc, arXiv:1510.04263,
arXiv:1507.05836, arXiv:1505.01987, arXiv:1504.05500,
arXiv:1503.08703, arXiv:1503.08219, arXiv:1503.06850} and
flavor physics \cite{arXiv:1509.05414, arXiv:1504.00930, arXiv:1501.02044}.

Even though the MSSM framework is predictive, its Lagrangian terms responsible for softly breaking SUSY contain over a hundred new parameters.  This impairs the practical predictivity of the model.  Mediation of
supersymmetry breaking by Planck-scale physics is a popular and viable motivation for reducing the free parameters to a small number at the Grand
Unified Theory (GUT) scale \cite{Chamseddine:1982jx,Barbieri:1982eh,Ibanez:1982ee,Hall:1983iz,Ellis:1982wr,AlvarezGaume:1983gj}. Here we analyse three scenarios motivated by gravity
mediation: the Constrained MSSM (CMSSM)~\cite{Nilles:1983ge} and two
of its Non-Universal Higgs Mass (NUHM1, NUHM2)
extensions~\cite{Matalliotakis:1994ft,Olechowski:1994gm,Berezinsky:1995cj,Drees:1996pk,Nath:1997qm}.

Opinions differ regarding the phenomenological feasibility of these
models, especially the CMSSM.  Global fits of the CMSSM after
Run I of the Large Hadron Collider (LHC) indicate that its
experimentally-viable parameter space has been pushed to regions with
superpartners heavier than 1\,TeV.  The relic density of DM in these scenarios is
set by neutralino-stau co-annihilation, resonant annihilation through a heavy
Higgs, or a large Higgsino component~\cite{Han:2016gvr, Bechtle:2014yna, arXiv:1405.4289, arXiv:1402.5419, MastercodeCMSSM, arXiv:1312.5233, arXiv:1310.3045, arXiv:1309.6958, arXiv:1307.3383, arXiv:1304.5526, arXiv:1212.2886, Strege13, Gladyshev:2012xq, Kowalska:2012gs, MasterCode12b, arXiv:1207.1839, arXiv:1207.4846, Roszkowski12, SuperbayesHiggs, Fittino12, MasterCode12, arXiv:1111.6098, Fittino, Trotta08, Ruiz06, Allanach06, Fittino06, Baltz04, SFitter}.  Relaxing the assumption of scalar soft-mass
universality at the GUT scale provides much more flexible
phenomenology \cite{arXiv:1509.02929,
Mastercode15,arXiv:1502.04127, arXiv:1412.3403,
Buchmueller:2014yva,arXiv:1407.1481, arXiv:1404.2277, arXiv:1312.5233,
arXiv:1309.2984, arXiv:1309.0036, arXiv:1307.0782,
arXiv:1306.0344,Fittino12}. Global fits of the NUHM1 show that
the tension that exists in the CMSSM between the measured Higgs mass
and precision/flavour observables is reduced due to the decoupling of
the Higgs sector from the squark and slepton sectors, and a new region
of chargino co-annihilation opens up for dark
matter~\cite{Mastercode15, Buchmueller:2014yva, arXiv:1405.4289, MastercodeCMSSM, arXiv:1312.5233, Strege13, Fittino12}. Extending the parameter space to the
NUHM2~\cite{Mastercode15, Buchmueller:2014yva, arXiv:1405.4289} relaxes the constraints on the scalar masses even further.

In this work we use the \gambit framework~\cite{gambit,DarkBit,ColliderBit,FlavBit,SDPBit,ScannerBit} to scan and assess the
viability of the parameter spaces of each of these three GUT-scale
scenarios in detail. We also carry out a detailed comparison of our results with previous ones, to understand the impact
of the improved theoretical calculations and updated experimental data that we include, and as a verification of our new computational framework.  We have also carried out similar analyses of scalar singlet DM \cite{SSDM} and `phenomenological' (weak-scale) SUSY models \cite{MSSM} with the GAMBIT framework.

There are several important features of our study that make it the most definitive exploration of the CMSSM, NUHM1 or NUHM2 to date:
\begin{enumerate}
\item We apply the DM relic density constraint as an upper bound only.  This requires that the cosmological density of the lightest neutralino does not exceed the observed density of DM. This is a conservative option from the point of view of excluding a light mass spectrum, as it introduces more possibilities for light Higgsino and light Higgsino-bino DM than in studies where a lower bound is also applied.
\item We include a significantly higher number of observables in our combined likelihood than has been done before.  These include rates in multiple direct and indirect searches for DM, a wide range of LHC sparticle searches and Higgs observables, and an up-to-date set of flavour physics observables and electroweak precision measurements.
\item In addition to improving the quantity of data included in the fit, we have also improved the quality of the typical simulation treatments, including direct Monte Carlo simulation of LHC observables during the global fit, event-level indirect search likelihoods, and direct DM search limits based on rigorous simulation of the relevant experiments.
\item Using \gambit allows us to pursue a thorough theoretical and statistical approach, where theoretical assumptions are consistently treated across different observables and experimental searches.  This includes the accurate treatment, via nuisance parameters, of uncertainties associated with the local DM distribution, nuclear matrix elements relevant for direct detection, and SM parameters.
\item \gambit includes an interface to \diver~\cite{ScannerBit}, a new scanner based on differential evolution, which provides significantly improved sampling performance compared to conventional techniques.  This allows us to more accurately locate and more comprehensively map small regions of high likelihood.
\item The public, open-source nature of \gambit\footnote{\href{http://gambit.hepforge.org}{gambit.hepforge.org}} makes our study transparent, reproducible and extendible by the reader.
\end{enumerate}

In Section~\ref{sec:gen}, we introduce the CMSSM, NUHM1 and NUHM2, along with their parameters, the ranges and priors over which we vary those parameters, and the algorithms and settings that we use for sampling them.  Section~\ref{sec:lnL} contains a summary of the experimental data, observables and likelihood calculations that go into each fit.  We then present our results in Section~\ref{sec:current}, before looking at the implications of our scans for future searches for the models in question, and concluding in Section~\ref{sec:conc}.

All input files, samples and best-fit benchmarks produced for this paper are publicly accessible from \textsf{Zenodo} \cite{the_gambit_collaboration_2017_801642}.

\section{Models and scanning framework}
\label{sec:gen}

\subsection{Model definitions and parameters}
\label{sec:models}

From a statistical standpoint, there is no fundamental difference between models that describe SM physics (and astrophysics) and physics beyond the SM (BSM).  \GB therefore treats BSM models on exactly the same footing as models that describe nuisance parameters, which are designed to quantify uncertainties on better-constrained quantities. The only difference is that nuisance models are generally more strongly constrained by the likelihood than BSM models.

In this paper, we simultaneously sample from four models in each scan: one GUT-scale SUSY model (CMSSM, NUHM1 or NUHM2; Sec.~\ref{sec:susymodels}), and three specific nuisance models.  The first nuisance model includes the parameters of the SM (Sec.~\ref{sec:sm}), the second parameterises the density and velocity distribution of the DM halo (Sec.~\ref{sec:astro}), and the third encapsulates the nuclear uncertainties relevant for DM direct detection (Sec.~\ref{sec:nuclear}).

\subsubsection{SUSY models}
\label{sec:susymodels}

The definitions of the MSSM superpotential and soft-breaking Lagrangian that we use are specified in Sec.~5.4.3 of Ref.~\cite{gambit}, and we follow the conventions established there.  All the BSM models that we investigate in this paper are subsets of the \gambit model \textsf{MSSM63atMGUT} \cite{gambit}, which is the most general formulation of the $CP$-conserving MSSM, with the soft masses defined at the scale where the gauge couplings $g_1$ and $g_2$ unify (the GUT scale).

The complexity of the MSSM can be reduced considerably if one makes simplifying assumptions about the values of the soft masses at the GUT scale:
\begin{description}
\gsfitem{CMSSM}  The soft mass parameters at the GUT scale are fixed to a universal scalar mass $m_0$, a universal gaugino mass $m_{1/2}$ and a universal trilinear coupling $A_0$. The diagonal elements in the sfermion mass-squared matrices $\mathbf{m}^\mathbf{2}_Q$, $\mathbf{m}^\mathbf{2}_u$, $\mathbf{m}^\mathbf{2}_d$, $\mathbf{m}^\mathbf{2}_L$ and $\mathbf{m}^\mathbf{2}_e$ are set to $m^2_0$, all off-diagonal elements are set to zero, and the scalar Higgs mass-squared parameters $m_{H_u}^2$ and $m_{H_d}^2$ are set to $m^2_0$. The gaugino masses $M_1$, $M_2$ and $M_3$ are set to $m_{1/2}$ and all trilinear couplings, $(\mathbf{A}_u)_{ij}, (\mathbf{A}_d)_{ij}$ and $(\mathbf{A}_e)_{ij}$ are set to $A_0$. Electroweak symmetry breaking (EWSB) conditions fix the soft-breaking bilinear $b$ and the magnitude of the superpotential bilinear $\mu$ at the SUSY scale. The remaining free parameters in the Higgs sector are the sign of $\mu$ and the ratio of the vacuum expectation values of the two Higgs doublets $\tan\beta\equiv v_\mathrm{u}/v_\mathrm{d}$, which is defined at the scale $m_Z$.  The CMSSM is defined by $m_0$, $m_{1/2}$, $A_0$, $\tan\beta(m_Z)$, and $\text{sgn}(\mu)$.
\gsfitem{NUHM1} The GUT-scale constraint on the soft scalar Higgs masses is relaxed, introducing the additional free parameter $m_H$. The soft Higgs masses $m_{H_u}$ and $m_{H_d}$ are not set equal to $m_0$, but instead obey the relation $m_{H_u}=m_{H_d}\equiv m_H$ at the GUT scale.  Here $m_H$ is treated as a real dimension-one parameter, ignoring scenarios where $m_H^2 < 0$. This means that at the GUT scale we require $m_{H_u}^2 = m_{H_d}^2  > 0$, which acts as the boundary condition under which the correct shape of the Higgs potential must be radiatively generated at the electroweak scale.  The parameters of the NUHM1 are $m_0$, $m_{1/2}$, $A_0$, $\tan\beta(m_Z)$, $\text{sgn}(\mu)$ and $m_{H}$.
\gsfitem{NUHM2} The constraint on the soft Higgs masses is further relaxed so that $m_{H_u}$ and $m_{H_d}$ become independent, real, dimension-one parameters at the GUT scale.  As in the NUHM1, $m_{H_u}^2$ and $m_{H_d}^2$ are always positive at the GUT scale, and the correct shape of the Higgs potential at the electroweak scale must be radiatively generated. The parameters are thus $m_0$, $m_{1/2}$, $A_0$, $\tan\beta(m_Z)$, $\text{sgn}(\mu)$, $m_{H_u}$ and $m_{H_d}$.
\end{description}

We assume throughout that $R$-parity is conserved, making the lightest supersymmetric particle (LSP) stable.  In this paper we consider only the possibility of neutralino LSPs, assigning zero likelihood to all parameter combinations where this is not the case.  Sneutrino DM in the MSSM \cite{Falk:1994es} is now essentially ruled out by direct detection, though it remains viable in MSSM extensions (see Ref.\ \cite{Arina:2007tm} for a review).  Gravitino LSP scenarios (e.g. \cite{Roszkowski:2004jd,Arvey:2015nra}) are still viable even in the CMSSM, so adding such models to the results that we present here would be an interesting future extension.

\begin{table}
\begin{center}
\begin{tabular}{l c c c}
\hline
Parameter & Minimum & Maximum & Priors    \\
\hline
\textbf{CMSSM}          &           &         &               \\
$m_0$                   & 50\,GeV   & 10\,TeV & flat, log     \\
$m_{1/2}$               & 50\,GeV   & 10\,TeV & flat, log     \\
$A_0$                   & $-$10\,TeV& 10\,TeV & flat, hybrid  \\
$\tan\beta$             & 3         & 70      & flat          \\
$\mathrm{sgn}(\mu)$     & $-$       & $+$     & binary        \\
\hline
\multicolumn{4}{l}{\textbf{NUHM1} -- as per CMSSM plus}       \\
$m_H$                   & 50\,GeV   & 10\,TeV & flat, log     \\
\hline
\multicolumn{4}{l}{\textbf{NUHM2} -- as per CMSSM plus}       \\
$m_{H_u}$               & 50\,GeV   & 10\,TeV & flat, log     \\
$m_{H_d}$               & 50\,GeV   & 10\,TeV & flat, log     \\
\hline
\end{tabular}
\caption{\label{tab:param} CMSSM, NUHM1 and NUHM2 parameters, ranges and priors adopted in the different scans contributing to the final results of this paper.  The ``hybrid'' prior for $A_0$ is flat where $|A_0| < 100$\,GeV, and logarithmic elsewhere. The ``binary'' prior for $\mathrm{sgn}(\mu)$ indicates that we repeated every scan for each sign. In addition to the listed priors, we also performed supplementary scans restricted to models with either $m_{\tilde{l}_1} < 1.5\,m_{\tilde{\chi}_1^0}$ or $m_{\tilde{u}_1} < 1.5\,m_{\tilde{\chi}_1^0}$. Details can be found in Sec.~\ref{sec:scan_alg}. \label{tab:SUSY_parameters}}
\end{center}
\end{table}

The parameter ranges that we scan over for the CMSSM, NUHM1 and NUHM2 can be found in Table~\ref{tab:SUSY_parameters}.  We allow the magnitudes of all dimensionful parameters to vary between 50\,GeV and 10\,TeV.  The lower cutoff is motivated by the constraints on sparticle masses from existing searches. The upper cutoff is somewhat arbitrary, but designed to encompass the mass range interesting for solving the hierarchy problem, and for leading to potentially-observable phenomenology.  We consider both positive and negative $\mu$, and the full range of $\tan \beta$ over which particle spectra can be consistently calculated and EWSB achieved in such models.

\subsubsection{Standard Model}
\label{sec:sm}

Here we define the SM as per SLHA2 \cite{Allanach:2008qq}, sampling from the \gambit model \textsf{StandardModel\_SLHA2} \cite{gambit}.  We identify the strength of the strong coupling at the scale of the $Z$ mass, $\alpha_s(m_Z)$, and the top quark pole mass, $m_t$, as the most relevant nuisance parameters within this model. Both affect the running of soft-breaking masses from the GUT scale.  The mass of the SM-like Higgs boson is also very sensitive to the top quark mass, and has a strong influence on the scan through the Higgs likelihood (see Sec.~\ref{sec:higgs_like}).

In all our fits, we allow both these parameters to vary within $\pm3\sigma$ of their observed central values \cite{ATLAS:2014wva,PDB}.  The resulting parameter ranges are shown in Table~\ref{tab:nuisance}.  We adopt flat priors on both $\alpha_s$ and $m_t$; their values are sufficiently well-determined that the prior has no impact on results.  The values of other SM parameters that we keep fixed in our scans are also shown in Table~\ref{tab:nuisance}.

\begin{table}[tp]
\begin{center}
\begin{tabular}{l@{\hspace{-5mm}}c@{\,}r}
\hline
Parameter & & Value($\pm$Range) \\
\hline
\textbf{Varied} & & \\
Strong coupling & $\alpha_s^{\MSBar}(m_Z)$      & $0.1185(18)$   \\
Top quark pole mass  & \phantom{$^{\MSBar}$}$m_t$\phantom{$^{\MSBar}$}  &  $173.34(2.28)$\,GeV\\
Local DM density & \phantom{$^{\MSBar}$}$\rho_0$\phantom{$^{\MSBar}$} &  0.2--0.8\,GeV\,cm$^{-3}$\\
Nuclear matrix el. (strange)  & \phantom{$^{\MSBar}$}$\sigma_s$\phantom{$^{\MSBar}$} & $43(24)$\,MeV \\
Nuclear matrix el. (up + down) & \phantom{$^{\MSBar}$}$\sigma_l$\phantom{$^{\MSBar}$} & $58(27)$\,MeV \\
\hline
\textbf{Fixed} & & \\
Electromagnetic coupling & $1/\alpha^{\MSBar}(m_Z)$        & $127.940$       \\
Fermi coupling $\times$ $10^{5}$ & \phantom{$^{\MSBar}$}$G_{F,5}$\phantom{$^{\MSBar}$} & $1.1663787$ \\
Z pole mass  & \phantom{$^{\MSBar}$}$m_Z$\phantom{$^{\MSBar}$}  &  $91.1876$\,GeV\\
Bottom quark mass & $m_b^{\MSBar}(m_b)$    & $4.18$\,GeV \\
Charm quark mass & $m_c^{\MSBar}(m_c)$ & $1.275$\,GeV \\
Strange quark mass & $m_s^{\MSBar}(2\,\text{GeV})$  & $95$\,MeV\\
Down quark mass & $m_d^{\MSBar}(2\,\text{GeV})$  &   $4.80$\,MeV  \\
Up quark mass & $m_u^{\MSBar}(2\,\text{GeV})$     & $2.30$\,MeV \\
$\tau$ pole mass  & \phantom{$^{\MSBar}$}$m_\tau$\phantom{$^{\MSBar}$}  &  $1.77682$\,GeV\\
CKM Wolfenstein parameters: & \phantom{$^{\MSBar}$}$\lambda$\phantom{$^{\MSBar}$} & 0.22537 \\
 & \phantom{$^{\MSBar}$}$A$\phantom{$^{\MSBar}$} & 0.814 \\
 & \phantom{$^{\MSBar}$}$\bar\rho$\phantom{$^{\MSBar}$} & 0.117 \\
 & \phantom{$^{\MSBar}$}$\bar\eta$\phantom{$^{\MSBar}$} & 0.353 \\
Most probable halo speed & \phantom{$^{\MSBar}$}$v_0$\phantom{$^{\MSBar}$} & 235\,km\,s$^{-1}$ \\
Local disk circular velocity & \phantom{$^{\MSBar}$}$v_{\rm rot}$\phantom{$^{\MSBar}$} & 235\,km\,s$^{-1}$ \\
Local escape velocity & \phantom{$^{\MSBar}$}$v_{\rm esc}$\phantom{$^{\MSBar}$} & 550\,km\,s$^{-1}$ \\
Up contribution to proton spin & \phantom{$^{\MSBar}$}$\Delta^{(p)}_u$\phantom{$^{\MSBar}$} & 0.842 \\
Down contrib. to proton spin & \phantom{$^{\MSBar}$}$\Delta^{(p)}_d$\phantom{$^{\MSBar}$} & $-$0.427 \\
Strange contrib. to proton spin & \phantom{$^{\MSBar}$}$\Delta^{(p)}_s$\phantom{$^{\MSBar}$} & $-$0.085 \\
\hline\end{tabular}
\caption{Standard Model, dark matter halo and nuclear nuisance parameters and ranges.  We vary each of the parameters in the first section of the table simultaneously with CMSSM, NUHM1 or NUHM2 parameters in all of our fits, employing flat priors on each.  The parameters listed in the second section of the table are constant in all scans.} \label{tab:nuisance}
\end{center}
\end{table}

\subsubsection{Dark matter halo model}
\label{sec:astro}

The density and velocity distributions that characterise the DM halo of the Milky Way constitute an important source of uncertainty for astrophysical observations, particularly direct and indirect searches for DM.  In this paper, we employ the \GB model \textsf{Halo\_gNFW\_rho0}~\cite{gambit} to describe the halo.  This consists of a generalised NFW \cite{Navarro:1995iw} spatial profile, tied to a locally Maxwell-Boltzmann velocity distribution by a specific input local density $\rho_0$.

Because we do not employ any observables in our fits that depend on the Milky Way density profile, the spatial part of this model plays no role. The local distribution of DM velocities $\vec{v}$ is given by
\begin{equation}
\label{eq:MB}
\tilde{f}(\mathbf{v}) = \frac{1}{N_{\rm esc}} (\pi v_0^2)^{-3/2} e^{-\mathbf{v}^2/v_0^2} \, ,
\end{equation}
where $v_{\rm esc}$ is the local Galactic escape velocity, $v_0$ is the most probable particle speed and
\begin{equation}
N_{\rm esc} \equiv \erf \left(\frac{v_{\rm esc}}{v_0} \right) - \frac{2 v_{\rm esc}}{\sqrt{\pi} v_0} \exp \left(- \frac{v_{\rm esc}^2}{v_0^2} \right) \,,
\end{equation}
is the normalisation factor induced by truncating the distribution at $v_{\rm esc}$.

In the Earth's rest frame, DM particles have a velocity distribution given by:
\begin{equation}
f(\mathbf{u}, t) = \tilde{f}(\mathbf{v}_{\rm obs}(t) + \mathbf{u}) \, ,
\end{equation}
where $\mathbf{v}_{\rm obs}$ is the velocity of the Earth relative to the Milky Way DM halo. This is given by:
\begin{equation}
\mathbf{v}_{\rm obs} = \mathbf{v}_{\rm LSR} + \mathbf{v}_{\rm \odot, pec} + \mathbf{V}_\oplus(t) \, ,
\end{equation}
where $\mathbf{v}_{\rm \odot, pec} = (11, 12, 7)$\,km\,s$^{-1}$ is the peculiar velocity of the Sun, which is known with very high precision~\cite{Schoenrich:2009bx}. The Local Standard of Rest (LSR) in Galactic coordinates moves with a velocity $\mathbf{v}_{\rm LSR} = (0, v_{\rm rot}, 0)$, while $\mathbf{V}_\oplus(t) = 29.78$\,km\,s$^{-1}$ \cite{Freese:2012xd} denotes the speed of the Earth in the solar rest frame.

For an NFW profile $v_0$ is within 10\% of $v_{\rm rot}$.  As shown in Table~\ref{tab:nuisance}, we set both these parameters to 235\,km\,s$^{-1}$ \cite{DarkBit,Reid:2009nj,Bovy:2009dr} in all our scans.  Similarly, we adopt a fixed value of 550\,km\,s$^{-1}$ for the local escape speed \cite{Smith:2006ym}.

Because it has a substantial impact on direct detection and high-energy solar neutrino signals from DM, we vary the local density of DM as a nuisance parameter in all scans (Table \ref{tab:nuisance}).  Here we adopt an asymmetric range of $+0.4$ $-0.2$\,GeV\,cm$^{-3}$ around the canonical value of $\rho_0 = 0.4$\,GeV\,cm$^{-3}$, reflecting the log-normal form of the likelihood that we apply to this parameter (see Sec.\ \ref{halo_likelihood}).  The prior on $\rho_0$ has no impact because it is sufficiently well-constrained by the associated nuisance likelihood; we choose to make it flat.

See Refs.\ \cite{DarkBit,Akrami:2010dn} for further discussion and details of the DM halo model, parameters and uncertainties.

\subsubsection{Nuclear model}
\label{sec:nuclear}

A final class of uncertainty relevant for direct detection and neutralino capture by the Sun is due to the effective nuclear couplings in WIMP-nucleon cross-sections. For spin-independent interactions, these depend on the light-quark composition of the proton and the neutron.  We scan the \GB model \textsf{nuclear\_params\_sigmas\_sigmal}, parameterising the 6 individual hadronic matrix elements in terms of just two nuclear matrix elements
\begin{align}
\sigma_l &\equiv \tfrac12 (m_u + m_d) \langle N | \bar{u}u + \bar{d}d | N \rangle, \\
\sigma_s &\equiv m_s \langle N | \bar{s}s | N \rangle \, ,
\end{align}
which we take to be identical for $N=p$ and $N=n$~\cite{Young:2013nn}.  These two parameters, respectively, describe the light-quark and strange-quark contents of the nucleus.  We vary $\sigma_l$ and $\sigma_0$ over their $\pm3\sigma$ ranges in all fits.  Discussion of the values and uncertainties of these parameters can be found in Sec.\ \ref{nuclear_likelihood} and the \darkbit paper \cite{DarkBit}.  Like all other nuisance parameters listed in Table\ \ref{tab:nuisance}, the nuclear matrix elements are sufficiently well constrained that the prior is irrelevant, so we choose it to be flat.

The spin-dependent couplings are described by the spin content of the proton and neutron $\Delta_q^{(N)}$ for each light quark $q\in\{u,d,s \}$. As the values for the proton and neutron are related, only three of these parameters are independent.  As listed in Table\ \ref{tab:nuisance}, we specify the values for the proton, and set them to the central values discussed in Ref.\ \cite{DarkBit}.

\subsection{Scanning methodology}
\label{sec:scan_alg}

In this paper we carry out a number of different scans of each of the three GUT-scale models, employing multiple priors, sampling algorithms and settings.  We then merge the results of all scans for each model, in order to obtain the most complete sampling of the profile likelihood possible.  We leave discussion and presentation of Bayesian posteriors for a future paper, as they remain strongly dominated by the choice of prior even in such low-dimensional versions of the MSSM, and a detailed analysis of the their implications for fine-tuning and naturalness (e.g. \cite{Allanach:2007qk,Cabrera:2008tj}) is beyond the scope of the current paper.

\begin{table}
\begin{center}
\begin{tabular}{l l l}
\hline
Scanner & Parameter & Setting \\
\hline
\diver & \fortran{NP} & 19\,200  \\
                & \fortran{convthresh} &  $10^{-5}$ \\ \hline
\diver & \fortran{NP}& 6000  \\
(co-annihilation)  & \fortran{convthresh} & $10^{-4}$ \\ \hline
\multinest & \fortran{nlive}  &  5000 \\
                       & \fortran{tol } & 0.1 \\
\hline
\end{tabular}
\caption{\label{tab:scans} Samplers and their settings for the different scans of this paper. }
\end{center}
\end{table}

The parameter ranges and priors that we employ in scans of each model are listed in Table~\ref{tab:SUSY_parameters}.  We repeat every scan for positive and negative values of $\mu$.  We carry out scans with both flat and logarithmic priors on all dimensionful parameters.  In the case of the trilinear coupling $A_0$, which may be positive or negative, in our log-prior scans we employ a hybrid prior (\textsf{log\_flat\_join} in the language of Ref.~\cite{ScannerBit}), consisting of a symmetric logarithmic prior at large $|A_0|$, truncated to a flat prior at $|A_0|< 100$\,GeV.

We use two different samplers for our scans: \diver \textsf{1.0.0}~\cite{ScannerBit} and \multinest \textsf{3.10}~\cite{MultiNest}.  The settings that we use for each can be found in Table~\ref{tab:scans}.

\diver is a self-adaptive sampler based on differential evolution \cite{StornPrice95}. It samples the profile likelihood far more efficiently than traditional algorithms \cite{ScannerBit}, allowing high-quality profile likelihoods to be computed in a fraction of the time of previous SUSY global fits, using significantly fewer likelihood samples.  As a result, the majority of our results are driven by the \diver scans.  Following the extensive tests discussed in Ref.\ \cite{ScannerBit}, for most scans\footnote{For the special case of flat-prior CMSSM scans, where less stringent parameters already provide quite sufficient sampling, we use \fortran{NP} = 14\,400, \fortran{convthresh} = 10$^{-4}$.} we choose a population size of \fortran{NP} = 19\,200 and a convergence threshold of \fortran{convthresh} = 10$^{-5}$. The latter is defined in terms of the smoothed fractional improvement in the mean likelihood across the entire population.  Other than these two parameters, we employ \diver with the default settings defined in \scannerbit \cite{ScannerBit}.  In particular, this includes the $\lambda$jDE version (introduced in Ref.~\cite{ScannerBit}) of the self-adaptive jDE algorithm \cite{Brest06}.

\multinest is an implementation of nested sampling \cite{Skilling04}, a method optimised for the calculation of the Bayesian evidence.  As a by-product, it also produces posterior samples, which it obtains via likelihood evaluation.  It can therefore be very useful for sampling profile likelihoods as well, especially for smoothly mapping isolikelihood contours.  However, it typically requires a rather long runtime to properly find the global best fit and any highly-localised likelihood modes \cite{Akrami09,SBSpike,ScannerBit}.  We employ it here mainly to bulk out our sampling of the main likelihood mode of each scan a little, in regions where the profile likelihood is comparatively flat.  For this purpose, we run \multinest with relatively loose settings, choosing \fortran{nlive} = 5000 live points and a stopping tolerance of \fortran{tol} = 0.1.  The tolerance is given in terms of the estimated fractional remaining unsampled evidence.

For more details on the performance of the two scanning algorithms, and comparisons to others, please see the \scannerbit paper~\cite{ScannerBit}.

To more densely sample the narrow strips in parameter space where neutralino-sfermion co-annihilations play an important role in determining the relic density of DM, we also carry out two specially-targeted versions of each log-prior scan.  In these scans, we restrict the mass of either the lightest slepton or the lightest squark to within 50\% of the mass of the lightest neutralino, i.e.\ $m_{\tilde l_1} \le 1.5\,m_{\tilde \chi_1^0}$ or $m_{\tilde q_1} \le 1.5\,m_{\tilde \chi_1^0}$.  Although these additional scans are not necessary for \textit{finding} the sfermion co-annihilation regions (\diver typically uncovers these regions anyway in untargeted scans), they are useful for ensuring that the boundaries of these regions are mapped thoroughly.

We carry out the additional scans using \diver only, building the initial population exclusively from models that satisfy the mass-ratio cut, before evolving it as usual.  As it takes many random draws from the prior to successfully build such an initial population, we run these scans with a reduced population of \fortran{NP} = 6000, and a looser convergence criterion (\fortran{convthresh} = 10$^4$) than the untargeted equivalents.

This results in a total of 3 models $\times$ 2 $\mathrm{sgn}(\mu)$ $\times$ (2 priors $\times$ 2 scanners $+$ 2 targeted co-annihilation scans) = 36 separate scans.  For the CMSSM, NUHM1 and NUHM2, this results in a total of 71, 94 and 117 million viable samples, respectively.  Each of these 36 scans typically took 1--3 days to run on 2400 modern (Intel Core i7) supercomputer cores.

In all profile likelihood plots that we show in the paper, we sort the samples of our scans into 60 bins across the range of data values that they cover in each direction.  We then interpolate with a bilinear scheme to a finer resolution of 500 when plotting \cite{pippi}.  This expressly avoids any smoothing of the resulting profile likelihoods, which would amount to manipulation of the likelihoods of our samples.  The binning and interpolation process can produce some cosmetic artefacts, in particular a sawtooth pattern in regions where the likelihood drops off sharply.  Using a fixed number of bins across the data range (rather than the plot range) can also sometimes produce surprising effects when plotting multiple regions on the same axes, as the same region will typically appear smaller and `smoother' for subsets of the data where all samples lie within a small range of parameter values than for subsets with samples spread across the entire plot plane (as the size of an individual bin is much larger in the latter case).  This should be kept in mind especially when viewing plots of mass correlations for multiple co-annihilation mechanisms and comparing preferred regions to LHC sensitivity curves (e.g.\ Fig.\ \ref{fig:colour-stop}).

\section{Observables and likelihoods}
\label{sec:lnL}

\subsection{Nuisance likelihoods}

\subsubsection{Standard Model}
We include independent Gaussian likelihoods for each of the two SM nuisance parameters in our scans.  We evaluate the strong coupling $\alpha_s$ at the scale $\mu=m_Z$ in the \MSbar scheme, and compare with $\alpha_{s}(m_{Z}) = 0.1185 \pm 0.0006$ from lattice QCD~\cite{PDB}. We interpret the quoted uncertainty as a $1\sigma$ confidence interval, and do not incorporate any additional theoretical uncertainty.

For the top quark pole mass $m_t$, we compare with the combined measurements of experiments at the Tevatron and LHC: $m_t=173.34\pm0.27\text{(stat)}\pm0.71\text{(syst)}$\,GeV, with a total uncertainty of 0.76\,GeV~\cite{ATLAS:2014wva}. We do not assign any separate systematic error to our interpretation of the experimental result as the top pole mass.

\subsubsection{Local halo model}
\label{halo_likelihood}

The canonical local density of DM extracted from fits to stellar kinematic data is $\bar{\rho}_0=0.4$\,GeV\,cm$^{-3}$ (see e.g.\ \cite{Catena:2009mf,Pato:2015dua}).  Because arbitrarily small or negative densities are unphysical, we adopt a log-normal distribution for the likelihood of $\rho_0$,
\begin{equation}
\mathcal{L}_{\rho_0} =  \frac{1}{\sqrt{2\pi} \sigma'_{\rho_0} \rho_0} \exp \left(- \frac{\ln(\rho_0 / \bar\rho_0)^2}{2 {\sigma^{\prime 2}_{\rho_0}}} \right) \, ,
\end{equation}
where $\sigma'_{\rho_0} = \ln(1 + \sigma_{\rho_0}/\rho_0)$ and $\sigma_{\rho_0}$ is taken to be $0.15 \ \mathrm{GeV}/\mathrm{cm}^3$.  We refer the reader to Ref.~\cite{gambit} for additional implementation details of the \GB log-normal likelihood, and Refs.~\cite{DarkBit,Read:2014qva,Akrami:2010dn} for a more extended discussion of the central value and uncertainty on this parameter.

\subsubsection{Nuclear matrix elements}
\label{nuclear_likelihood}

We constrain the nuclear matrix elements $\sigma_s$ and $\sigma_l$ using Gaussian likelihood functions, with central $\pm$ $1\sigma$ values of $43\pm8$\,MeV and $58\pm9$\,MeV, respectively.  The former is based on lattice calculations \cite{Lin:2011ab}, whereas the latter is a weighted average of a number of different results in the literature \cite{DarkBit}.

\subsection{Spectrum calculation}
\label{sec:spectrum}

We use \FS \textsf{1.5.1} \cite{Athron:2014yba} to compute the mass spectrum of the MSSM.  This code obtains model-dependent information
from \SARAH \cite{Staub:2008uz,Staub:2010jh}, and borrows some numerical
routines from \SOFTSUSY \cite{Allanach:2001kg,Allanach:2013kza}.  \FS
employs full three-family, two-loop renormalisation group equations (RGEs) and
full one-loop self-energies and tadpoles.  In addition, it computes the Higgs mass
using two-loop corrections at $\oatas$, $\oabas$,
$\oatq$, $\oabq$, $\oatauq$ and
$\oatab$~\cite{Degrassi:2001yf,Brignole:2001jy,Brignole:2002bz,Dedes:2003km}.

Large logarithms appear when the supersymmetric spectrum is very heavy. To improve precision, these can be
resummed using techniques from effective field theory (EFT) \cite{Draper:2013oza,Bagnaschi:2014rsa,Vega:2015fna,Lee:2015uza,
Bahl:2016brp,Athron:2016fuq}.  This method has been implemented in several public
codes \cite{Vega:2015fna,Bahl:2016brp,Athron:2016fuq}. However, the
hierarchical spectrum assumed in the EFT calculation only appears in small subspaces of the
models over which we scan in this paper.  The public codes that implement this calculation, and that are
suitable for cluster-scale parameter scans, are rather new\footnote{For example, during this work we
found a bug in the resummation that affected results at low
masses when testing with \FeynHiggs~\textsf{2.12.0} --- though this should be
corrected in later versions.}.  It was also pointed out in
Ref.~\cite{Athron:2016fuq} that the accuracy of the fixed-order Higgs
mass prediction in \FS is much better than one would naively expect
at large sparticle masses, due to accidental cancellations. We have therefore retained
the fixed-order calculations in \FS for calculating the Higgs mass in this paper.

\subsection{Relic density of dark matter}
\label{sec:Loh2}

The thermal relic abundance of the lightest neutralino is a strong constraint on the MSSM. Many parameter combinations lead to more DM than the cosmological abundance observed by \textit{Planck}, which is $\Omega_{\rm c}h^2 = 0.1188\pm0.0010$ \cite{Planck15cosmo}.  For the relic density not to exceed this value, if the lightest
neutralino is heavier than $\sim$100\,GeV,  typically one or more specific depletion mechanisms must be active.  These include co-annihilation of light sfermions or charginos with the lightest neutralino, and resonance or `funnel' effects, where the lightest neutralino has a mass very close to half that of another neutral species.

We compute the relic density of each model taking into account DM annihilation to all two-body final states, 
 including full co-annihilation \cite{Edsjo:1997bg,Edsjo:2003us}, thermal and resonance effects, using the native \darkbit relic density calculator \cite{DarkBit}, connected to various subroutines of \darksusy \textsf{5.1.3} \cite{darksusy}.  We obtain the effective annihilation rate $W_{\rm eff}$ from \darksusy, passing all spectrum, decay and SM information from \GB, and considering co-annihilations with particles up to 60\% heavier than the lightest neutralino.  We also employ the \darksusy Boltzmann solver, setting the option \cpp{fast = 1}.  This ensures that the relic density calculation for most models takes less than a second.  This setting controls the convergence criteria of the Boltzmann solver, and is the recommended option unless accuracy of better than 1\% is required.

The likelihood that we employ penalises only models that predict more than the observed relic density.  The likelihood function is a half Gaussian (see Ref.\ \cite{DarkBit} and Sec.\ 8.3 of Ref.\ \cite{gambit}), centered on the \textit{Planck} observation but treating it as an upper limit.  Consistent with our choice of the \cpp{fast} parameter for the Boltzmann solver, we retain the \darkbit default theoretical uncertainty of 5\% on the relic density, adding it in quadrature to the observational error.  Further discussion of this number in the context of higher-order corrections can be found in Refs.\ \cite{DarkBit,arXiv:1602.08103,Harz:2014gaa,Harz:2014tma,Harz:2012fz,arXiv:1510.02473,Baro:2007em,Baro:2009na}.

\subsection{Gamma rays from dark matter annihilation}
Neutralino annihilation in astrophysical objects would produce a variety of final states, leading to both prompt gamma rays and those produced as final decay products.  Dwarf spheroidal galaxies are particularly important targets, as they are strongly dominated by dark rather than visible matter, and exhibit little or no astrophysical gamma-ray emission.  Limits from gamma-ray observations of dwarf galaxies have therefore played an increasingly important role in global fits, e.g.\ \cite{Scott09c,Cheung:2012gi,Roszkowski12,Liem}.

For most neutralino masses, the most stringent gamma-ray limits on DM annihilation come from joint analyses of multiple Milky Way satellite galaxies \cite{Geringer-Sameth11,Ackermann:2011wa,Ackermann:2013yva,Geringer-Sameth:2014qqa,LATdwarfP8} using data from the \textit{Fermi} Large Area Telescope (\textit{Fermi}-LAT).  We employ likelihoods from the analysis of six years of \texttt{Pass 8} data in the direction of 15 dwarf spheroidal galaxies \cite{LATdwarfP8}, as implemented in \gamlike \cite{DarkBit}.

\gamlike constructs a composite likelihood
\begin{equation}
  \ln\mathcal{L}_\text{exp} = \sum_{k=1}^{N_\text{dSph}}\sum_{i=1}^{N_\text{ebin}}
  \ln\mathcal{L}_{ki}(\Phi_i \cdot J_k)\;,
  \label{eqn:FermiTabLike}
\end{equation}
from gamma-ray data sorted into $N_\text{dSph}$ fields of view (one for each dwarf) and $N_\text{ebin}$ energy bins. The partial likelihoods $\mathcal{L}_{ki}$ describe the likelihood of obtaining the observed number of photons in the $i$th energy bin from the $k$th dwarf. The energy-dependent factor
\begin{equation}
  \Phi_i = \frac{\langle\sigma v\rangle_0}{8\pi m_\chi^2}\int_{\Delta E_i} dE \,\frac{dN_\gamma}{dE}
  \label{eqn:gamLikePhi}
\end{equation}
depends on the MSSM model, whereas the astrophysical factor
\begin{equation}
  J_k = \int_{\Delta \Omega_k} d\Omega\int_\text{l.o.s.} ds \, \rho_\chi(s)^2
\end{equation}
is a model-independent property of each dwarf galaxy.  Here the differential gamma-ray multiplicity per annihilation is $dN_\gamma/dE$, the zero-velocity annihilation cross-section is $\langle\sigma v\rangle_0\equiv \sigma v|_{v\to0}$, $m_\chi$ is the DM mass, and $\rho_\chi(s)$ is the DM density along the line of sight parameter $s$ in a given dwarf.  $\Delta E_i$ is the width of the $i$th energy bin, and $\Delta \Omega_k$ is the solid angle around the position of the $k$th dwarf over which gamma-ray data are being considered.

As in~\cite{LATdwarfP8}, \gamlike profiles over the $J_k$-factors as nuisance parameters, giving a final likelihood of
\begin{equation}
  \ln \mathcal{L}_\text{dwarfs}^\text{prof.}(\Phi_i) = \max_{J_1\dots J_k}\left(\ln\mathcal{L}_\text{exp} + \ln\mathcal{L}_J\right)\;,
\end{equation}
where
\begin{equation}
  \ln\mathcal{L}_J = \sum_{k=1}^{N_\text{dSph}} \ln\mathcal{N}(\log_{10} J_k | \log_{10} \hat J_k,\sigma_k)\;.
\end{equation}
Here the probability distribution for each $J_k$ is assumed to follow an independent log-normal distribution with mean $\hat J_k$ and width $\sigma_k$.

We compute the predicted spectrum $dN_{\gamma}/dE$ for each model by combining tabulated two-body annihilation spectra from \darksusy~\cite{darksusy} with yields computed on the fly with the \darkbit Fast Cascade Monte Carlo \cite{DarkBit}.  To this, we add the dominant contribution from photon internal bremsstrahlung \cite{Bringmann:2007nk}.
For each parameter combination, we rescale the expected gamma-ray flux by the squared ratio of the predicted relic density to the observed value, allowing for the fact that neutralinos may only be a fraction of DM.  We limit this scaling factor to 1, not rescaling signals when the predicted relic density is greater than the observed value.

\subsection{High-energy neutrinos from dark matter annihilation in the Sun}
\label{sec:neutrinos}

The Sun is expected to capture neutralinos from the local halo by nuclear scattering.  Subsequent neutralino annihilation and interaction of the annihilation products in the solar core would produce GeV-energy neutrinos, which may be detectable at the Earth. The rate-limiting step for all MSSM models is the capture process, which depends sensitively on both the spin-dependent and spin-independent nuclear scattering cross-sections.  The dominant constraints on intermediate and high-mass neutralino annihilation in the solar interior currently come from the IceCube experiment~\cite{IC79,IC86}. We use the \darkbit interface \cite{DarkBit} to \nulike \textsf{1.0.4} \cite{IC22Methods,IC79_SUSY}, which computes an unbinned likelihood from the event-level energy and angular information contained in the three independent event selections of the 79-string IceCube dataset \cite{IC79}. We predict the neutrino spectra at Earth using \darksusy \textsf{5.1.3}, which contains tabulated results previously obtained from \wimpsim~\cite{Blennow08}.\footnote{\wimpsim is available at \href{http://www.fysik.su.se/~edsjo/wimpsim/}{www.fysik.su.se/$\sim$edsjo/wimpsim/}.} Slightly stronger limits are also available from the 86-string dataset \cite{IC86}, but not in a format that allows them to be accurately applied to MSSM models.

\subsection{Direct detection of dark matter}

The dominant direct DM constraints on the models in this paper come from the LUX~\cite{LUX2013,LUX2016,LUXrun2}, Panda-X \cite{PandaX2016} and PICO~\cite{PICO60,PICO2L} experiments.  We also include likelihoods from XENON100~\cite{XENON2013}, SuperCDMS~\cite{SuperCDMS} and SIMPLE~\cite{SIMPLE2014}. A new analysis from PICO-60 \cite{Amole:2017dex} appeared after much of this paper was already finalised, but the majority of MSSM models susceptible to that limit are already probed in our scans by the IceCube 79-string likelihood (Sec.~\ref{sec:neutrinos}).  We do not include the recent XENON1T result \cite{Aprile:2017iyp}, but given that its sensitivity improvement relative to LUX is smaller than the error in our likelihood approximation \cite{DarkBit}, this will not impact our results.

For each experimental search and combination of MSSM, halo and nuclear parameters, we use the likelihood functions contained in \ddcalc \cite{DarkBit} to compute a Poisson likelihood,
\begin{equation} \label{eqn:Poisson}
  \mathcal{L}_i(N_{\mathrm{p},i}|N_{\mathrm{o},i})
  = \frac{(b_i+N_{\mathrm{p},i})^{N_{\mathrm{o},i}} \, e^{-(b_i+N_{\mathrm{p},i})}}{N_{\mathrm{o},i}!}.
\end{equation}
Here $N_{\mathrm{o},i}$ is the number of observed events in the analysis region of the $i$th experiment, $b_i$ is the expected number of background events, and $N_{\mathrm{p},i}$ is the expected number of signal events. \ddcalc computes the latter by interpolating in pre-computed efficiency tables, which include both detector and acceptance effects.  The signal prediction takes into account both the spin-dependent and spin-independent interactions expected from each MSSM model. We compute the DM-nucleon couplings for each MSSM model using \darksusy \textsf{5.1.3}~\cite{darksusy}.

We scale the direct detection yields for each parameter combination by the ratio of the predicted relic density to the value observed by \textit{Planck} \cite{Planck15cosmo}, allowing for the fact that neutralinos may not constitute all of DM.  We do not rescale direct detection rates when the predicted relic density is larger than the observed value.

\subsection{Electroweak precision observables}
We include likelihooods from \precisionbit \cite{SDPBit} for the $W$ mass and the anomalous magnetic moment of the muon $a_\mu$. These functions construct a basic Gaussian likelihood based on the difference between the calculated and measured value, and combine theoretical and experimental uncertainties in quadrature.

The $W$ mass must be recalculated using the details of the SUSY spectrum. In the present scans, the value of $m_W$ comes from \flexiblesusy.  \specbit assigns a theoretical uncertainty of 10\,MeV to this quantity, based on the size of two-loop corrections \cite{SDPBit}.  \precisionbit compares these to $m_W=80.385\pm 0.015$\,GeV~\cite{PDB}, based on mass measurements and uncertainties from the Tevatron and LEP experiments.

For $a_\mu$, we assume an SM contribution of $a_{\mu,\mathrm{SM}}  = (11659180.2 \pm 4.9) \times 10^{-10}$, which comes from theoretical calculations based on $e^+e^-$ data \cite{1010.4180}. We evaluate the supersymmetric contribution using \gmtwocalc \textsf{1.3.0}~\cite{gm2calc}, which determines an uncertainty on its result by estimating the magnitude of neglected higher-order corrections using the two-loop Barr-Zee corrections \cite{barrzee}. The total predicted value is the sum of the SM and MSSM contributions, and the total uncertainity the sum in quadrature of their individual uncertainties.  We compare this with the experimental measurement of $a_\mu  = (11659208.9 \pm 6.3) \times 10^{-10}$ \cite{PDG10,gm2exp}, where the experimental error is the sum in quadrature of the systematic ($3.3\times10^{-10}$) and statistical ($5.4\times10^{-10}$) contributions.

\subsection{Flavour physics likelihoods}
Scans in this paper include 59 flavour observables from \flavbit \cite{FlavBit}.  These are sorted into four different categories for likelihood calculation:\begin{enumerate}
\item Tree-level leptonic and semi-leptonic $B$ and $D$ meson decays (8 observables).  Branching fractions for $B\to D\mu\nu_\mu$, $B\to D^*\mu\nu_\mu$, $B\to \tau\nu_\tau$, $D\to \mu\nu_\mu$, $D_s\to \mu\nu_\mu$ and $D_s\to \tau\nu_\tau$, as well as ratios $R_D\equiv{\cal B}(B\to D\tau \nu_\tau)/{\cal B}(B\to D\ell \nu_\ell)$ and $R_{D^{*}}\equiv{\cal B}(B\to D^{*}\tau \nu_\tau)/{\cal B}(B\to D^{*}\ell \nu_\ell)$. Here either $\mu$ or $e$ may be substituted for $\ell$, as both are effectively massless in the $B$-meson system.
\item Electroweak penguin decays (48 observables).  Eight observables ($AFB, FL, S_3, S_4, S_5, S_7, S_8$ and $S_9$) for the decay $B^0\to K^{*0}\mu^+\mu^-$, each in six different angular ($q^2$) bins.
\item Rare leptonic $B$ decays (2 observables).  Branching fractions for $B^0\to\mu^+\mu^-$ and $B_s^0\to\mu^+\mu^-$.
\item The branching fraction for $B\to X_s\gamma$, for photon energies $E_\gamma>1.6$\,GeV (1 observable).
\end{enumerate}
All observable predictions draw on \superiso \textsf{3.6} \cite{Mahmoudi:2007vz,Mahmoudi:2008tp}.  We have not included the $B_s$--$\bar{B}_s$ meson mass difference $\Delta M_s$, owing to the fact that it is only calculable within \flavbit via \feynhiggs, which we otherwise avoided for the scans of this paper in the interests of speed, and due to worries about its most recent versions' accuracy in parts of the parameter space (some details of which have been mentioned earlier in this Section).

Recent LHCb results in the exclusive modes have already provided substantial additional constraints as compared to the available inclusive results from the $B$ factories. In particular, several angular observables in the $B^0\to K^{*0} \mu^+\mu^-$ decay have been measured for the first time.

We construct a separate likelihood function for observables in each of the four categories above, including correlated uncertainties on observables within each category wherever warranted.  The likelihood functions consider correlations between experimental measurement errors separately from correlations between theoretical errors (arising from e.g.\ common scale or form factor uncertainties), and then sum them to obtain the final covariance matrix.  \flavbit then computes the likelihood within each category using a $\chi^2$ approximation,
\begin{equation}
\ln\mathcal{L} = -\frac12\chi^2 = -\frac12\sum_{i,j=1}^N (y_i - x_i) (V^{-1})_{ij} (y_j-x_j)\;,
\end{equation}
where $x_i$ and $y_i$ are the experimental measurements and theoretical predictions, respectively, and $V$ is the covariance matrix.

In the first likelihood category, \flavbit includes experimental measurements, correlations and combinations from Refs.\ \cite{Olive:2016xmw,Amhis:2016xyh,HFAG17_moriond} and theoretical uncertainties from Refs.\ \cite{PhysRevD.85.094025,Lattice:2015tia}, supplemented by our own additional calculations with a beta version of \superiso \textsf{3.7}.  The experimental measurements and correlated uncertainties of $B^0\to K^{*0}\mu^+\mu^-$ angular observables come from LHCb \cite{Aaij:2015oid}, and the theoretical errors and correlations from Refs.\ \cite{Hurth:2016fbr,Mahmoudi:2016mgr}.  Data for rare leptonic decays are the latest from LHCb and CMS \cite{Aaij:2017vad,CMS:2014xfa}, and theoretical uncertainties come from Ref.\ \cite{Buras:2012ru}.  For $B\to X_s\gamma$, we use the latest average \cite{Misiak:2017bgg} of measurements by Belle \cite{Saito:2014das,Belle:2016ufb} and Babar \cite{Aubert:2007my,Lees:2012wg,Lees:2012ym}, and a theoretical uncertainty of 7\% \cite{Misiak:2015xwa,Czakon:2015exa}.  More details can be found in the \flavbit paper \cite{FlavBit}.

\begin{table}[tb]
  \centering
  \begin{tabular}{lll}
   \toprule
    Production & Decay & Experiment  \\
    \midrule
    $\tilde l \tilde l^*$ & $\tilde l \to l \tilde\chi^0_1$ +c.c.\ & ALEPH~\cite{Heister:2001nk}, L3~\cite{L3:sleptons_squarks} \\
    $\tilde\chi^0_i\tilde\chi^0_1$ & $\tilde\chi^0_i\to q\bar q\tilde\chi^0_1$  & OPAL~\cite{OPAL:gauginos}\\
    ($i=2,3,4$) &  $\tilde\chi^0_i\to l\bar l \tilde\chi^0_1$ & L3~\cite{L3:gauginos}\\
    $\tilde\chi^+_i\tilde\chi^-_i$ & $\tilde\chi^+_i\tilde\chi^-_i\to q\bar q'q\bar q'\tilde\chi^0_1 \tilde\chi^0_1$ & OPAL~\cite{OPAL:gauginos}\\
    ($i=1,2$)  & $\tilde\chi^+_i\tilde\chi^-_i\to q\bar q' l\nu\tilde\chi^0_1\tilde\chi^0_1$ & OPAL~\cite{OPAL:gauginos}\\
    & $\tilde\chi^+_i\tilde\chi^-_i\to l\nu l\nu\tilde\chi^0_1\tilde\chi^0_1$ & OPAL~\cite{OPAL:gauginos}, L3~\cite{L3:gauginos}\\
    \bottomrule
  \end{tabular}
  \caption{Results from LEP on sparticle pair production used in the scans of this paper. Here $\tilde l = \tilde e_L,\tilde e_R, \tilde\mu_L, \tilde\mu_R, \tilde\tau_1$ or $\tilde\tau_2$.}
  \label{tab:SUSY_LEP}
\end{table}

\subsection{Searches for superpartners at LEP}
Even though they are typically overshadowed by constraints from the LHC, LEP searches can have a significant impact in some parts of the parameter spaces that we consider in this paper. This is especially true for light, highly-degenerate spectra. Direct limits on sparticle production at LEP have typically taken the form of hard lower limits on sparticle masses, at e.g.\ 95\% CL, computed with model-dependent assumptions~\cite{darksusy,Belanger:2001fz}. \colliderbit instead uses the individual cross-section limits on pair production of neutralinos, charginos and sleptons from the ALEPH, L3 and OPAL experiments, as a function of the sparticle masses.

For each MSSM parameter combination, we compute the pair-production cross-sections at LEP for the processes given in Table~\ref{tab:SUSY_LEP}, using the cross-section calculations included in \colliderbit and based on the results of Refs.\ \cite{Dawson:1983fw,Bartl:1985fk,Bartl:1986hp,Bartl:1987zg}. We take the relevant sparticle decay branching fractions from \decaybit (choosing to obtain widths from a suitably-patched \SUSYHIT \textsf{1.5} \cite{SDPBit,Djouadi:2006bz}), and calculate the product of the cross-section and branching fraction for each process. This number can then be compared to digitised LEP cross-section limits in the plane of $m_{\tilde\chi_1^0}$ and the mass of the directly-produced sparticle, interpolating when the masses do not fall exactly on a grid point. This takes care of the mass-dependent experimental acceptance for each parameter point. We then calculate the likelihood of the experimental result assuming a Gaussian form, accounting for the dominant theoretical uncertainty on the signal prediction by varying the mass of the pair-produced sparticles within the uncertainties provided by \specbit. Finally, we multiply the likelihoods from the various experiments and channels, taking them as independent measurements.

Further details of the cross-section and likelihood calculations can be found in the \colliderbit paper~\cite{ColliderBit}.

\subsection{Searches for supersymmetry at the LHC}
Many searches for supersymmetric particles have been performed at the LHC by ATLAS and CMS, in a variety of final states arising from proton--proton collisions at $\sqrt{s}=7$, 8 and 13\,TeV \cite{ATLAS-SUSY-pub,CMS-SUSY-pub}. The results of all searches to date are consistent with the predictions of the SM, placing strong, model-dependent constraints on the mass spectrum of the MSSM. Taking into account the complete list of LHC searches is impractical; here we implement the most constraining analyses:
\begin{enumerate}
\item 0-lepton supersymmetry searches (ATLAS \& CMS, Run I \& Run II). These provide the best constraints on models with a light gluino or one or more light squarks. The analyses look for an excess of events in final states with jets and missing energy, using a variety of kinematic variables~\cite{ATLAS:0LEP_20invfb,ATLAS-CONF-2016-078,CMS-PAS-SUS-16-014}.
\item Third generation squark searches (ATLAS \& CMS, Run I). These searches target stop pair production, with subsequent decay to either a top quark and the lightest neutralino, or to a $b$ quark and a chargino. We include the results of ATLAS searches in 0-, 1- and 2-lepton final states~\cite{ATLAS:0LEPStop_20invfb,ATLAS:1LEPStop_20invfb,ATLAS:2LEPStop_20invfb}, and CMS searches for 1- and 2-lepton final states~\cite{CMS:1LEPDM_20invfb,CMS:2LEPDM_20invfb}. We also include the ATLAS search for direct sbottom production in final states with $b$-jets and missing energy~\cite{ATLAS:2bStop_20invfb}.
\item Multilepton supersymmetry searches (ATLAS and CMS, Run I).  We include 2- and 3-lepton searches by ATLAS~\cite{ATLAS:2LEPEW_20invfb,ATLAS:3LEPEW_20invfb} and the 3-lepton search by CMS~\cite{CMS:3LEPEW_20invfb}. These are typically the most constraining searches for direct production of charginos and neutralinos, and the 2 lepton search is also sensitive to slepton pair production and decay.
\item Dark matter searches (CMS, Run I). We include the CMS monojet search \cite{CMS:MONOJET_20invfb}, which constrains supersymmetric particle production in the case of compressed mass spectra.
\end{enumerate}

We use \colliderbit to calculate the expected signal yield for each combination of model parameters, in each analysis region, using the external Monte Carlo (MC) event generator \pythiaeight~\cite{Sjostrand:2006za,Sjostrand:2014zea}, the native \colliderbit detector parameterisation \buckfast \cite{ColliderBit}, and the \colliderbit implementation of the analysis cuts applied in each LHC paper.  \colliderbit contains a number of code optimisations of the \pythiaeight routines, including parallelisation of the main event loop via \omp.  These modifications make it feasible to run 20\,000 MC events per parameter combination during the global fit itself, as we do here. Due to the computational cost of calculating next-to-leading order (NLO) cross sections, we normalise the signal yields using leading-order (LO) plus leading-log (LL) cross-sections only, as provided by \pythiaeight. For a more exhaustive discussion of this choice see the \colliderbit paper \cite{ColliderBit}.

In a specific signal region with a predicted number of signal events $s$ and an expected number of background events $b$, the likelihood of observing $n$ events is described in \colliderbit by a marginalised form of the Poisson likelihood~\cite{Conrad03,Scott09c,IC22Methods},
\begin{equation}
\label{likelihood}
\mathcal{L}(n|s,b) = \int_0^\infty \frac{\left[\xi(s+b)\right]^{n}e^{-\xi(b+s)}}{n!}P(\xi)\mathrm{d}\xi\,,
\end{equation}
where $\xi$ is a scaling variable with a probability distribution centred on 1, designed to describe the effective rescaling of the signal + background prediction due to systematic uncertainties. Marginalising over $\xi$ this way, it is possible to include the effects of fractional systematic uncertainties on both the signal prediction ($\sigma_s$) and the background estimate ($\sigma_b$).\footnote{Due to our use of LO cross-sections, including a signal systematic associated with finite MC statistics is in practice rather pointless, as with 20\,000 simulated events this is basically always dwarfed by the systematic error associated with neglecting NLO corrections.  Considering that the LO cross-sections in the MSSM are known to almost always lie significantly below the NLO cross-section, our approach is in any case very conservative. In the present scan we have thus set $\sigma_s=0$. For details, see the \colliderbit paper~\cite{ColliderBit}.}

We assume a log-normal distribution for $\xi$,
\begin{equation}
\label{lognormal}
P(\xi|\sigma_\xi) = \frac{1}{\sqrt{2\pi}\sigma_\xi}\frac{1}{\xi}\exp\left[-\frac{1}{2}\left(\frac{\ln\xi}{\sigma_\xi}\right)^2\right],
\end{equation}
where $\sigma_\xi^2 = \sigma_s^2 + \sigma_b^2$.  We compute this integral using the highly-optimised implementation in \nulike \textsf{1.0.4} \cite{IC22Methods, IC79_SUSY}.

The analyses listed above are statistically independent, either because they use a completely independent dataset (based on collisions at ATLAS versus CMS, or during Run I versus Run II), or because they utilise signal regions that have no overlap with the signal regions of any of the other searches. This allows us to simply multiply the likelihoods of all analyses in order to arrive at a combined likelihood.

However, within each analysis, signal regions are not always orthogonal, i.e. some contain events or significant systematics in common.  Given that there is no public information describing the correlations across these signal regions,\footnote{This is at least true for most analyses; Refs.\ \cite{Collaboration:2242860,CMS:2017kmd} are notable recent exceptions.} we calculate the likelihood for an analysis based on the signal region that is \textit{expected} to give the strongest limit. We determine the expected limit from each signal region by computing the expected ratio between the signal plus background and background-only likelihoods, in the hypothetical scenario where the observed number of events is exactly equal to the background expectation,
\begin{equation}
\Delta\ln\mathcal{L}_\mathrm{pred} = \ln \mathcal{L}(n=b|s,b) - \ln \mathcal{L}(n=b|s=0,b).
\label{delta_pred}
\end{equation}
Taking the difference with respect to the background log-likelihood prevents erroneous model-to-model jumps in the likelihood function (see Ref.\ \cite{ColliderBit} for more details).

Given the absence of published correlations between the yields (and uncertainties) in the various signal regions, this is arguably the best possible treatment, and it has the added merit of giving conservative results.  Because no significant excess has been observed in any of the LHC searches that we include, we restrict the combined LHC Run I and combined Run II log-likelihood each to a maximum of 0, i.e.\ forbidding mildly better fits than the SM (which are achievable via statistical fluctuations in the data or Monte Carlo simulation, at a little less than the $1\sigma$ level).

We included all Run I searches listed above directly in our main scans of the CMSSM, NUHM1 and NUHM2.  We then applied the likelihoods associated with the 13\,TeV, 13 fb$^{-1}$ Run II ATLAS and CMS 0-lepton searches in a postprocessing step, using the \scannerbit \textsf{postprocessor} scanner (see Sec.\ 6 of Ref.~\cite{ScannerBit}).  These searches uncovered no excesses, and therefore do not change the regions preferred by our scans except to disfavour a strip of additional models (compared to the Run I searches) at sparticle masses of a few hundred GeV. The accuracy of our sampling is therefore unaffected by their inclusion via postprocessing rather than in the original scans.\footnote{We applied the Run II searches this way not for reasons of computational speed, but just as a matter of practicality, given when supercomputing time, Run II results and different components of \GB respectively became available.}

\subsection{Higgs physics}
\label{sec:higgs_like}

We use likelihoods from \higgsbounds \textsf{4.3.1} \cite{Bechtle:2008jh,Bechtle:2011sb,Bechtle:2013wla} and \higgssignals \textsf{1.4.0} \cite{HiggsSignals}, as interfaced via \colliderbit \cite{ColliderBit}. These provide two likelihood terms: one based on limits from LEP, and the other on measurements of Higgs masses and signal strengths at the LHC (plus some subdominant contributions from the Tevatron).

The combined LEP Higgs likelihood is an approximate Gaussian likelihood, valid in the asymptotic limit.  \higgsbounds constructs this from the full \CLsb distribution, accounting for the effect of varying production cross-sections and Higgs masses by interpolating in a grid of pre-calculated values.

The LHC Higgs likelihood is based on mass and signal-strength measurements reported by ATLAS and CMS. The mass and signal-strength data contribute separate $\chi^2$ terms to the overall LHC Higgs log-likelihood. For each channel where a mass measurement is available, a $\chi^2$ contribution is calculated for the hypothesis that each neutral Higgs particle is responsible for the observed 125\,GeV boson~\cite{Chatrchyan2012,Aad:2012tfa}.  Only the minimum value enters the final likelihood. This minimisation allows for the possibility that multiple resonances exist at 125\,GeV with near-degenerate masses. The signal-strength contribution to the $\chi^2$ uses a covariance matrix that contains all published experimental uncertainties on all measurements of signal strengths, including their correlations.

As discussed in Section~\ref{sec:spectrum}, we obtain theoretical predictions of Higgs masses from \FlexibleSUSY, adopting an uncertainty of 2\,GeV on the mass of the lightest neutral Higgs, and 3\% on all other Higgses \cite{SDPBit}.  We compute Higgs decay rates and branching fractions using \SUSYHIT \textsf{1.5} \cite{Djouadi:2006bz} via \decaybit~\cite{SDPBit}.  To obtain the neutral Higgs boson production cross sections, we employ an effective coupling approximation, assuming that the BSM-to-SM ratios of Higgs production cross sections are equal to the ratios of the relevant squared couplings.  We determine the coupling ratios using the partial width approximation, in which the ratios of squared BSM-to-SM couplings are taken to be equal to the ratios of the equivalent partial decay widths.  To obtain branching fractions for SM-like Higgs bosons of equivalent mass to those in our MSSM models, we use lookup tables computed with \HDECAY \textsf{6.51} \cite{Djouadi:1997yw,Butterworth:2010ym}.  More details can be found in the \decaybit paper \cite{SDPBit}.

\begin{figure*}[tbp]
 \centering
 \includegraphics[width=0.49\textwidth]{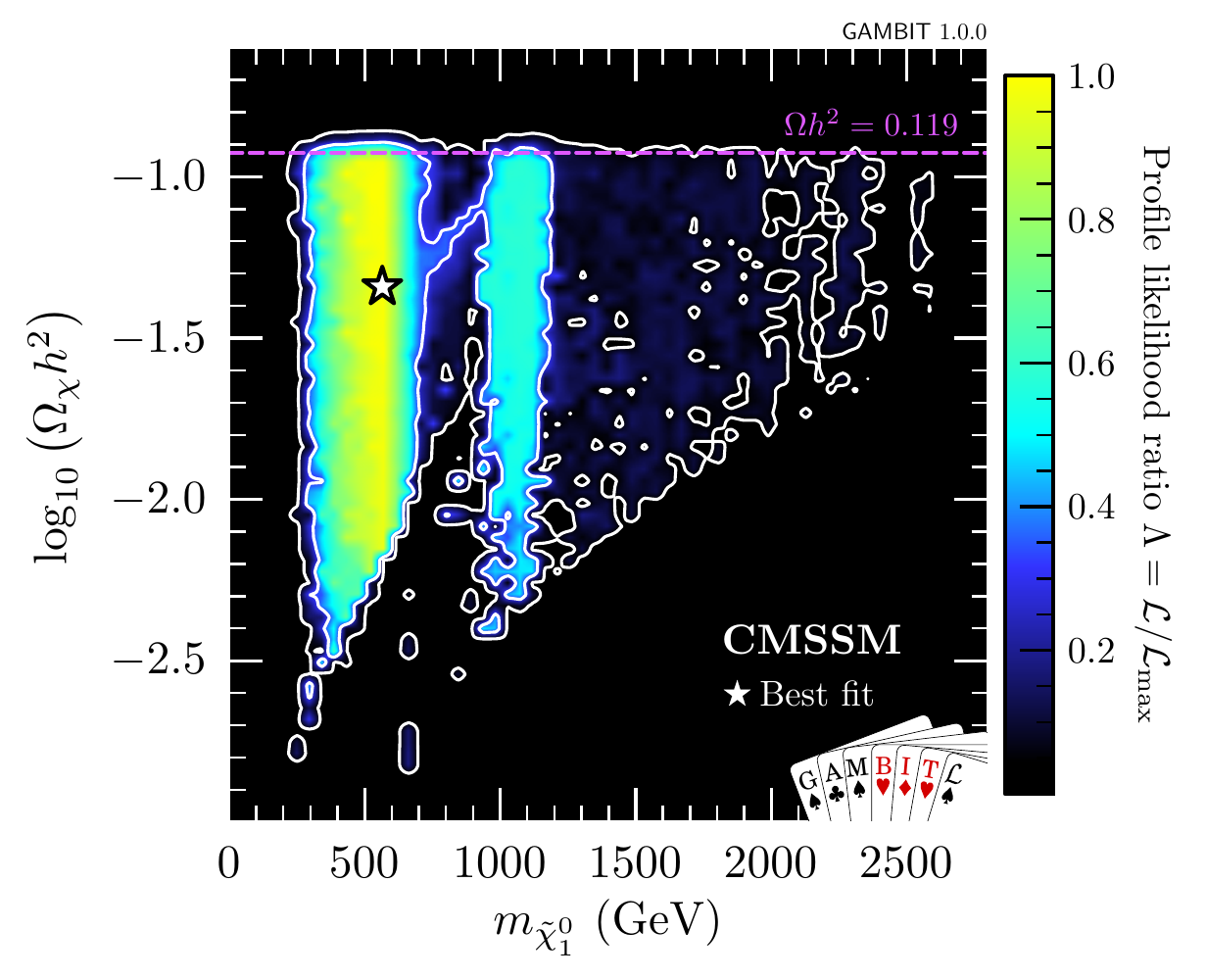}
 \includegraphics[width=0.49\textwidth]{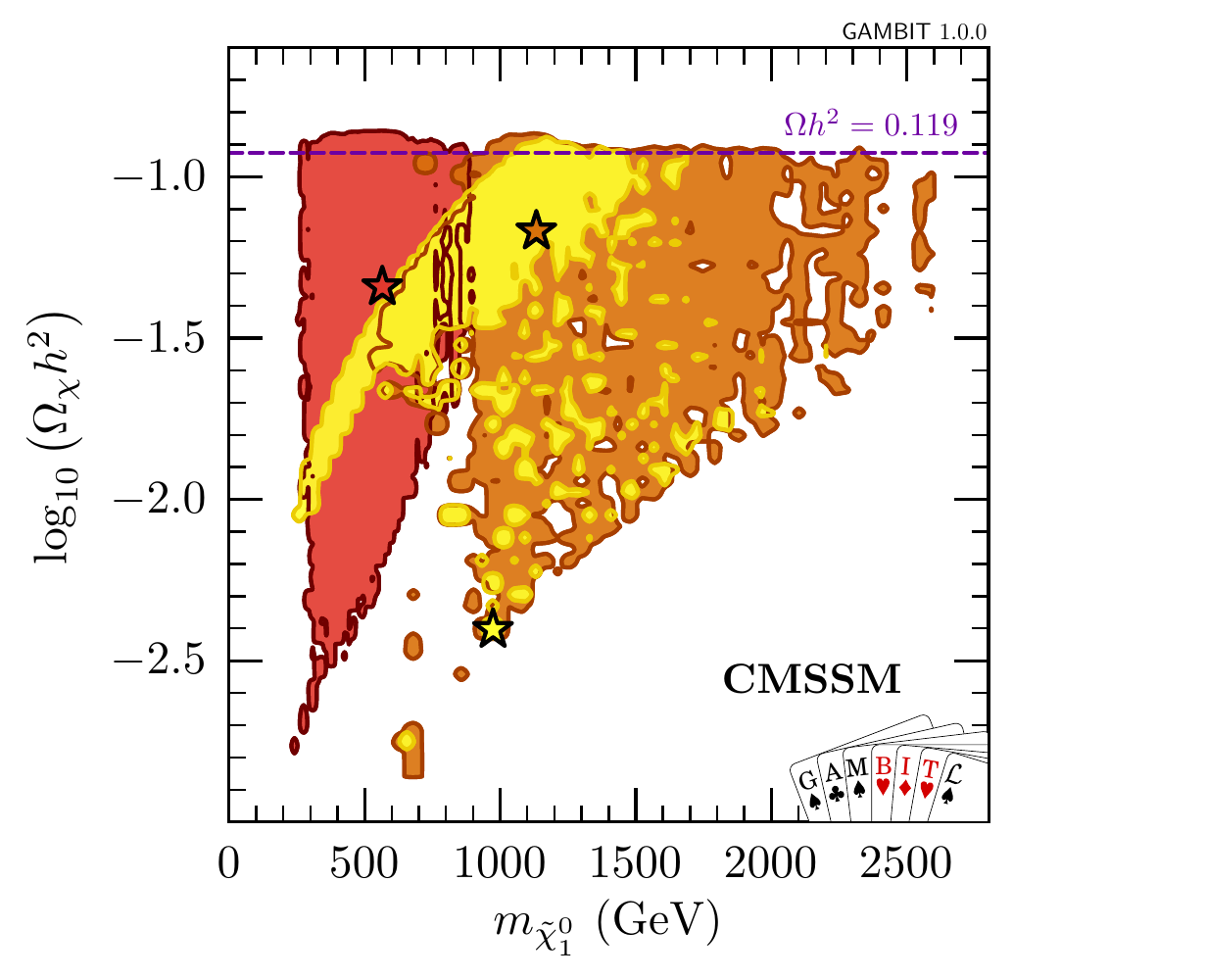}\\
 \includegraphics[height=4mm]{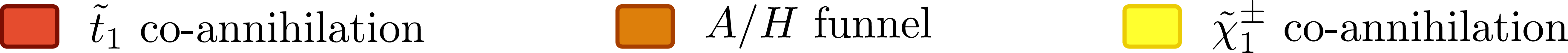}
 \caption{Profile likelihoods and confidence regions for the CMSSM, in terms of the mass and thermal relic abundance ($\Omega_\chi h^2$) of the lightest neutralino. \textit{Left:}  The profile likelihood ratio, plotted with $1 \sigma$ and $2 \sigma$ contour lines drawn in white, and the best fit point indicated by a star. \textit{Right:} Mechanisms for ensuring that the relic density of DM does not exceed the measured value, through either chargino co-annihilation, resonant annihilation via the $A/H$-funnel, or stop co-annihilation. Other potential mechanisms (e.g.\ stau co-annihilation) are not shown, as they do not lie within $2 \sigma$ of the best-fit point of the entire sample. $2 \sigma$ contours for each mechanism are plotted using darker lines, and best-fit points are indicated by a correspondingly coloured star.
 }
 \label{fig:mchi_oh2}
\end{figure*}

\section{Results}
\label{sec:current}

\subsection{CMSSM}
\label{sec:CMSSM}

In the left panel of Fig.\ \ref{fig:mchi_oh2}, we show the joint profile-likelihood ratio for the mass of lightest neutralino and the relic density in the CMSSM.  In the right panel, we show the same 95\% CL regions colour-coded according to the possible mechanisms by which different models may avoid exceeding the observed relic density of DM. We classify these regions as follows:
\begin{itemize}
\item stau co-annihilation: $m_{\tilde{\tau}_1} \leq 1.2\,m_{\tilde\chi^0_1} $,
\item stop co-annihilation: $m_{\tilde{t}_1} \leq 1.2\,m_{\tilde\chi^0_1}$,
\item chargino co-annihilation: $\tilde\chi_1^0$ $\ge50\%$ Higgsino,
\item $A/H$-funnel: $1.6\,m_{\tilde\chi^0_1} \leq m_\textrm{heavy} \leq 2.4\,m_{\tilde\chi^0_1}$,
\end{itemize}
where `heavy' may be $A^0$ or $H^0$, i.e. a model qualifies if either Higgs is in range.

We emphasise that this classification is not exclusive.  The labels that we give to these regions are merely a convenient shorthand for the precise mass/composition relations that we give above.  In particular, they should \textit{not} be interpreted as definitive indications that a specific mechanism is solely (nor even predominantly) responsible for setting the relic density of the neutralino.  These relations indicate \textit{necessary but not sufficient} conditions for a given mechanism to play a significant role in setting the relic density.  The colour-coding in Fig.~\ref{fig:mchi_oh2} (right) is done on the basis of the subset of the points in the 2$\sigma$ region of the full scan that fulfil each of the mass/composition relations, and the resulting shading of regions is overlaid.  In many cases, as we will show, single parameter combinations can satisfy two or more of the mass/composition conditions, and can thus be classified as members of multiple regions.  In these cases, one of the mechanisms sometimes dominates over the others.  Hybrid sub-regions also exist where the relic density is controlled by two or more mechanisms. For clarity, we make no attempt to show any of these cases as separate regions, nor to colour according to which (if any) mechanism dominates in overlapping regions.  For specific cases of interest, we do, however, attempt to clarify these finer issues in our discussion of the results that we show.

Even within individual regions, readers should be wary of the need for nuance in interpreting the ``relic density mechanism'' labels.  Points labelled ``chargino co-annihilation'' will typically exhibit co-annihilation of the lightest neutralino with both the lightest chargino \textit{and} the next-to-lightest neutralino, as small $\tilde\chi_1^0$--$\tilde\chi_2^0$ and $\tilde\chi_1^0$--$\tilde\chi_1^\pm$ mass splittings are an automatic consequence of a predominantly Higgsino LSP.  Nevertheless, \textit{both} these co-annihilation processes are outweighed in many models simply by boosted $\tilde\chi_1^0$--$\tilde\chi_1^0$ annihilation, brought about by the dominance of the Higgsino component in the lightest neutralino.  Similarly, $A/H$-funnel points will exhibit resonant annihilation through both the CP-odd Higgs, $A^0$, and the heavy CP-even Higgs, $H^0$, which are close to degenerate in mass in the CMSSM (and NUHM models).  The CP-odd Higgs resonance dominates at the present day, however, as s-channel annihilation via the CP-even state is velocity suppressed.

In contrast to previous studies of the CMSSM, we apply the relic density measurement as an upper limit only, allowing for the possibility that thermal neutralinos do not constitute all of DM.  This has important consequences for the resulting phenomenology.

Higgsino LSPs are automatically nearly degenerate with the lightest chargino and next-to-lightest neutralino, leading to efficient co-annihilation and an under-abundant relic density for $m_\chi \lesssim 1$\,TeV.  In isolation, this effect naturally gives the observed relic density at neutralino masses of about a TeV, and lower and higher values at smaller and larger neutralino masses, respectively.\footnote{Note that the Sommerfeld effect can be important in the context of pure Higgsino DM; see Sec.\ \ref{sec:ID} for details.} This effect can be seen in the low-mass yellow strip in Fig.\ \ref{fig:mchi_oh2}. If the LSP is instead a ``well-tempered'' \cite{ArkaniHamed:2006mb} admixture of Higgsino and bino\footnote{In the CMSSM, this well-tempered mixture is realised within the ``focus point'' region \cite{Feng:1999zg,Feng:2000gh,Feng:2000bp,Feng:2005hw,Feng:2011aa,Feng:2012jfa,Draper:2013cka}.}, then the efficiency of the co-annihilation effect can be tuned to give the exact observed relic density, even at very low neutralino masses.  Such scenarios are, however, heavily constrained by recent LUX \cite{LUX2016,LUXrun2} and Panda-X \cite{PandaX2016} limits on the spin-independent scattering cross-section \cite{Baer:2016ucr,Athron:2016gor,Badziak:2017the}.  As we see in the low-mass section of Fig.\ \ref{fig:mchi_oh2} however, relaxing the demand that the neutralino must explain all of DM allows models to be more Higgsino-dominated, leading to subdominant neutralino DM.  The reduced relic density also helps Higgsino models avoid limits from spin-dependent nuclear scattering, which would otherwise prove rather constraining.

Similarly, at masses above 1\,TeV, the not-quite-efficient-enough Higgsino co-annihilation can be supplemented by additional resonant annihilation through the heavy Higgs funnel, bringing the relic density down to the observed value, or lower.  These models can be seen as overlapping yellow and orange regions at $m_\chi \gtrsim 1$\,TeV in the right panel of Fig.\ \ref{fig:mchi_oh2}.

\begin{figure*}[tbp]
  \centering
  \includegraphics[width=0.49\textwidth]{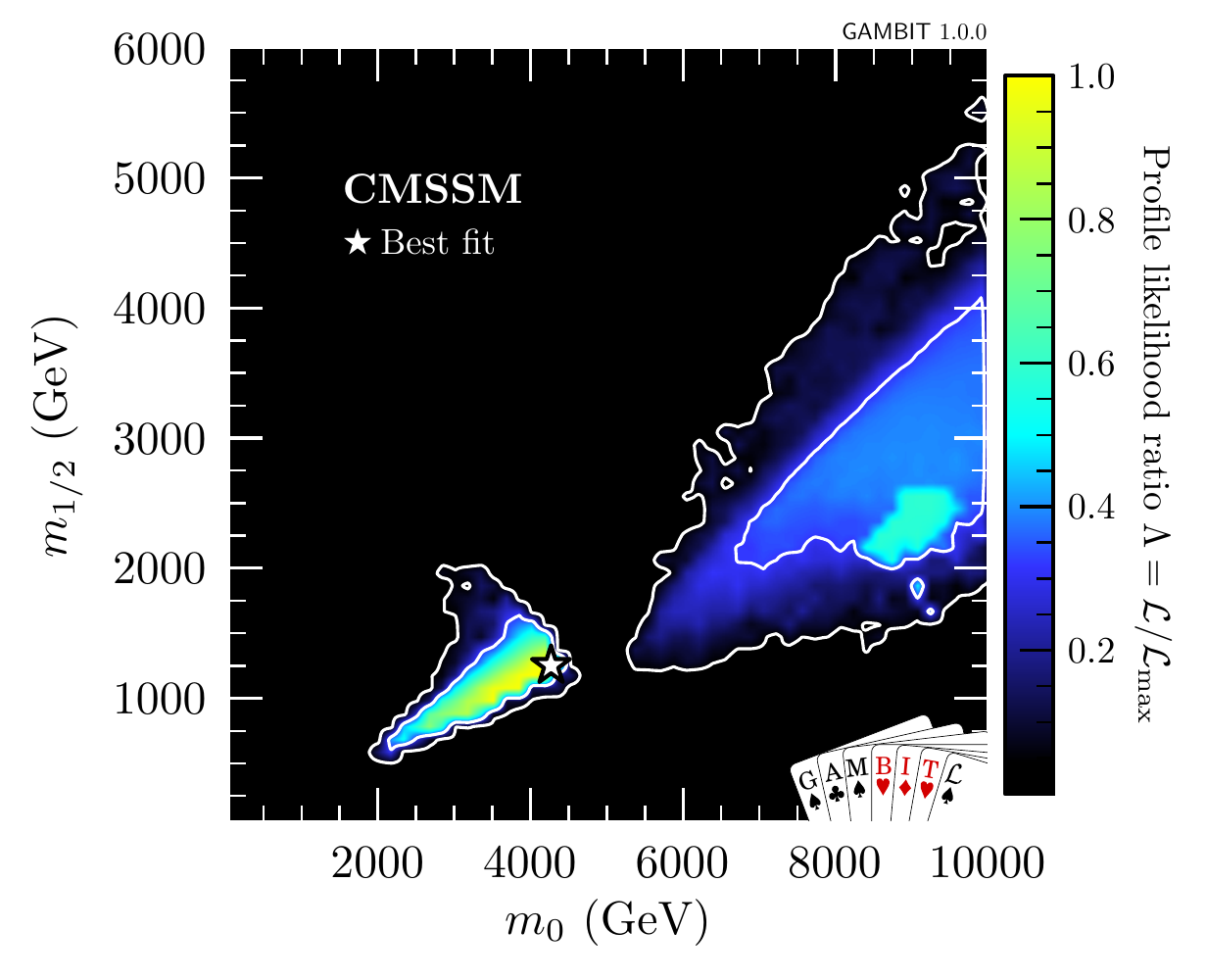}
  \includegraphics[width=0.49\textwidth]{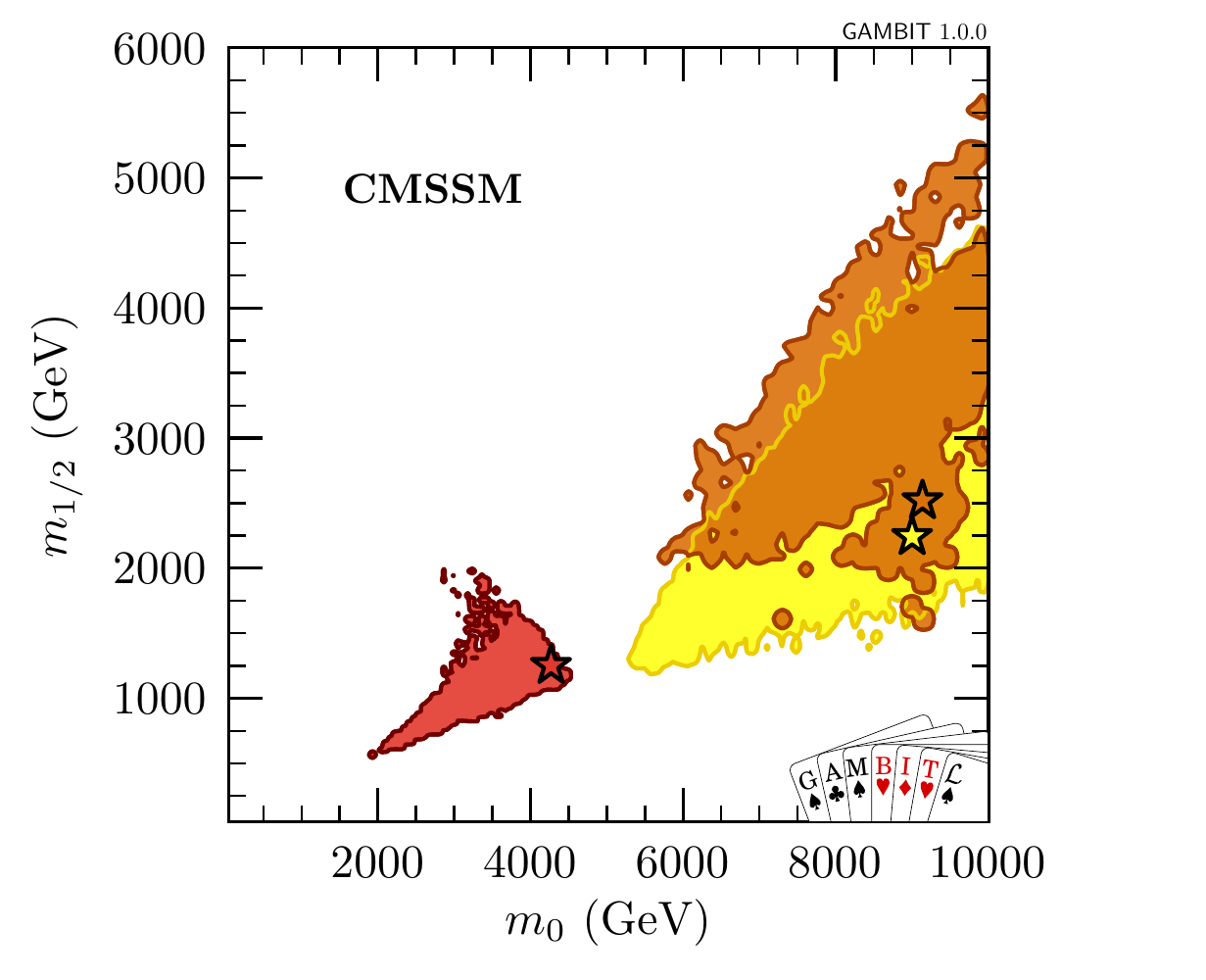}
  \includegraphics[width=0.49\textwidth]{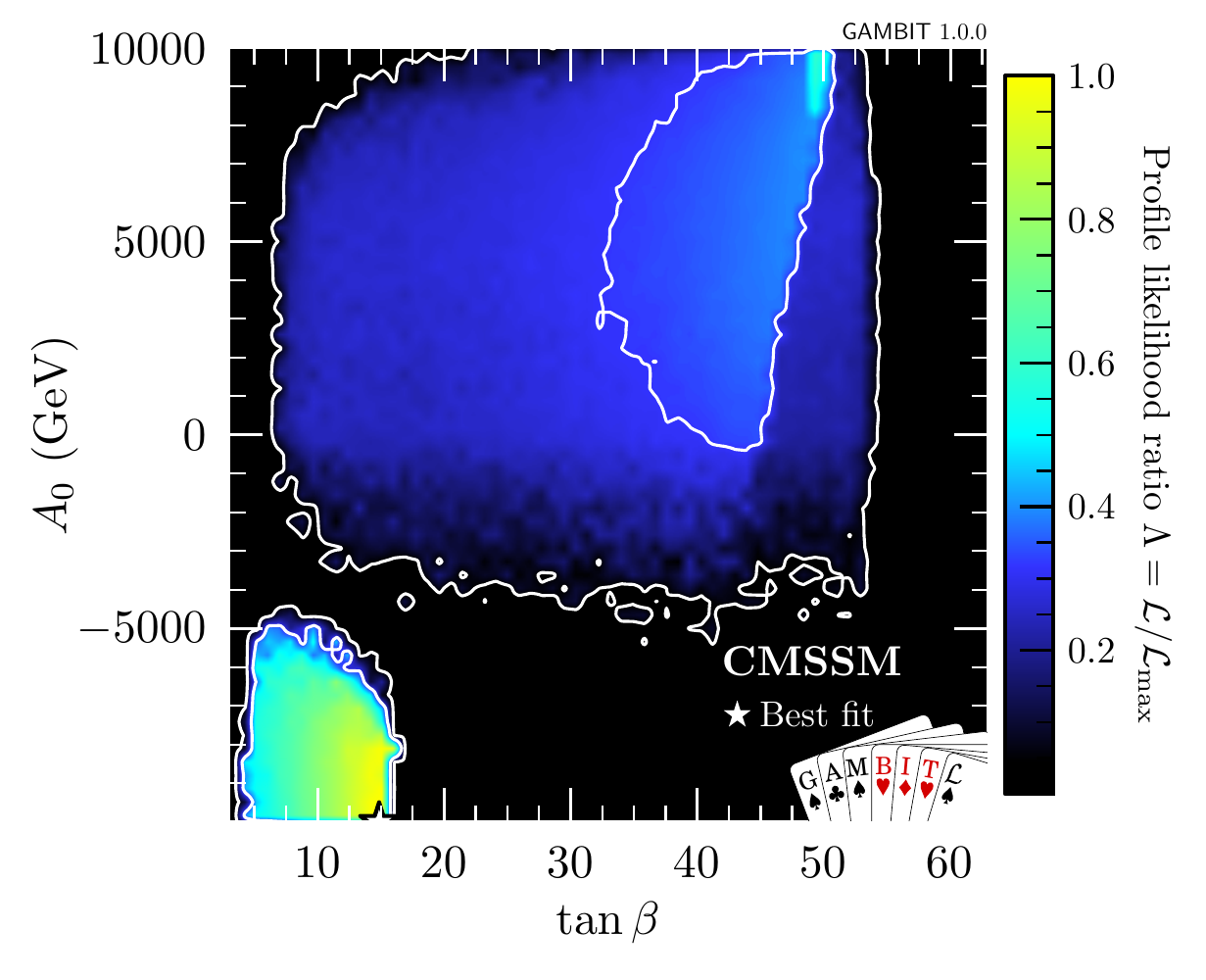}
  \includegraphics[width=0.49\textwidth]{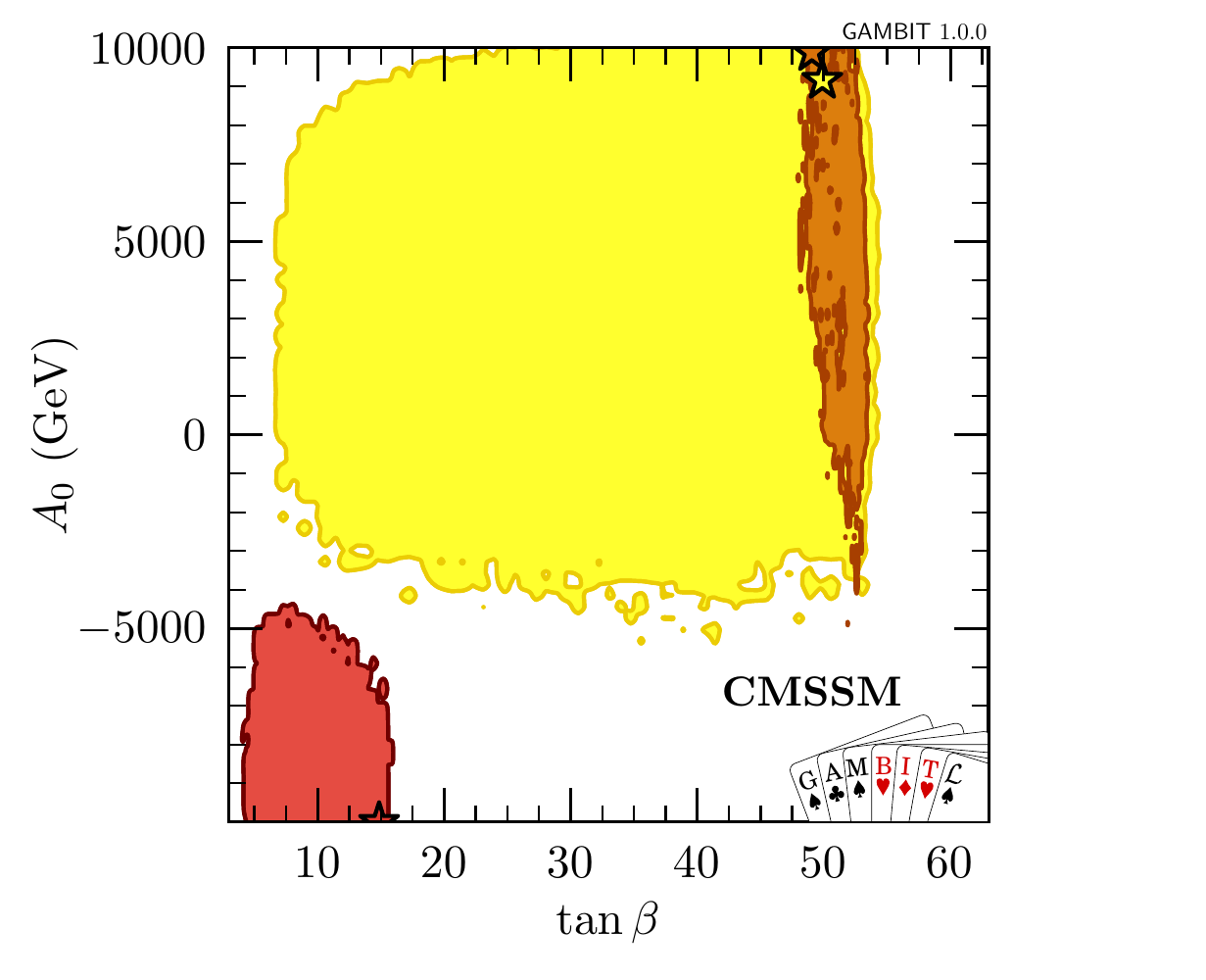}\\
  \includegraphics[height=4mm]{figures/rdcolours3.pdf}
  \caption{ \textit{Left:} The profile likelihood ratio in the CMSSM, for $m_0$ and $m_{1/2}$ (top) and $\tan\beta$ and $A_{0}$ (bottom), with explicit 68\% and 95\% CL contour lines drawn in white, and the best fit point indicated by a star. \textit{Right:} Colour-coding shows the mechanisms active in models within the 95\% CL contour for avoiding thermal overproduction of neutralino dark matter, through either chargino co-annihilation, resonant annihilation via the $A/H$ funnel, or stop co-annihilation. Other potential mechanisms (e.g. stau co-annihilation) are not present, as they do not lie within the 95\% CL contour.}
  \label{fig:2d_parameter_plots_cmssm}
\end{figure*}

We now see that relaxing the relic density constraint to an upper limit opens up a much richer set of phenomenologically-viable scenarios, with lighter Higgsino or mixed Higgino-bino
LSPs.  From the perspective of global fits, treating the relic density as an upper bound is a conservative approach, and allows us to test whether the preference for heavy spectra found in recent studies \cite{Fittinocoverage,Mastercode15,Han:2016gvr} persists even when a greater variety of light LSPs is permitted.

The right panel of Fig.\ \ref{fig:mchi_oh2} shows that at 95\% CL, all of the identified annihilation mechanisms (stop
co-annihilation, $A/H$-funnel and chargino co-annihilation) permit solutions where the measured relic
density is fully accounted for, as well as scenarios where only a very small fraction of the DM relic abundance is explained in the CMSSM.
The fit does not demonstrate any clear preference for the relic density to be under-abundant or very close to the measured value.  Looking at the top of this plot, we indeed see the established picture for
chargino co-annihilation discussed above, where a pure Higgsino DM candidate should have a mass of around 1~TeV to fit the observed relic density.

In Fig.\ \ref{fig:2d_parameter_plots_cmssm}, we show 2D CMSSM joint profile likelihoods for $m_0$ and $m_{1/2}$, as well as for $\tan\beta$ and $A_{0}$.  Here the plots include both positive and negative $\mu$, and are again coloured by relic density mechanism.  We see a large region of high likelihood at large $m_0$ and $m_{1/2}$, consisting of overlapping chargino co-annihilation and $A/H$-funnel points. The $A/H$-funnel region is concentrated at high $\tan\beta$, as is well known from previous studies of the CMSSM (e.g.\ Ref.\ \cite{Roszkowski:2001sb}).  The chargino co-annihilation region disfavours large negative $A_0$, in agreement with existing results in the literature.\footnote{See for example Fig.\ 2d of Ref.\ \cite{Fittinocoverage}, and the middle panels of Fig.\ 2 of Ref.\ \cite{Han:2016gvr}.}

At lower $m_0$ and $m_{1/2}$, a stop co-annihilation region appears, with a light stop very close in mass to the lightest neutralino.  Due to constraints from direct searches, as well as Higgs-mass measurements at the LHC, which push up the sfermion masses, these scenarios can only be obtained through very large stop mixing. This restricts the stop co-annihilation region to very large and negative $A_0$ values, and low-to-moderate $\tan \beta$, as can be seen in the bottom panels of Figure \ref{fig:2d_parameter_plots_cmssm}. This region has not been seen in most of the recent global fit literature, as revealing it requires not only consideration of large, negative $A_0$ values, but also very careful scanning of the parameter space.\footnote{As this manuscript was undergoing final editing, an updated version of Ref.\ \cite{Han:2016gvr} was released, showing a stop co-annihilation region in good agreement with ours.}

The preference for large and negative $A_0$ in stop co-annihilation could lead to colour- or charge-breaking minima in the scalar potential.  We have investigated the presence of such problems for points in the stop co-annihilation region, using several conditions that have been proposed in the literature:\begin{enumerate}
\item $A^2_t < 3.0 ( \mathbf{m}^\mathbf{2}_{Q_{3,3}} + \mathbf{m}^\mathbf{2}_{u_{3,3}} + \mu^2 + m_{H_u}^2)$ \cite{Casas:1996de},
\item $A^2_t < 7.5 ( \mathbf{m}^\mathbf{2}_{Q_{3,3}} + \mathbf{m}^\mathbf{2}_{u_{3,3}})  - 3 \mu^2$ \cite{Kusenko:1996jn}, and
\item $A^2_t < 3.4 (\mathbf{m}^\mathbf{2}_{Q_{3,3}} + \mathbf{m}^\mathbf{2}_{Q_{3,3}}) + 60 (m_{H_u}^2 + \mu^2 )$, based on the results in Ref.\ \cite{Blinov:2013fta}.
\end{enumerate}
We found that whilst some points in this region do violate one or more of these conditions, removing all points that do so neither modifies the shapes of the likelihood contours in our plots, nor the fact that the best-fit occurs in the stop co-annihilation region. This question could in principle be investigated further by calculating the tunnelling probability for each point, e.g.\ using \textsf{Vevacious} \cite{Camargo-Molina:2013qva}.  However, it is  not possible to do this in a reasonable amount of time with the large number of points in our scans.  Even though the conditions above are not definitive, being neither necessary nor sufficient to establish that the vacuum of the theory breaks gauge invariance, neither is studying stability with tools such as \textsf{Vevacious}, due to the large number of scalar fields in the MSSM and the resulting difficulty of finding all relevant minima of the potential.  We therefore leave detailed investigation of such issues for a future paper.

\begin{figure*}[tb]
  \centering
\includegraphics[width=0.49\textwidth]{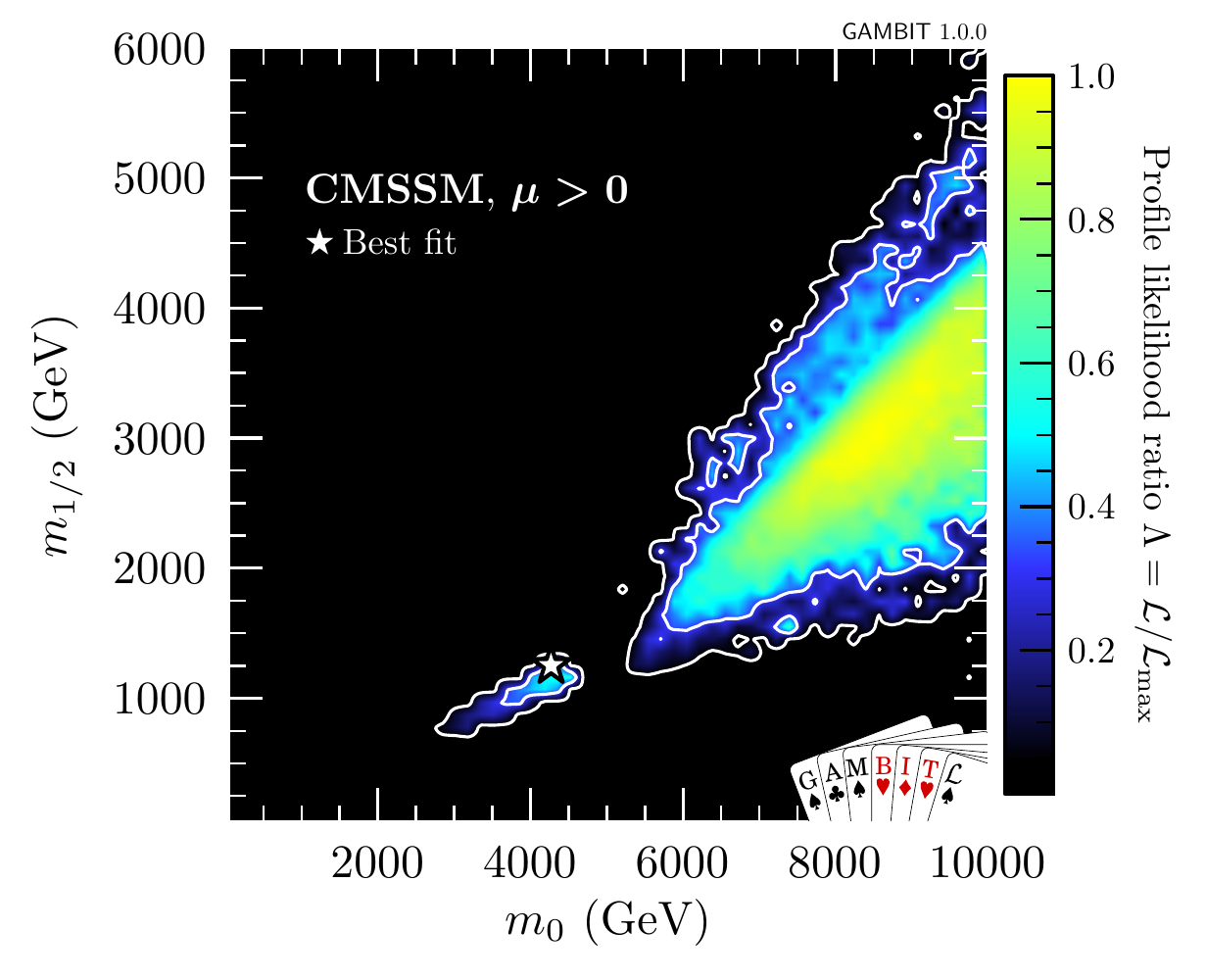}
\includegraphics[width=0.49\textwidth]{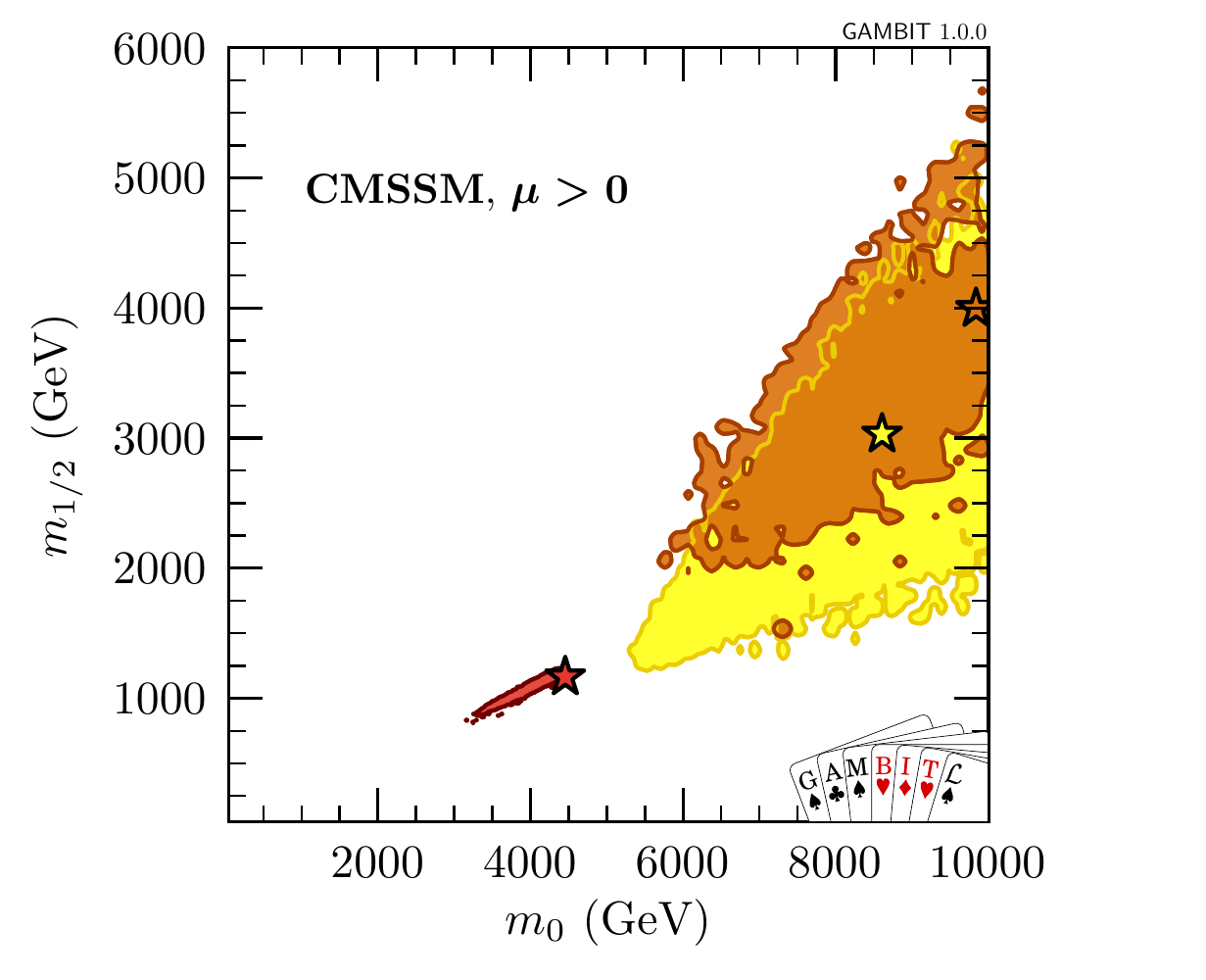}
\includegraphics[width=0.49\textwidth]{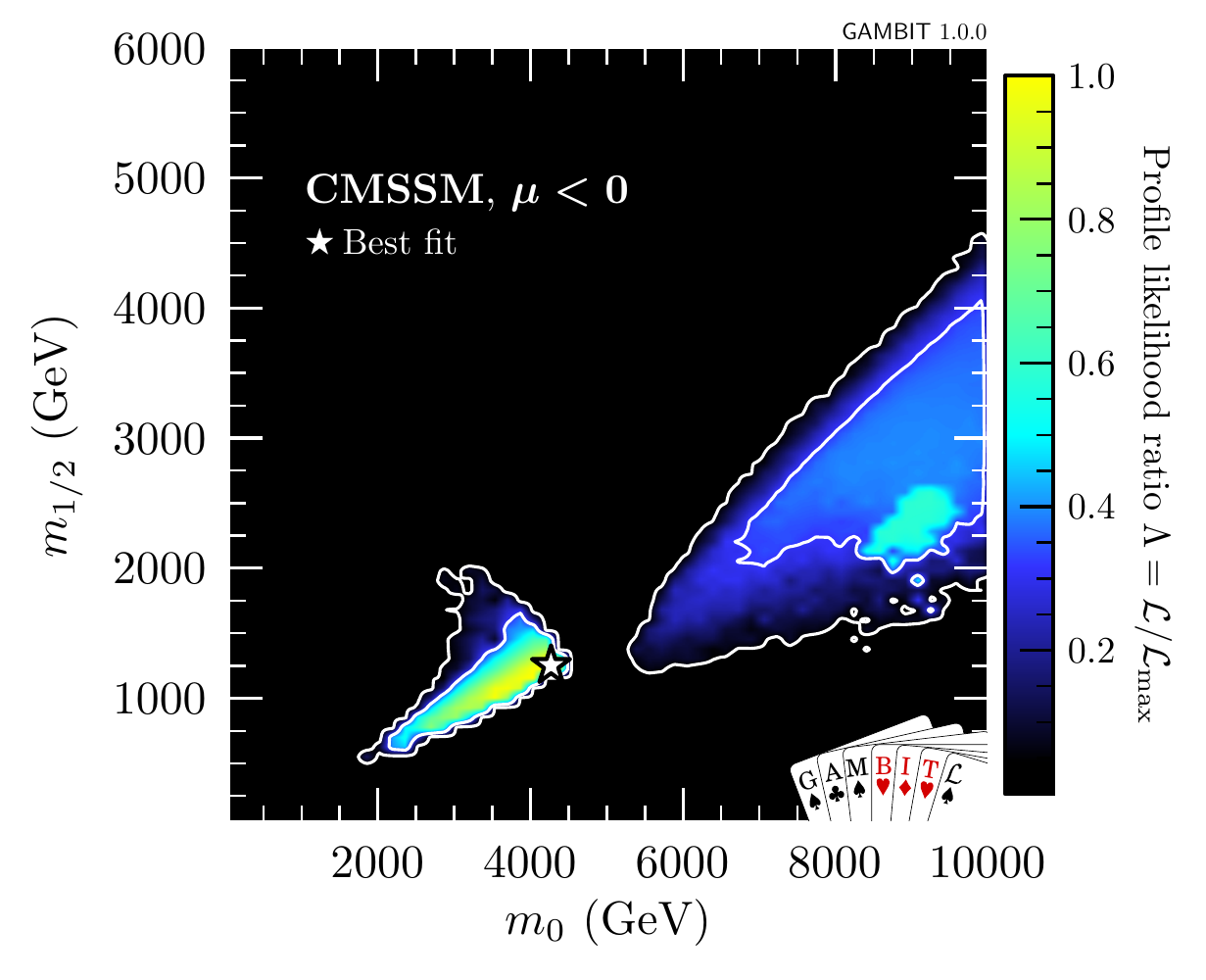}
\includegraphics[width=0.49\textwidth]{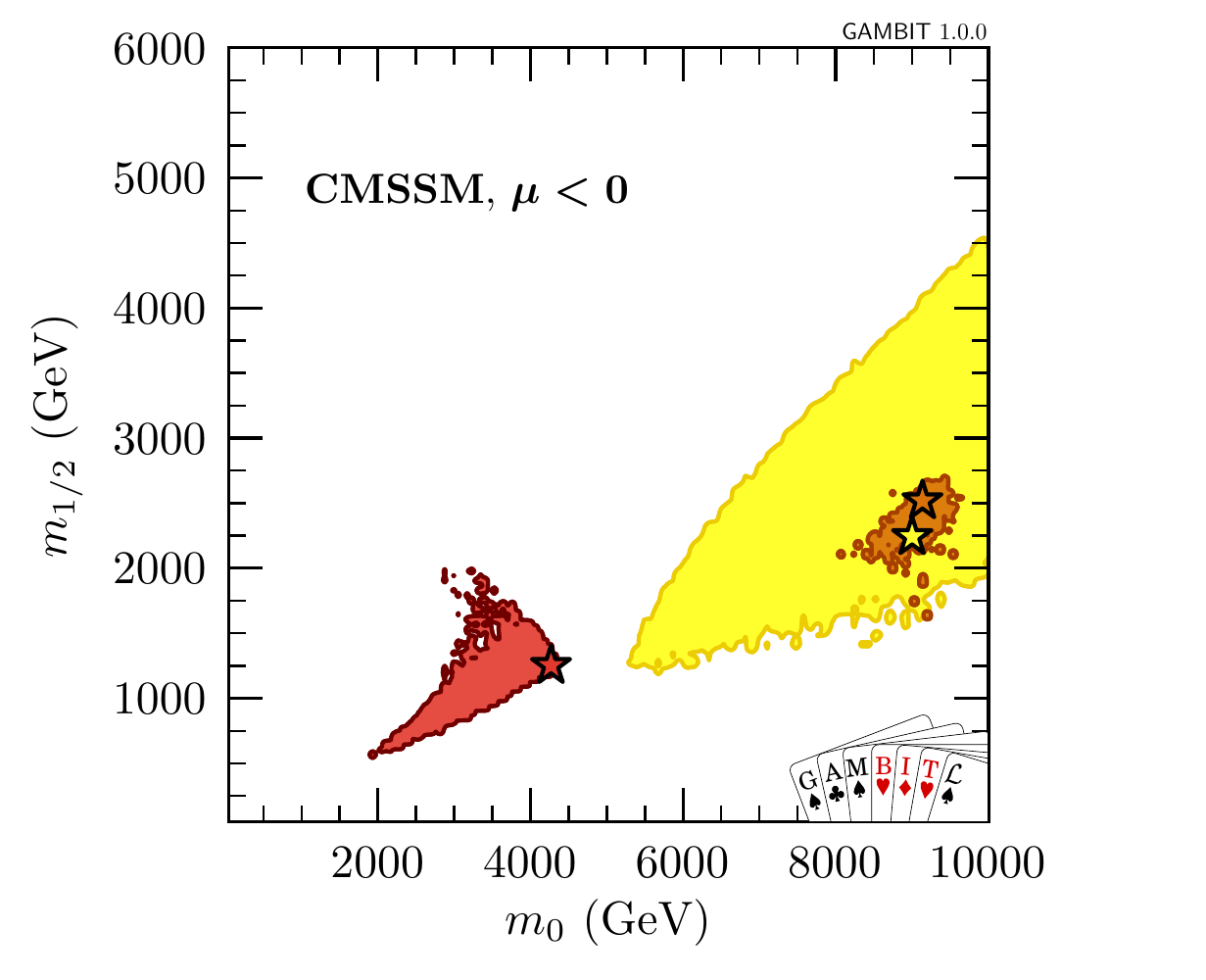}\\
\includegraphics[height=4mm]{figures/rdcolours3.pdf}
\caption{\textit{Left:} Profile likelihood ratio in the $m_0-m_{1/2}$ plane of the CMSSM, for $\mu \geq 0$ (top) and $\mu < 0$ (bottom). \textit{Right:} Colour-coding showing the mechanisms for avoiding a relic density of DM that exceeds the observed value.}
  \label{fig:2d_m0_m12_pm_mu}
\end{figure*}

Looking at the lower-right panel of Fig.\ \ref{fig:2d_parameter_plots_cmssm}, the stop co-annihilation region undoubtedly extends to even lower values of $A_0$ than we have considered here.  Combined with possible impacts of Sommerfeld enhancement on the relic density \cite{Ellis:2014ipa}, this would have the effect of allowing stop co-annihilation to extend to very large values of $m_0$ (Ref.\ \cite{Ellis:2014ipa} found stop co-annihilation models with $m_0$ as large as 13\,TeV).  However, as $A_0$ becomes more negative, colour- and charge-breaking vacua become an ever-increasing concern.

In contrast with previous results, we do not find a stau co-annihilation region inside the 95\% CL region surrounding our overall best fit. We \textit{do} find stau co-annihilation solutions in the same region of parameter space as seen in the literature\footnote{See for example Fig.\ 3 of Ref.\ \cite{Fittinocoverage}, Fig.\ 2 of Ref.\ \cite{Han:2016gvr} and Fig.\ 1 of \cite{Mastercode15}.} when we look at $4 \sigma$ confidence regions. In addition to this, we also see small islands of stau co-annihilation appear inside the $2 \sigma$ contours if we remove the LHC Run II likelihood (leaving only Run I analyses), although these are much smaller than seen in the previous literature. Therefore, the likelihood of the stau co-annihilation strip is suppressed in our results relative to those in the literature.  Beyond the LHC Run II likelihood, which has also been shown to impact this region in Ref.\ \cite{Han:2016gvr}, the suppression compared to other regions of good fit comes mostly from the LHC Higgs likelihood. This is influenced by the following differences in our analysis compared to existing analyses (for recent examples see Refs.\cite{Fittinocoverage, Mastercode15,Han:2016gvr}): i) relaxation of the relic density constraint to an upper bound, allowing light Higgsino DM scenarios and consequentially relaxing the constraint on $\mu$; ii) differences in the Higgs mass calculation (\FlexibleSUSY rather than \FH) and branching ratio calculations (\HDECAY rather than \FH); iii) a wider prior mass range than some previous scans; and iv) an improved scanning technique, which finds a modestly better fit in the other regions, relative to the stau co-annihilation region.

The plots in Fig.~\ref{fig:2d_parameter_plots_cmssm} combine scans for $\mu<0$ and $\mu>0$.  As the sign of $\mu$ is a discrete parameter, it is useful to also investigate each sign independently. These plots are shown in
Fig.~\ref{fig:2d_m0_m12_pm_mu}. A preference for positive $\mu$, from the anomalous magnetic moment of the muon $a_\mu$, has been reported previously (e.g.\ \cite{Mastercode15}).  In our results Higgs observables and LHC sparticle search likelihoods (and the large allowed range for dimensionful parameters) push up the mass scale of the preferred sparticle spectrum, minimising the impact of the $a_\mu$ likelihood and removing the preference for positive $\mu$. We see a mild preference for $\mu < 0$, which has a best fit log-likelihood of $-263.75$, as compared to $-265.00$ for $\mu>0$. The negative $\mu$ results also exhibit an enhanced stop co-annihilation region at low mass, and a reduced $A/H$-funnel region at higher mass, relative to the positive $\mu$ results.

The larger, better-fitting stop co-annihilation region at $\mu<0$ is driven entirely by the Higgs signal likelihood, in particular the fit to the gauge boson signal strengths.  Positive $\mu$ in this region suppresses the $h\to WW,ZZ,\gamma\gamma$ branching fractions to below the observed values, leading to a best-fit likelihood worse than the $\mu<0$ equivalent by $\Delta\ln\mathcal{L} = 1.3$.  Indeed, the $\mu<0$ fit is actually slightly \textit{better} than the fit of the SM to the Higgs data, by $\Delta\ln\mathcal{L} = 0.9$ units.  Although the implied preference for $\mu<0$ over $\mu>0$ is weak, at just $1.1\sigma$ (in 2D), this demonstrates that precision Higgs physics has now reached the stage where it can directly constrain the parameters of supersymmeteric scenarios.

The vastly different size of the heavy Higgs funnel for $\mu<0$ and $\mu>0$ is due to differences in LSP composition.  For $\mu>0$, the $A/H$ funnel contains many Higgsino LSP points, which combine with the chargino co-annihilation mechanism to form an extensive hybrid region.  In contrast, the bino solutions that do exist in this region are somewhat more concentrated in the $m_0$--$m_{1/2}$ plane. For $\mu<0$, the heavy Higgs funnel region is almost exclusively bino, leaving the chargino co-annihilation to exist mostly as a pure mechanism, and resulting in an upper limit on the mass of the LSP of $\sim$1.2\,TeV.  Although there are some relatively isolated points in the $\mu<0$ scan exhibiting hybrid funnel-chargino co-annihilation behaviour, it seems difficult to obtain valid solutions to the RGEs with such spectra when $\mu$ is negative.

\begin{table*}
\small
\center
\sisetup{round-mode = places, round-precision = 3,round-integer-to-decimal}
\begin{tabular}{l S[table-format=+3.3] S[table-format=+3] S[table-format=+3] S[table-format=+3] S[table-format=+3] S[table-format=+3]}
Likelihood term & {\ \ Ideal} & {$A/H$-funnel} & {\ \ \ $\tilde{\tau}$ co-ann.} & {\ \ \ \ $\tilde{t}$ co-ann.} & {\ $\tilde{\chi}_1^{\pm}$ co-ann.} & {\ \ \ $\Delta\ln\mathcal{L}_\mathrm{BF}$} \\
\hline
LHC sparticle searches & 0 &  0.000  & 0.000 & 0.000 & 0.000  &  0.000 \\
LHC Higgs & -37.7339 &  -37.960  &  -41.296  &  -38.042  &  -38.069  &  .3081  \\
LEP Higgs & 0 &  0.000  &  0.000  &  0.000  &  0.000  &  0 \\
ALEPH selectron & 0 &  0.000  & 0.000  & 0.000  & 0.000  &  0 \\
ALEPH smuon & 0 &  0.000  &  0.000  &  0.000  &  0.000  &  0  \\
ALEPH stau & 0 &  0.000  &  0.000  &  0.000  &  0.000  &  0 \\
L3 selectron & 0 &  0.000  &  0.000  &  0.000  &  0.000  &  0  \\
L3 smuon & 0 & 0.000  &  0.000  &  0.000  &  0.000  &  0 \\
L3 stau & 0 & 0.000  &  0.000  &  0.000  &  0.000  &  0 \\
L3 neutralino leptonic & 0 & 0.000  & 0.000  & 0.000  & 0.000  &  0  \\
L3 chargino leptonic & 0 &  0.000  &  0.000  &  0.000  &  0.000  &  0 \\
OPAL chargino hadronic & 0 &  0.000  &  0.000  &  0.000  &  0.000  &  0 \\
OPAL chargino semi-leptonic & 0 &  0.000  &  0.000  &  0.000  &  0.000  &  0 \\
OPAL chargino leptonic & 0 &  0.000  &  0.000  &  0.000  &  0.000  &  0 \\
OPAL neutralino hadronic & 0 &  0.000  &  0.000  &  0.000  &  0.000  &  0  \\
$B_{(s)}\to \mu^+\mu^-$ & 0 &  -1.939  &  -2.739  &  -2.029  &  -1.939  &  2.029  \\
Tree-level $B$ and $D$ decays & 0 &  -15.515  &  -15.491  &  -15.283  &  -15.610  &  15.283 \\
$B^0\to K^{*0}\mu^+\mu^-$ & -184.260 &  -196.506  &  -197.469  &  -196.088  &  -196.309  &  11.828  \\
$B\to X_s\gamma$ & 9.799 &  9.258  &  9.525  &  9.106  &  9.184  &  .693  \\
$a_\mu$ & 20.266 &  13.915  &  14.556  &  13.977  &  13.903  &  6.289  \\
$W$ mass & 3.281 &  3.084  & 3.093  & 3.050  & 3.095  &  .231  \\
Relic density & 5.989 &  5.989  &  5.984  &  5.989  &  5.989  &  0 \\
PICO-2L & -1 &  -1.000  &  -1.000  &  -1.000  &  -1.000  &  0 \\
PICO-60 F & 0 &  0.000  &  0.000  &  0.000  &  0.000  &  0  \\
SIMPLE 2014 & -2.972 &  -2.972  &  -2.972  &  -2.972  &  -2.972  &  0  \\
LUX 2015  & -0.64 &  -0.676  &  -0.642  &  -0.640  &  -0.727  &  0  \\
LUX 2016  & -1.467 &   -1.539  &  -1.472  &  -1.467  &  -1.646  &  0 \\
PandaX 2016  & -1.886 &  -1.936  &  -1.889  &  -1.886  &  -2.009  &  0  \\
SuperCDMS 2014 & -2.248 &  -2.248  &  -2.248  &  -2.248  &  -2.248  &  0 \\
XENON100 2012 & -1.693 &   -1.675  &  -1.692  &  -1.693  &  -1.651  &  0  \\
IceCube 79-string & 0.0 &  0.000  &  0.000  &  0.000  &  0.000  &  0 \\
$\gamma$ rays (\textit{Fermi}-LAT dwarfs) & -33.244 &  -33.421  &  -33.393  &  -33.381  &  -33.394  &  .137 \\
$\rho_0$ & 1.142 &  1.141  &  1.142  &  1.141 &  1.141  &  .001 \\
$\sigma_s$ and $\sigma_l$ & -6.115 &  -6.115  &  -6.116  &  -6.115  &  -6.117  &  0 \\
$\alpha_s(m_Z)(\MSBar)$ & 6.500 &  6.487  &  6.479  &  6.481  &  6.479  &  .019 \\
Top quark mass & -0.645 &  -0.645  &  -0.645  &  -0.649  &  -0.645  &  .004 \\
\hline
Total & -226.927 & -264.273	& -268.287 & -263.747 & -264.546 & 36.820 \\
\hline
&  &  &  &  & \\
Quantity &  & {$A/H$-funnel} & {\ \ \ $\tilde{\tau}$ co-ann.} & {\ \ \ \ $\tilde{t}$ co-ann.} & {\ $\tilde{\chi}_1^{\pm}$ co-ann.} &  \\
\cmidrule{1-6}
$A_0$ &  &  9924.435  &  -1227.154  &  -9965.036  &  9206.079 \\
$m_0$ &  &  9136.379  &  1476.893  &  4269.402  &  9000.628 \\
$m_{1/2}$ &  &  2532.163  &  2422.340  &  1266.043  &  2256.472 \\
$\tan\beta$ &  &  49.048  &  48.594  &  14.857  &  49.879 \\
$\mathrm{sgn}(\mu)$ &  &  \multicolumn{1}{r}{$-$}  &  \multicolumn{1}{r}{$+$}  &  \multicolumn{1}{r}{$-$}  &  \multicolumn{1}{r}{$-$} \\
$m_t$ & &  173.366  &  173.358  &  173.267  &  173.329 \\
$\alpha_s(m_Z)(\MSBar)$ & &  0.119  &  0.119  &  0.119  &  0.119 \\
$\rho_0$ & &  0.394  &  0.401  &  0.403  &  0.394 \\
$\sigma_s$ &  &  42.950  &  43.031  &  42.975  &  43.503 \\
$\sigma_l$ &  &  57.976  &  58.544  &  57.887  &  58.155 \\
\cmidrule{1-6}
$M_1$ & &  1140.417  &  1089.994  &  556.554  &  1011.999 \\
$\mu$ & &  -1409.433  &  2621.118  &  -4073.398  &  -983.112 \\
$m_{\tilde{t}_1}$ & &  6554.967  &  3594.650  &  592.052  &  6279.661 \\
$m_{\tilde{\tau}_1}$ & &  6590.901  &  1076.748  &  4071.458  &  6407.136 \\
$m_{A}$ & &  2292.366  &  2182.200  &  5612.268  &  1953.735 \\
$m_{h}$ & &  124.896  &  124.054  &  125.007  &  124.797 \\
$m_{\tilde{\chi}_1^0}$ & &  1133.191  &  1076.738  &  565.069  &  973.418 \\
$($\%bino, \%Higgsino$)$ & & \multicolumn{1}{r}{$(99,1)$} & \multicolumn{1}{r}{$(100,0)$} & \multicolumn{1}{r}{$(100,0)$} & \multicolumn{1}{r}{$(44,56)$} \\
$m_{\tilde{\chi}_2^0}$ & &  1432.774  &  1999.921  &  1083.062  &  -1005.489 \\
$($\%bino, \%Higgsino$)$ & & \multicolumn{1}{r}{$(1,98)$} & \multicolumn{1}{r}{$(0,1)$} & \multicolumn{1}{r}{$(0,0)$} & \multicolumn{1}{r}{$(0,100)$} \\
$m_{\tilde{\chi}_1^\pm}$ & &  1430.811  &  2000.084  &  1083.224  &  1002.018 \\
$($\%wino, \%Higgsino$)$ & & \multicolumn{1}{r}{$(1,99)$} & \multicolumn{1}{r}{$(99,1)$} & \multicolumn{1}{r}{$(100,0)$} & \multicolumn{1}{r}{$(1,99)$} \\
$m_{\tilde{g}}$ & &  5545.587  &  5017.077  &  2926.857  &  5002.109 \\
$\Omega h^2$ & &  {$6.88\times10^{-2}$}  &  {$1.06\times10^{-1}$} &  {$4.62\times10^{-2}$} & {$4.00\times10^{-3}$}  \\
\cmidrule{1-6}
\end{tabular}
\caption{\label{tab:cmssm-bf-1} Best-fit points in the CMSSM, for each of the regions characterised by a specific mechanism for suppressing the relic density of dark matter. Here we show the likelihood contributions,  parameter values at each point, and some quantities relevant for the interpretation of mass spectra at the different best fits.  We also give likelihood components for a canonical `ideal' likelihood (see text), along with its offset from the global best-fit point. SLHA1 and SLHA2 files corresponding to the best-fit point in each region can be found in the online data associated with this paper \cite{the_gambit_collaboration_2017_801642}.}
\end{table*}

\begin{figure}[t]
\centering
\includegraphics[width=\columnwidth]{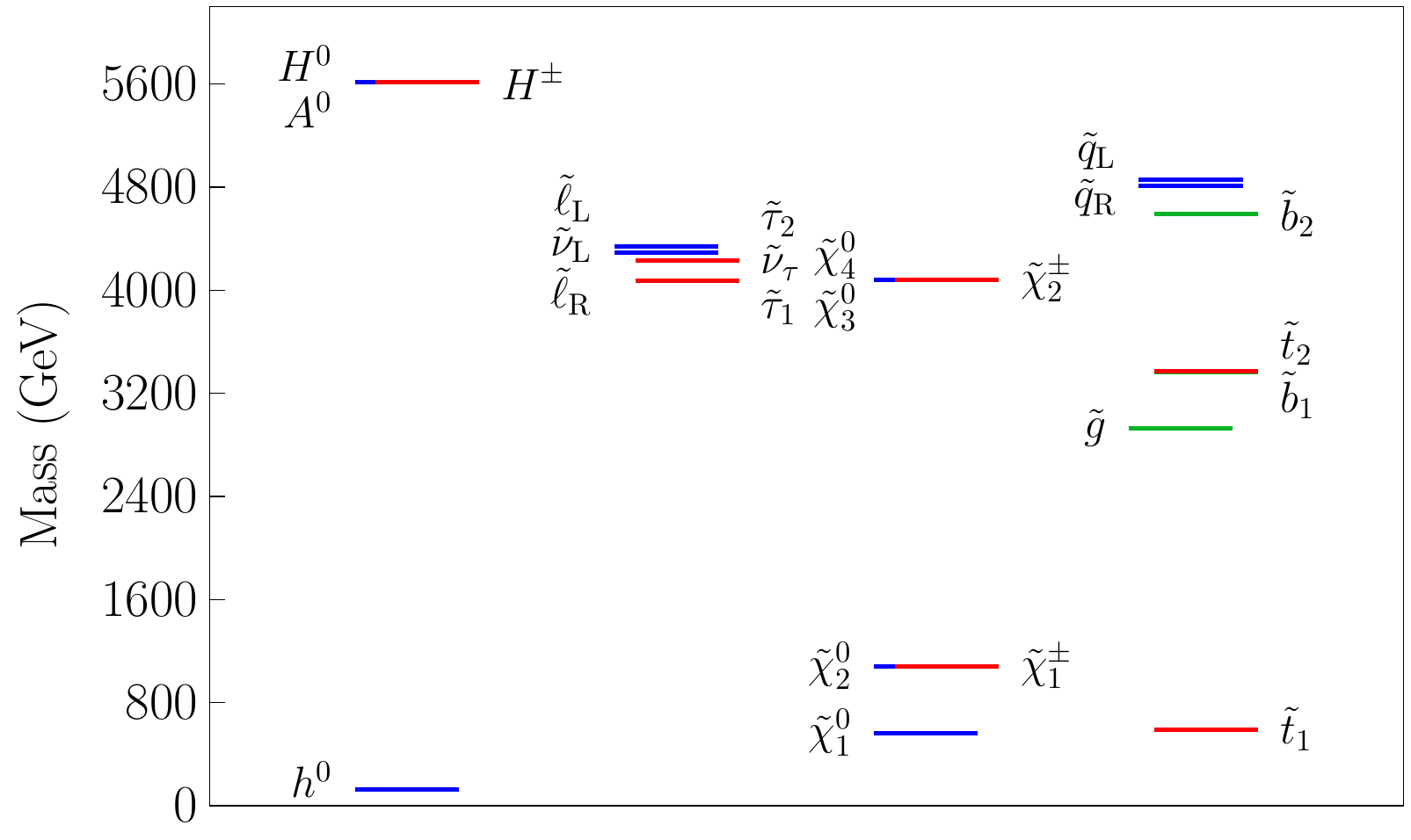}
\caption{Sparticle mass spectrum of the CMSSM best-fit point.}
\label{fig:cmssm-bf-spectrum}
\end{figure}

The best-fit points for each relic density mechanism (with positive and negative $\mu$ results combined) are given in Table \ref{tab:cmssm-bf-1}.  These are also shown in the figures of this section, as stars coloured by their corresponding region. We also give the mass spectrum for the global best-fit CMSSM point (column 5 of Table~\ref{tab:cmssm-bf-1}) in Fig.~\ref{fig:cmssm-bf-spectrum}, demonstrating that the only light superpartners are the lightest stop, the lightest two neutralinos and the lightest chargino, which is almost exactly degenerate in mass with the $\tilde{\chi}_2^0$. The $\tilde{\chi}_1^0$ is a pure bino for this point, whereas the $\tilde{\chi}_2^0$ and  $\tilde{\chi}_1^\pm$  are pure wino. The point generates a relic density within the allowed range through stop co-annihilation, but with a $\tilde{t}_1-\tilde{\chi}_1^0$ mass difference of $\simeq 40$\,GeV.  This mass difference should ensure prompt stop decay and potential visibility in future compressed spectrum searches at the LHC.

In Table \ref{tab:cmssm-bf-1}, we also give a detailed breakdown of the likelihood contributions from the different searches discussed in Sec.\ \ref{sec:lnL}, and compare to an `ideal' reference likelihood.  The ideal likelihood is defined as the best likelihood that a model could be expected to achieve, were it to perfectly predict all detections, and make no additional contribution beyond that predicted from background for all other searches.  Computing this is straightforward for most likelihood components, as it follows directly from setting the model prediction to either the observed value (e.g. $m_W$, $\Omega_\mathrm{c}h^2$, $a_\mu$, any nuisance parameters) or the background-only prediction (e.g. direct DM, LHC and neutrino searches).  In some cases however, where multiple sub-observables are involved and the background-only or SM prediction can be improved on by including a BSM contribution, a more nuanced calculation is required.  This is the case for the LHC Higgs and electroweak penguin ($B^0\to K^{*0}\mu^+\mu^-$) likelihoods.  For these components, we define the ideal likelihood to be the highest value possible in a more general phenomenological scenario.  In the flavour sector, we use the $B^0\to K^{*0}\mu^+\mu^-$ likelihood at the best fit point of the scan of the flavour EFT shown in Ref.\ \cite{FlavBit}.  In the Higgs sector, we take the best-fit likelihood obtainable by allowing the mass, width and decay branching fractions of a single scalar to vary freely in order to fit the full set of data contained in \higgssignals.

\begin{figure*}[t]
  \centering
  \includegraphics[width=0.49\textwidth]{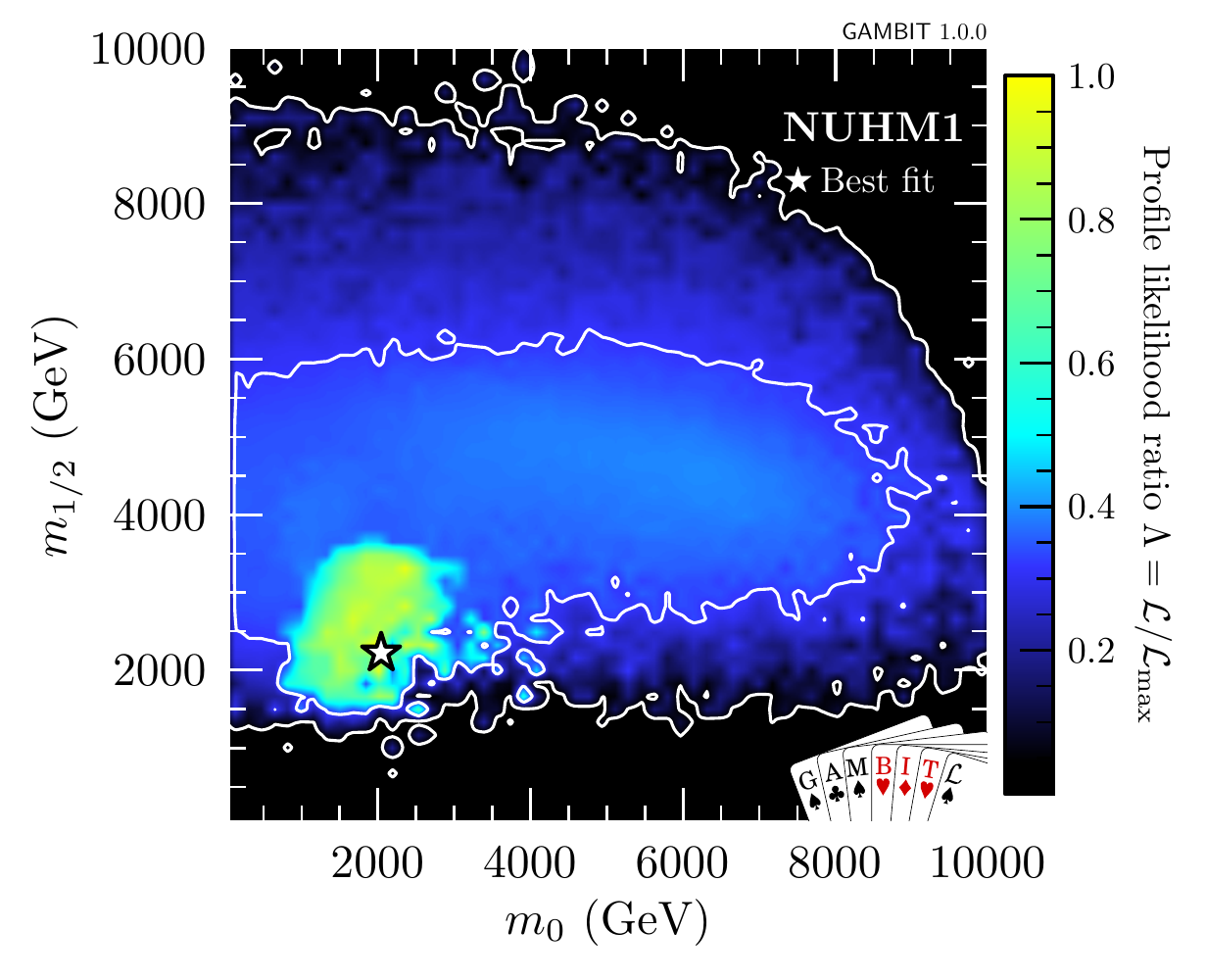}
  \includegraphics[width=0.49\textwidth]{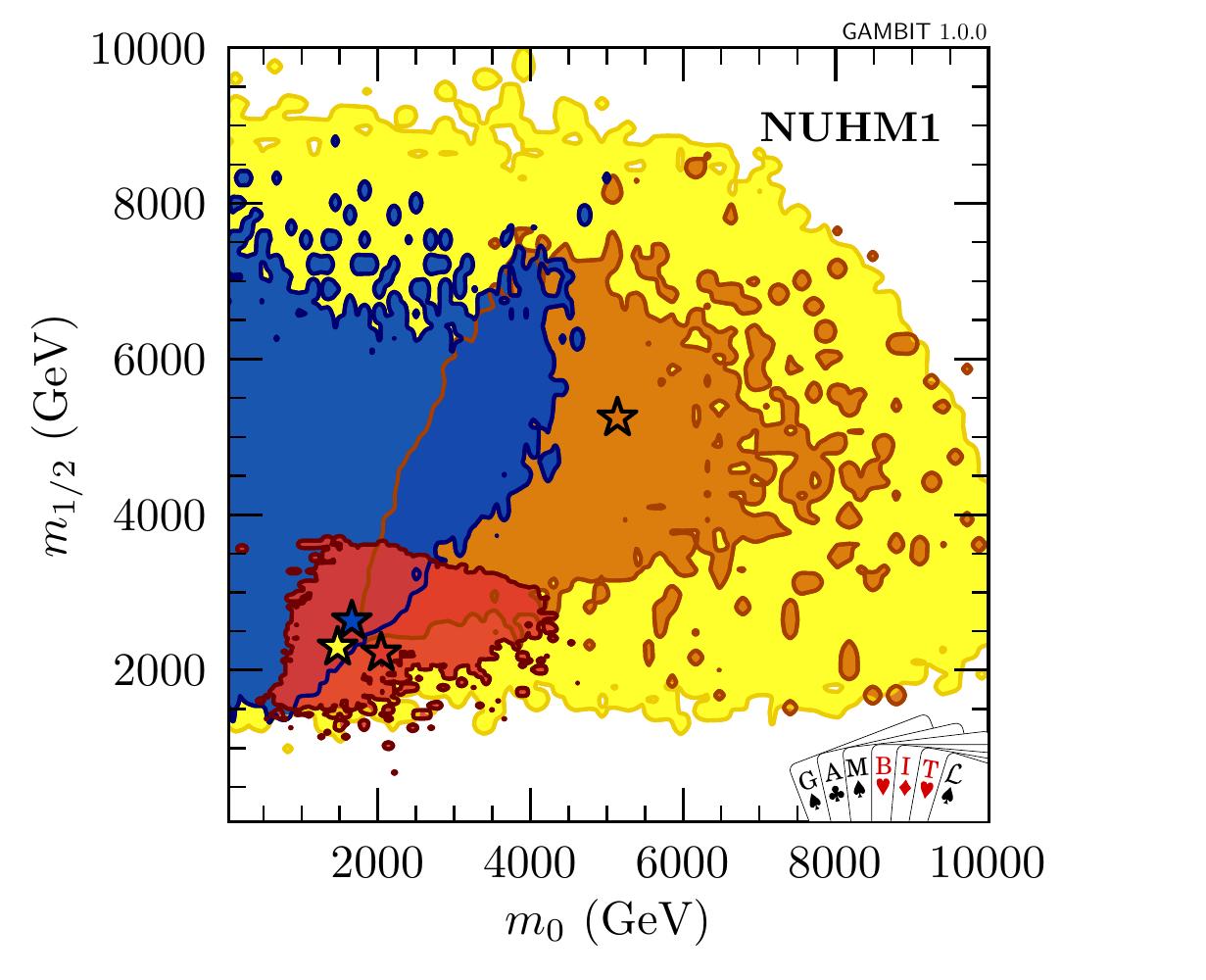}\\
  \includegraphics[width=0.49\textwidth]{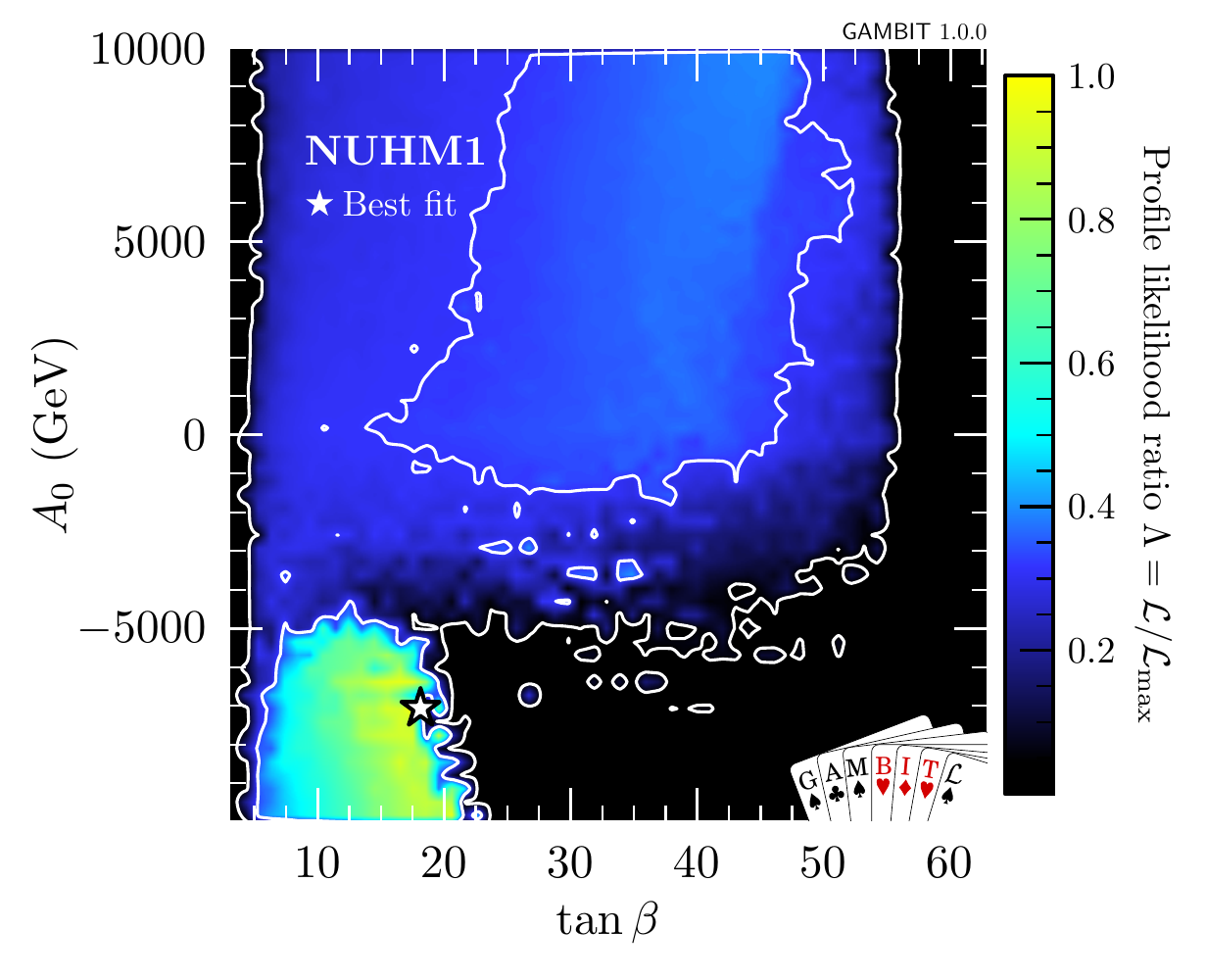}
  \includegraphics[width=0.49\textwidth]{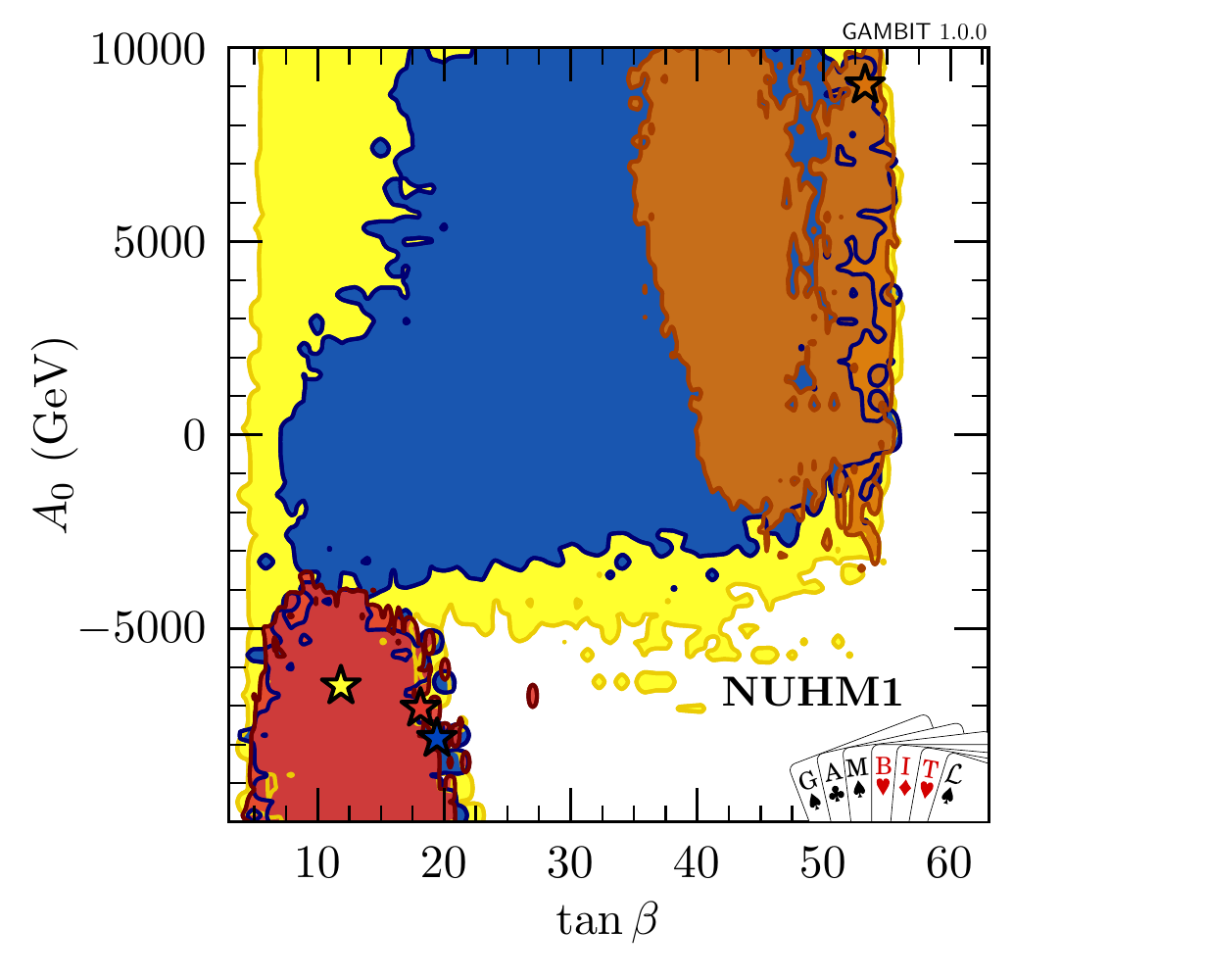}\\
  \includegraphics[height=4mm]{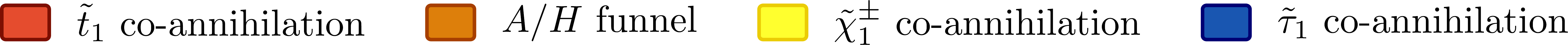}
  \caption{\textit{Left:} Profile likelihood ratio in the planes of the NUHM1 parameters $m_0$ and $m_{1/2}$ (top), and $\tan\beta$ and $A_{0}$ (bottom).  Explicit contour lines for 68\% and 95\% CL are drawn in white and the best fit point is indicated with a star. Right: Colour-coding shows the mechanisms to avoid exceeding the observed relic density of DM.}
  \label{fig:2d_parameter_plots_nuhm1}
\end{figure*}

\begin{figure*}[t]
  \centering
  \includegraphics[width=0.49\textwidth]{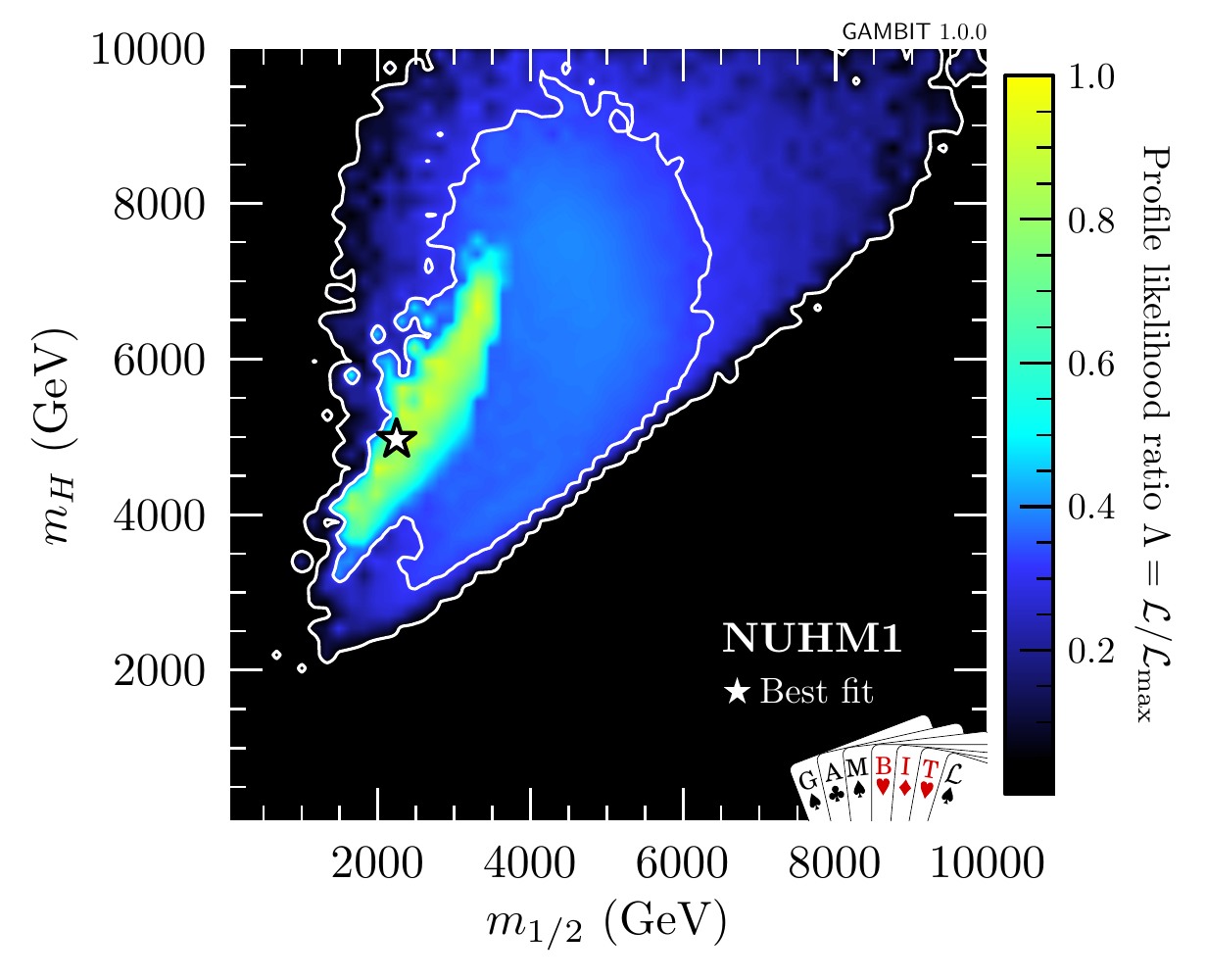}
  \includegraphics[width=0.49\textwidth]{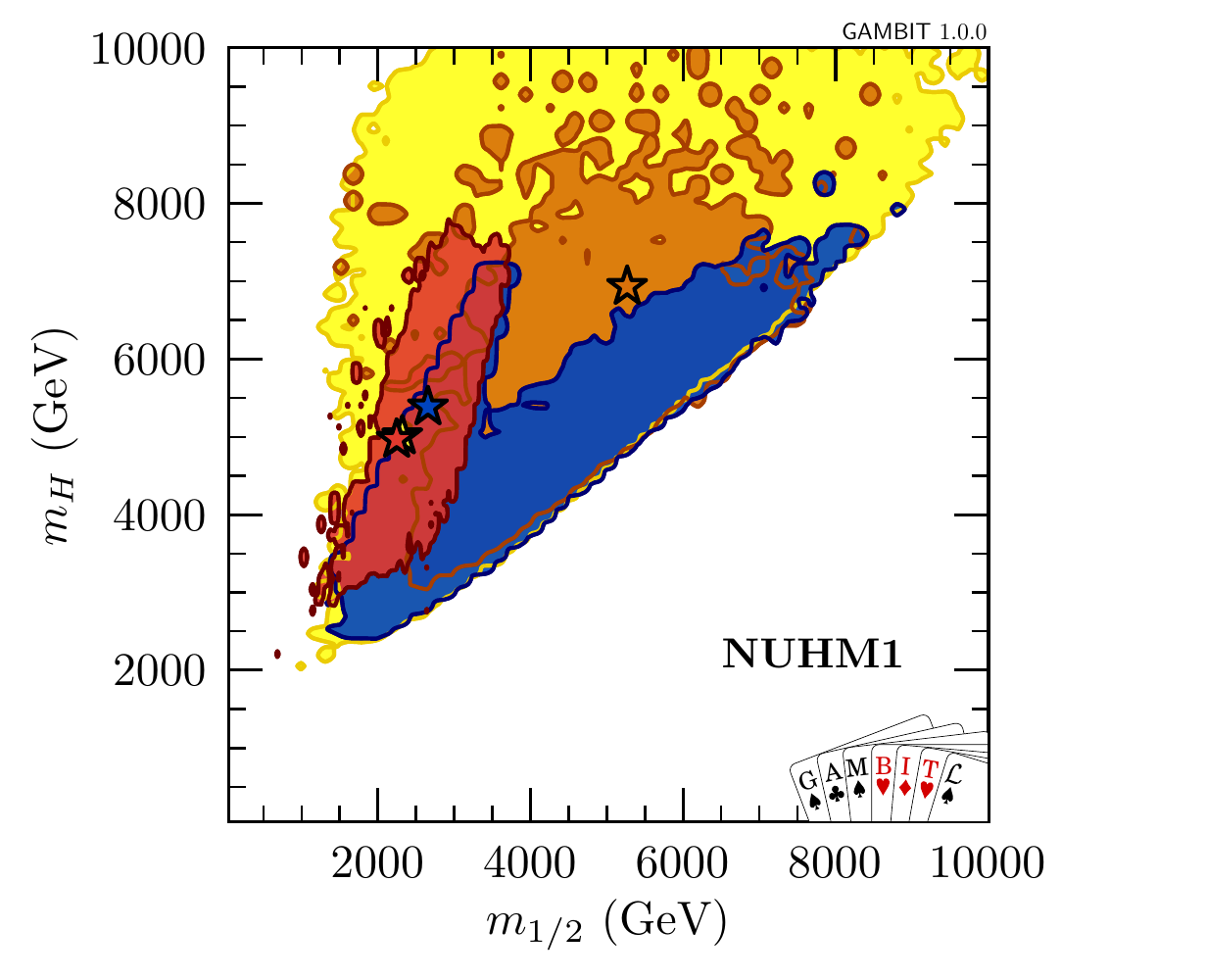}\\
  \includegraphics[width=0.49\textwidth]{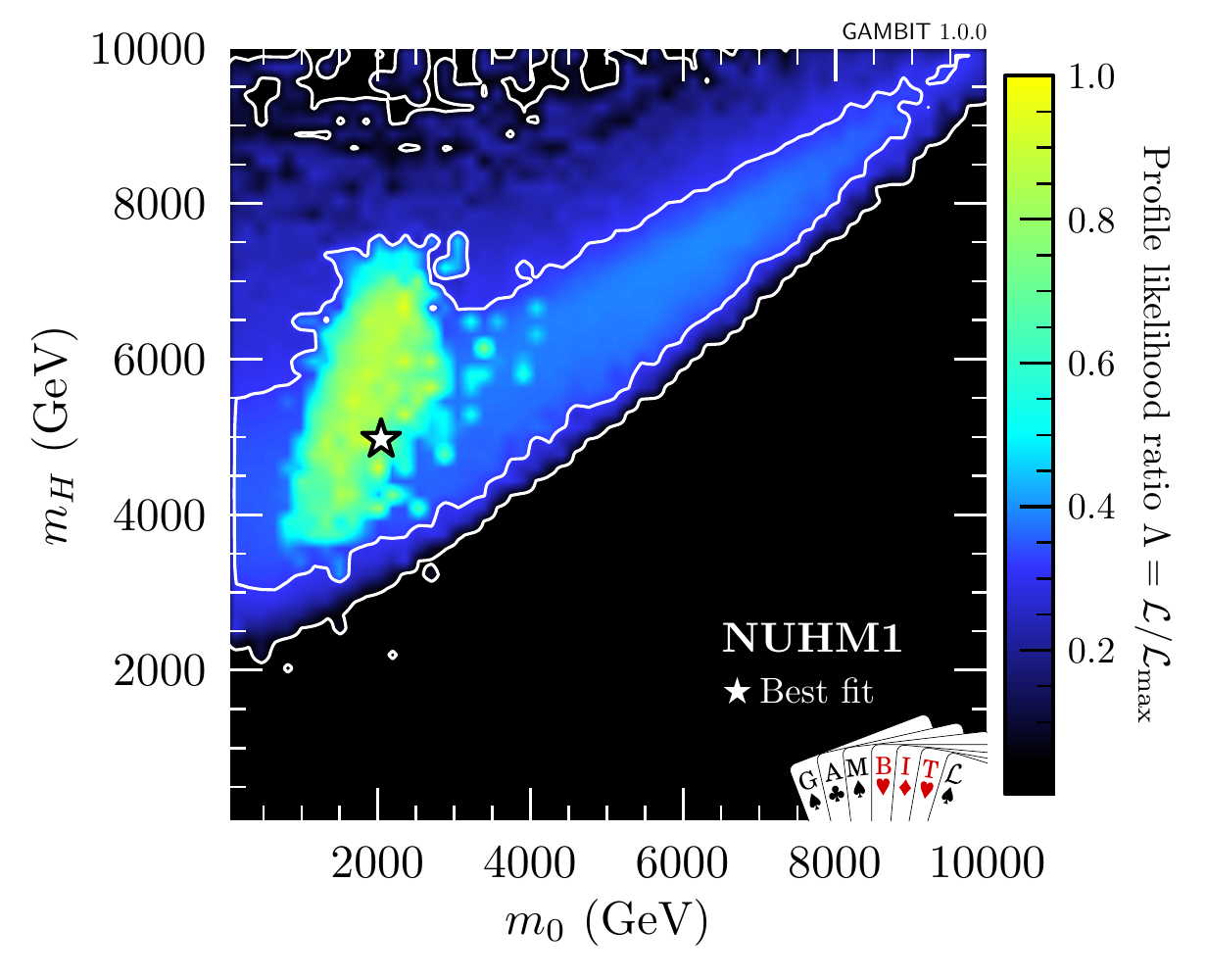}
  \includegraphics[width=0.49\textwidth]{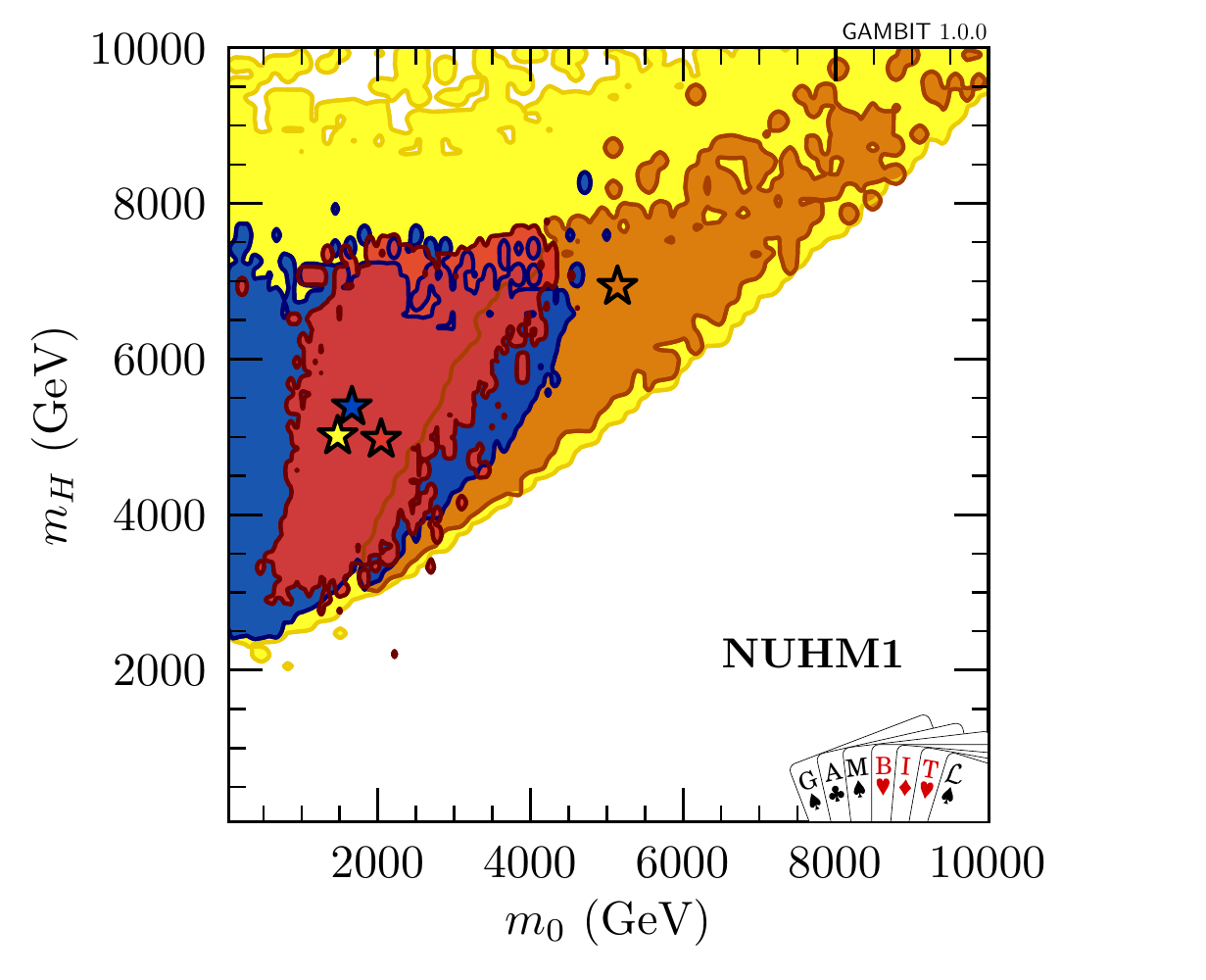}\\
  \includegraphics[height=4mm]{figures/rdcolours4.pdf}
  \caption{As per Fig.\ \ref{fig:2d_parameter_plots_nuhm1}, but for the $m_{1/2}$--$m_H$ (top) and $m_{0}$--$m_H$ (bottom) planes. }
  \label{fig:2d_parameter_plots_nuhm1_set2}
\end{figure*}

The log-likelihood difference of the best fit in the CMSSM to the ideal likelihood is $\Delta\mathcal{L}_\mathrm{BF} = 36.820$.  This difference is largely driven by known anomalies that cannot be explained by either the SM or the MSSM, including the magnetic moment of the muon ($\Delta\mathcal{L} = 6.289$; see Ref.\ \cite{SDPBit}) and the $B^0\to K^{*0}\mu^+\mu^-$ angular observables ($\Delta\mathcal{L} = 11.828$; see Ref.\ \cite{FlavBit}).  The largest contribution to $\Delta\mathcal{L}_\mathrm{BF}$ comes from anomalies in tree-level $B$ and $D$ decays, in particular the partially-correlated branching fraction ratios $R_D$ and $R_{D^*}$.  These have values of 0.308 and 0.248 respectively at the best-fit point of the scan.  For comparison, the SM predictions are 0.300 and 0.252, whereas the observed values are $0.403\pm0.047$ and $0.310\pm0.017$.  Further discussion and details can be found in the \flavbit paper \cite{FlavBit}.

The log-likelihood difference of a point relative to the ideal log-likelihood can be used to give some indication of the goodness of fit, as its definition is very similar to half of the ``likelihood $\chi^2$'' of Baker \& Cousins \cite{Baker:1983tu}.  The likelihood $\chi^2$ is known to follow a $\chi^2$ distribution in the asymptotic limit.  The main difficulty in using this fact is estimating the effective degrees of freedom of the fit, as carrying out simulations to find the true distribution of our test statistic is computationally intractable.\footnote{We note that a similar thing \textit{has} been done in the CMSSM \cite{Fittinocoverage}, but using a likelihood function far quicker to compute than ours, based on interpolation in a 2D grid of LHC signal yields rather than explicit simulation for each parameter combination.}  Given the number of observables that actively constrain the fit, a reasonable guess for the effective degrees of freedom is probably something in the range of 30--50, leading to a $p$-value of between $2\times10^{-5}$ and 0.02 for the CMSSM; neither a particularly good fit nor catastrophically bad, given the uncertainties involved in the estimate of the $p$-value.  Taking 40 as a canonical estimate of the degrees of freedom, for the sake of later comparison with the NUHM1 and NUHM2, the $p$-value would be $9.4\times10^{-4}$.

\subsection{NUHM1}
\label{sec:NUHM1}
The main results from the NUHM1 scan are shown in Figs.~\ref{fig:2d_parameter_plots_nuhm1} and~\ref{fig:2d_parameter_plots_nuhm1_set2}. Fig.\ \ref{fig:2d_parameter_plots_nuhm1} shows results in the
$m_0$--$m_{1/2}$ and $\tan\beta$--$A_{0}$ planes, with plots of the
profile likelihood ratio on the left and the DM annihilation mechanisms, defined as
in the previous subsection, on the right. In comparison to the CMSSM
equivalent, Fig.\ \ref{fig:2d_parameter_plots_cmssm}, one can see that
the additional freedom in the NUHM1 substantially extends the
likelihood contours, so that much of the parameter space is now
allowed.

In particular, we now find a stau co-annihilation region, which was
absent in the CMSSM results.  The extra freedom present in the Higgs
sector in the NUHM1 avoids the penalty from the LHC Higgs likelihood
seen in stau co-annihilation models in the CMSSM.  Furthermore, we now
see chargino co-annihilation solutions within the $2 \sigma$ contours
that extend to arbitrarily low $m_0$.  These are both a consequence of
the fact that once $m_0$ is decoupled from $m_H$, the former can be
pushed low without impacting EWSB.  This allows light staus to exist, making
stau co-annihilation viable, and also means that $|\mu|$ can be low at
arbitrarily small $m_0$, leading to Higgsino LSPs.

Such low values of $m_0$ in chargino-coannihilation scenarios suggests
that the first- and second-generation squarks may be light enough
to be constrained directly by collider searches.  However, a detailed
examination reveals that their masses remain above 2\,TeV, and out of reach
of LHC limits, for all models within our 2$\sigma$ contours.

A similar expansion of the chargino co-annihilation region \footnote{The stau
co-annihilation region also extends to arbitrarily low $m_0$, but this
is because our definition of stau co-annihilation admits the
possibility of an under-abundant Higgsino DM candidate with hybrid
stau and chargino co-annihilation.} has been seen in the previous
literature comparing the CMSSM and NUHM1 models (see e.g.\ Fig.~6 of
Ref.\ \cite{MasterCode12b} and Fig.~1 of Ref.\ \cite{Mastercode15},
although the contours do not reach arbitrarily low $m_0$ for all
$m_{1/2}$ in those studies. This difference can be explained by the
additional freedom associated with only applying the relic density
measurement as a one-sided limit.  We checked that demanding
neutralinos make up all of DM removes some low-$m_0$ scenarios from
the $2 \sigma$ contours, such that a better agreement with the results
in the literature is obtained.

Another interesting feature of this is that these low $m_0$ values
mean that the NUHM1 admits significantly lighter squarks within the
chargino co-annihilation region than in the CMSSM.

We give results in the $m_H$--$m_{1/2}$ and $m_H$--$m_{0}$ planes for the
NUHM1 in Fig.~\ref{fig:2d_parameter_plots_nuhm1_set2}. These
plots show a sharp cut-off in the likelihood near the diagonals
$m_0 \approx m_H$ and $m_{1/2} \approx m_H$, such that $m_H$ must be greater than both $m_0$ and $m_{1/2}$. This structure emerges
from the combination of several effects. Reducing $m_H$ with respect
to $m_0$ and $m_{1/2}$ leads to $m_{H_u}^2$ running to a more negative
value, and this in turn leads to a larger $\mu$ value through the EWSB
conditions. Past this boundary in
Fig.~\ref{fig:2d_parameter_plots_nuhm1_set2}, $\mu$ is then always
significantly larger than the bino mass, $M_1$.  As a result the
neutralino LSP is always bino in these scenarios and requires either
an $A/H$-funnel or sfermion co-annihilation mechanism to reduce the
relic density to the measured value or below.  The $A/H$-funnel
mechanism also requires $\mu$ to be small, as $\mu^2$ gives a
contribution to the pseudoscalar mass.  As a result, if $\mu$ is much
larger than $2 M_1$, then the relation for the $A/H$-funnel mechanism, $m_A \approx 2 m_{\tilde\chi_1^0} \approx 2
M_1$, cannot be achieved.  Although we do find majority-bino LSPs
annihilating through an $A/H$-funnel, these have smaller values of
$\mu$ than can be achieved when $m_H$ is less than either $m_0$ or
$m_{1/2}$.  Finally, sfermion co-annihilation can be effective in
this region, but only for lower values of $m_0$ and $m_{1/2}$.  In
those scenarios, if $m_H \lesssim m_0, m_{1/2}$ the likelihood is
suppressed by the LHC Higgs likelihood, because it is
difficult to fit the $125$\,GeV Higgs there.

\begin{table*}[p]
\small
\center
\sisetup{round-mode = places, round-precision = 3,round-integer-to-decimal}
\begin{tabular}{l S[table-format=+3.3] S[table-format=+3] S[table-format=+3] S[table-format=+3] S[table-format=+3] S[table-format=+3]}
Likelihood term & {\ \ Ideal} & {$A/H$-funnel} & {\ \ \ $\tilde{\tau}$ co-ann.} & {\ \ \ \ $\tilde{t}$ co-ann.} & {\ $\tilde{\chi}_1^{\pm}$ co-ann.} & {\ \ \ $\Delta\ln\mathcal{L}_\mathrm{BF}$} \\
\hline
LHC sparticle searches & 0 &  0.000  & 0.000 & 0.000 & 0.000  &  0 \\
LHC Higgs & -37.7339 &  -38.646  &  -38.182  &  -38.271  &  -38.531  &  .5371  \\
LEP Higgs & 0 &  0.000  &  0.000  &  0.000  &  0.000  &  0 \\
ALEPH selectron & 0 &  0.000  & 0.000  & 0.000  & 0.000  &  0 \\
ALEPH smuon & 0 &  0.000  &  0.000  &  0.000  &  0.000  &  0  \\
ALEPH stau & 0 &  0.000  &  0.000  &  0.000  &  0.000  &  0 \\
L3 selectron & 0 &  0.000  &  0.000  &  0.000  &  0.000  &  0  \\
L3 smuon & 0 & 0.000  &  0.000  &  0.000  &  0.000  &  0 \\
L3 stau & 0 & 0.000  &  0.000  &  0.000  &  0.000  &  0 \\
L3 neutralino leptonic & 0 & 0.000  & 0.000  & 0.000  & 0.000  &  0  \\
L3 chargino leptonic & 0 &  0.000  &  0.000  &  0.000  &  0.000  &  0 \\
OPAL chargino hadronic & 0 &  0.000  &  0.000  &  0.000  &  0.000  &  0 \\
OPAL chargino semi-leptonic & 0 &  0.000  &  0.000  &  0.000  &  0.000  &  0 \\
OPAL chargino leptonic & 0 &  0.000  &  0.000  &  0.000  &  0.000  &  0 \\
OPAL neutralino hadronic & 0 &  0.000  &  0.000  &  0.000  &  0.000  &  0  \\
$B_{(s)}\to \mu^+\mu^-$ & 0 &  -1.985  &  -2.033  &  -2.032  &  -2.043  &  2.032  \\
Tree-level $B$ and $D$ decays & 0 &  -15.703  &  -15.286  &  -15.286  &  -15.282  &  15.286 \\
$B^0\to K^{*0}\mu^+\mu^-$ & -184.260 &  -196.553  &  -195.323  &  -194.855  &  -194.825  &  10.595  \\
$B\to X_s\gamma$ & 9.799 &  9.272  &  8.696  &  8.430  &  8.351  &  1.369  \\
$a_\mu$ & 20.266 &  14.158  &  13.837  &  13.819  &  13.836  &  6.447  \\
$W$ mass & 3.281 &  3.095  & 3.062  & 3.075  & 3.096  &  .206  \\
Relic density & 5.989 &  5.989  &  5.989  &  5.989  &  5.989  &  0 \\
PICO-2L & -1 &  -1.000  &  -1.000  &  -1.000  &  -1.000  &  0 \\
PICO-60 F & 0 &  0.000  &  0.000  &  0.000  &  -0.001  &  0  \\
SIMPLE 2014 & -2.972 &  -2.972  &  -2.972  &  -2.972  &  -2.972  &  0  \\
LUX 2015  & -0.64 &  -0.666  &  -0.646  &  -0.659  &  -0.676  &  .019  \\
LUX 2016  & -1.467 &   -1.519  &  -1.479  &  -1.504  &  -1.539  &  .037 \\
PandaX 2016  & -1.886 &  -1.921  &  -1.894  &  -1.912  &  -1.936  &  .026  \\
SuperCDMS 2014 & -2.248 &  -2.248  &  -2.248  &  -2.248  &  -2.248  &  0 \\
XENON100 2012 & -1.693 &   -1.680  &  -1.690  &  -1.684  &  -1.675  &  .009  \\
IceCube 79-string & 0.0 &  -0.014  &  0.000  &  0.000  &  -0.135  &  0 \\
$\gamma$ rays (\textit{Fermi}-LAT dwarfs) & -33.244 &  -33.384  &  -33.364  &  -33.373  &  -33.398  &  .129 \\
$\rho_0$ & 1.142 &  1.141  &  1.141  &  1.140 &  1.141  &  .002 \\
$\sigma_s$ and $\sigma_l$ & -6.115 &  -6.115  &  -6.135  &  -6.124  &  -6.117  &  .009 \\
$\alpha_s(m_Z)(\MSBar)$ & 6.500 &  6.491  &  6.488  &  6.493  &  6.494  &  .007 \\
Top quark mass & -0.645 &  -0.647  &  -0.673  &  -0.655  &  -0.645  &  .010 \\
\hline
Total & -226.927 & -264.907 & -263.712 & -263.629 & -264.115 & 36.702\\
\hline
&  &  &  &  & \\
Quantity &  & {$A/H$-funnel} & {\ \ \ $\tilde{\tau}$ co-ann.} & {\ \ \ \ $\tilde{t}$ co-ann.} & {\ $\tilde{\chi}_1^{\pm}$ co-ann.} &  \\
\cmidrule{1-6}
$A_0$ &  &  9084.348  &  -7798.283  &  -7016.861  &  -6439.114 \\
$m_0$ &  &  5139.563  &  1659.858  &  2042.775  &  1472.445 \\
$m_{1/2}$ &  &  5266.693  &  2656.510  &  2245.476  &  2319.968 \\
$m_{H}$ &  &  6954.864  &  5407.626  &  4990.078  &  5034.071 \\
$\tan\beta$ &  &  53.263  &  19.430  &  18.128  &  11.840 \\
$\mathrm{sgn}(\mu)$ &  &  \multicolumn{1}{r}{$+$}  &  \multicolumn{1}{r}{$-$}  &  \multicolumn{1}{r}{$-$}  &  \multicolumn{1}{r}{$-$} \\
$m_t$ & &  173.393  &  173.522  &  173.451  &  173.362 \\
$\alpha_s(m_Z)(\MSBar)$ & &  0.119  &  0.118  &  0.119  &  0.119 \\
$\rho_0$ & &  0.403  &  0.398  &  0.408  &  0.396 \\
$\sigma_s$ &  &  42.776  &  43.646  &  43.747  &  42.478 \\
$\sigma_l$ &  &  57.737  &  56.355  &  57.132  &  58.024 \\
\cmidrule{1-6}
$M_1$ & &  2419.401  &  1184.390  &  994.971  &  1023.177 \\
$\mu$ & &  836.283  &  -1753.895  &  -1462.491  &  -351.100 \\
$m_{\tilde{t}_1}$ & &  7902.945  &  1198.127  &  1032.608  &  1012.967 \\
$m_{\tilde{\tau}_1}$ & &  2231.113  &  1295.803  &  1819.486  &  1513.479 \\
$m_{A}$ & &  1805.767  &  5428.634  &  5002.455  &  5122.233 \\
$m_{h}$ & &  125.026  &  124.544  &  124.531  &  124.903 \\
$m_{\tilde{\chi}_1^0}$ & &  856.207  &  1179.991  &  993.716  &  358.905 \\
$($\%bino, \%Higgsino$)$ & & \multicolumn{1}{r}{$(0,100)$} & \multicolumn{1}{r}{$(100,0)$} & \multicolumn{1}{r}{$(100,0)$} & \multicolumn{1}{r}{$(0,100)$} \\
$m_{\tilde{\chi}_2^0}$ & &  -858.645  &  1760.580  &  1467.989  &  -364.815 \\
$($\%bino, \%Higgsino$)$ & & \multicolumn{1}{r}{$(0,100)$} & \multicolumn{1}{r}{$(0,98)$} & \multicolumn{1}{r}{$(0,98)$} & \multicolumn{1}{r}{$(0,100)$} \\
$m_{\tilde{\chi}_1^\pm}$ & &  857.791  &  1760.608  &  1467.887  &  362.366 \\
$($\%wino, \%Higgsino$)$ & & \multicolumn{1}{r}{$(0,100)$} & \multicolumn{1}{r}{$(2,98)$} & \multicolumn{1}{r}{$(2,98)$} & \multicolumn{1}{r}{$(0,100)$} \\
$m_{\tilde{g}}$ & &  10470.041  &  5462.593  &  4705.842  &  4823.285 \\
$\Omega h^2$ & &  {$7.03\times10^{-2}$}  &  {$5.24\times10^{-2}$} &  {$9.29\times10^{-2}$}  & {$1.59\times10^{-2}$} \\
\cmidrule{1-6}
\end{tabular}
\caption{\label{tab:nuhm1-bf-1} Best-fit points in the NUHM1, for each of the regions characterised by a specific mechanism for suppressing the relic density of dark matter. Here we show the likelihood contributions,  parameter values at each point, and some quantities relevant for the interpretation of mass spectra at the different best fits.  We also give likelihood components for a canonical `ideal' likelihood (see text), along with its offset from the global best-fit point. SLHA1 and SLHA2 files corresponding to the best-fit point in each region can be found in the online data associated with this paper \cite{the_gambit_collaboration_2017_801642}.}
\end{table*}

We investigated charge- and colour-breaking minima in the NUHM1 in the same way as in the CMSSM (Sec.\ \ref{sec:CMSSM}).  As in the CMSSM, a number of models within our 95\% CL regions are affected by one or more of the three proposed conditions, but removing all such parameter combinations does not move the best fit to a different region, nor substantially change the regions of parameter space preferred by our fits.

\begin{figure}[t]
\centering
\includegraphics[width=\columnwidth]{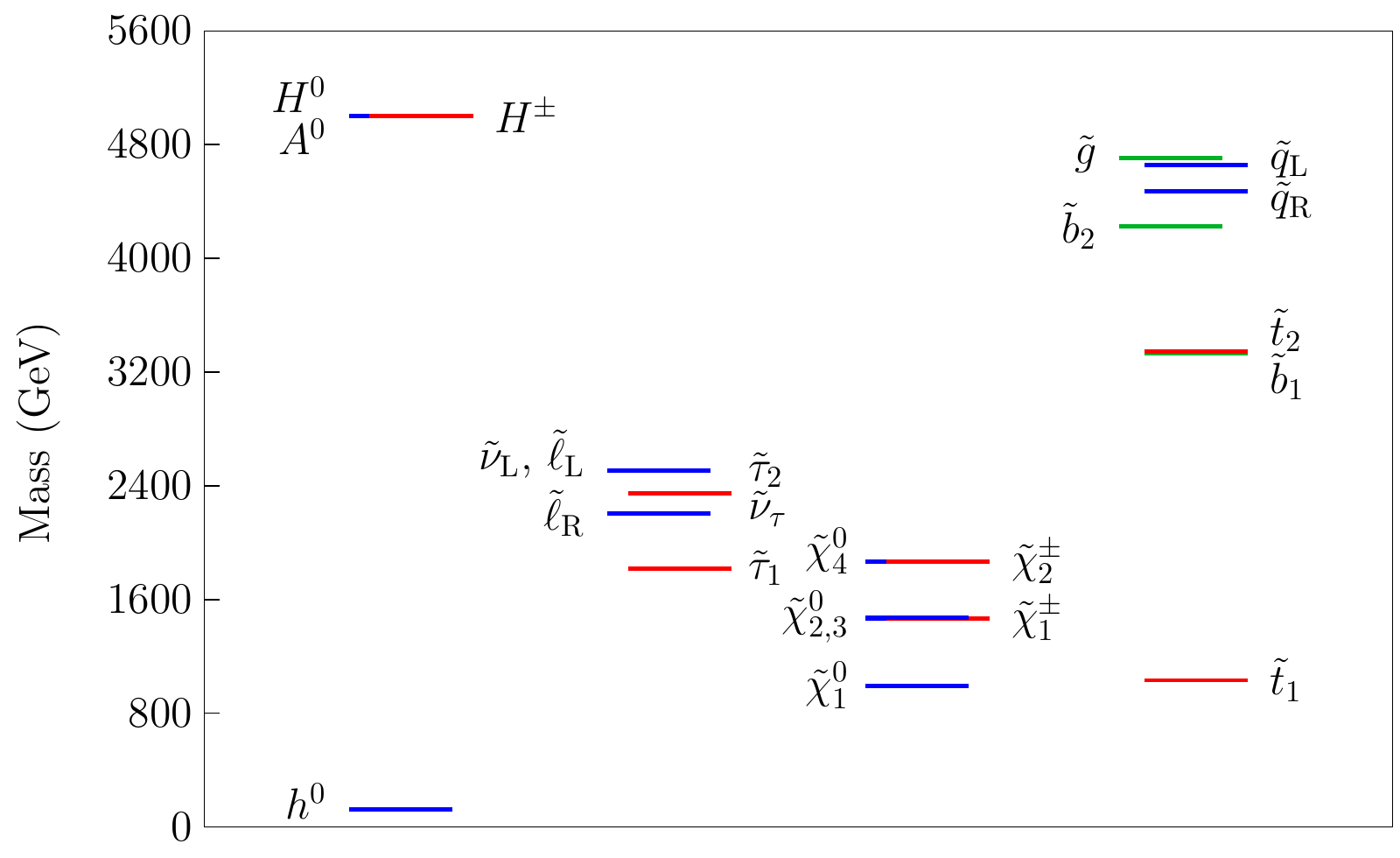}
\caption{Sparticle mass spectrum of the NUHM1 best-fit point.}
\label{fig:nuhm1-bf-spectrum}
\end{figure}

As we did for the CMSSM, in Table \ref{tab:nuhm1-bf-1} we show the best-fit points for each mechanism for depleting the relic density of DM. The best-fit point in the chargino co-annihilation region has small $\tilde{\chi}_1^0$, $\tilde{\chi}_2^0$ and $\tilde{\chi}_1^\pm$ masses, but escapes LHC exclusion due to the highly compressed mass spectrum for these sparticles. As in the CMSSM, the overall best-fit point lies in the stop co-annihilation region.  Its mass spectrum is shown in Figure~\ref{fig:nuhm1-bf-spectrum}. There are important differences to the CMSSM case, however. Firstly, the stop is heavier, now sitting just above 1\,TeV in mass. The $\tilde{t}_1-\tilde{\chi}_1^0$ mass difference is once again roughly 40\,GeV, ensuring prompt decay of the stop. The heavier stop mass is accompanied by a heavier mass spectrum in general, with no sparticles lighter than 800\,GeV in mass. The $\tilde{\chi}_1^0$ is pure bino, but the $\tilde{\chi}_2^0$, $\tilde{\chi}_3^0$ and $\tilde{\chi}_1^\pm$ are now predominantly Higgsino in character, leaving the $\tilde{\chi}_4^0$ and $\tilde{\chi}_2^\pm$ to be mostly wino. Discovery of this point would be very challenging at the LHC in the near future, due to the heavy weakly-coupled states, and the lack of light coloured states that have a large mass splitting with the $\tilde{\chi}_1^0$.

For the NUHM1, $\Delta\ln\mathcal{L}_\mathrm{BF} = 36.702$, slightly better than what we found in the CMSSM.  For the sake of comparison with the CMSSM ($p=9.4\times10^{-4}$ if computed with 40 degrees of freedom), we can compute a $p$-value assuming one less degree of freedom, i.e. 39.  This gives $7.1\times10^{-4}$, slightly \textit{worse} than the CMSSM.  We see that despite the improvement in the fit, the fact that it has not delivered a sufficiently large improvement in $\Delta\ln\mathcal{L}_\mathrm{BF}$ means that this is not enough to outweigh the penalty associated with the introduction of the additional parameter.

\begin{figure*}[t]
  \centering
  \includegraphics[width=0.49\textwidth]{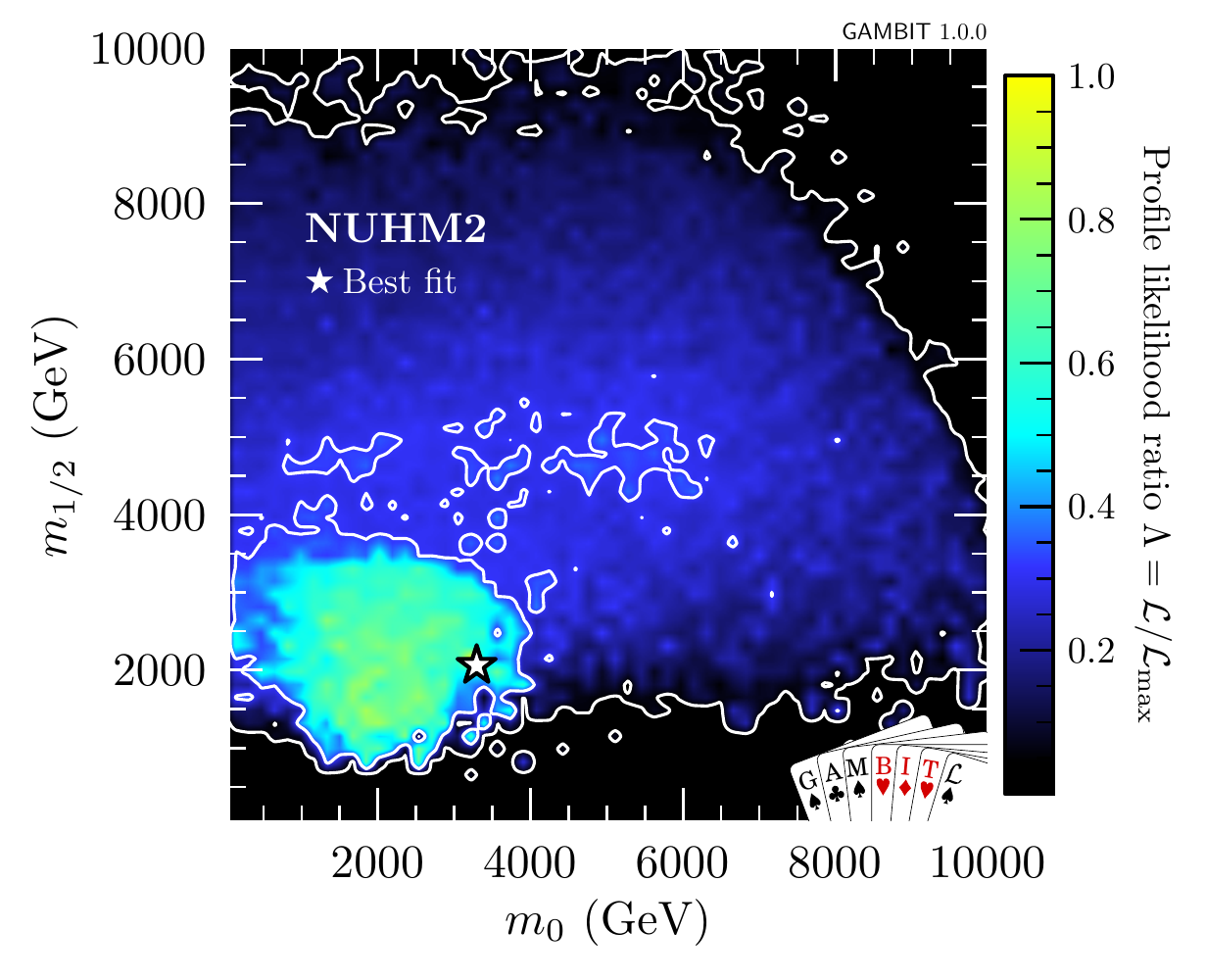}
  \includegraphics[width=0.49\textwidth]{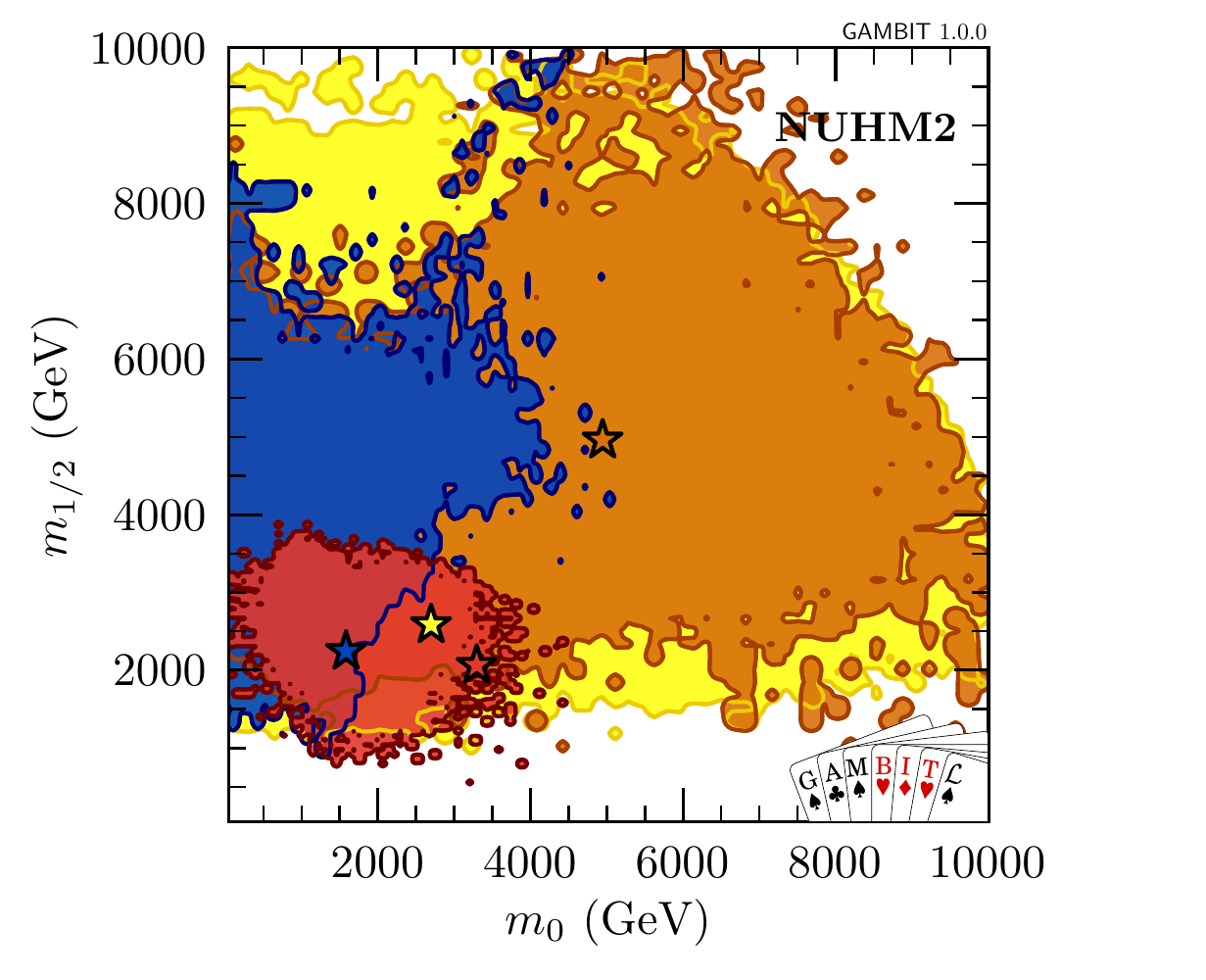}\\
  \includegraphics[width=0.49\textwidth]{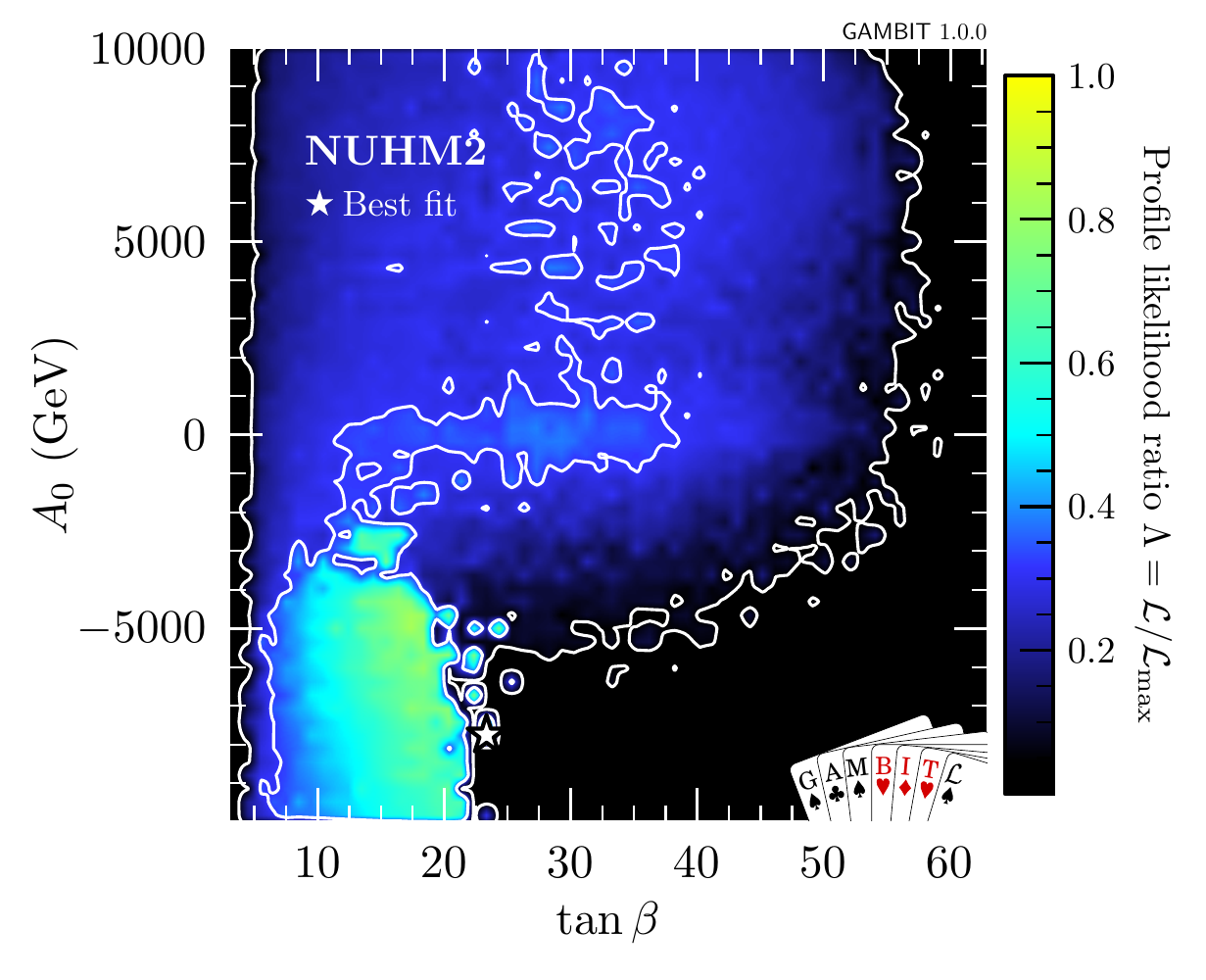}
  \includegraphics[width=0.49\textwidth]{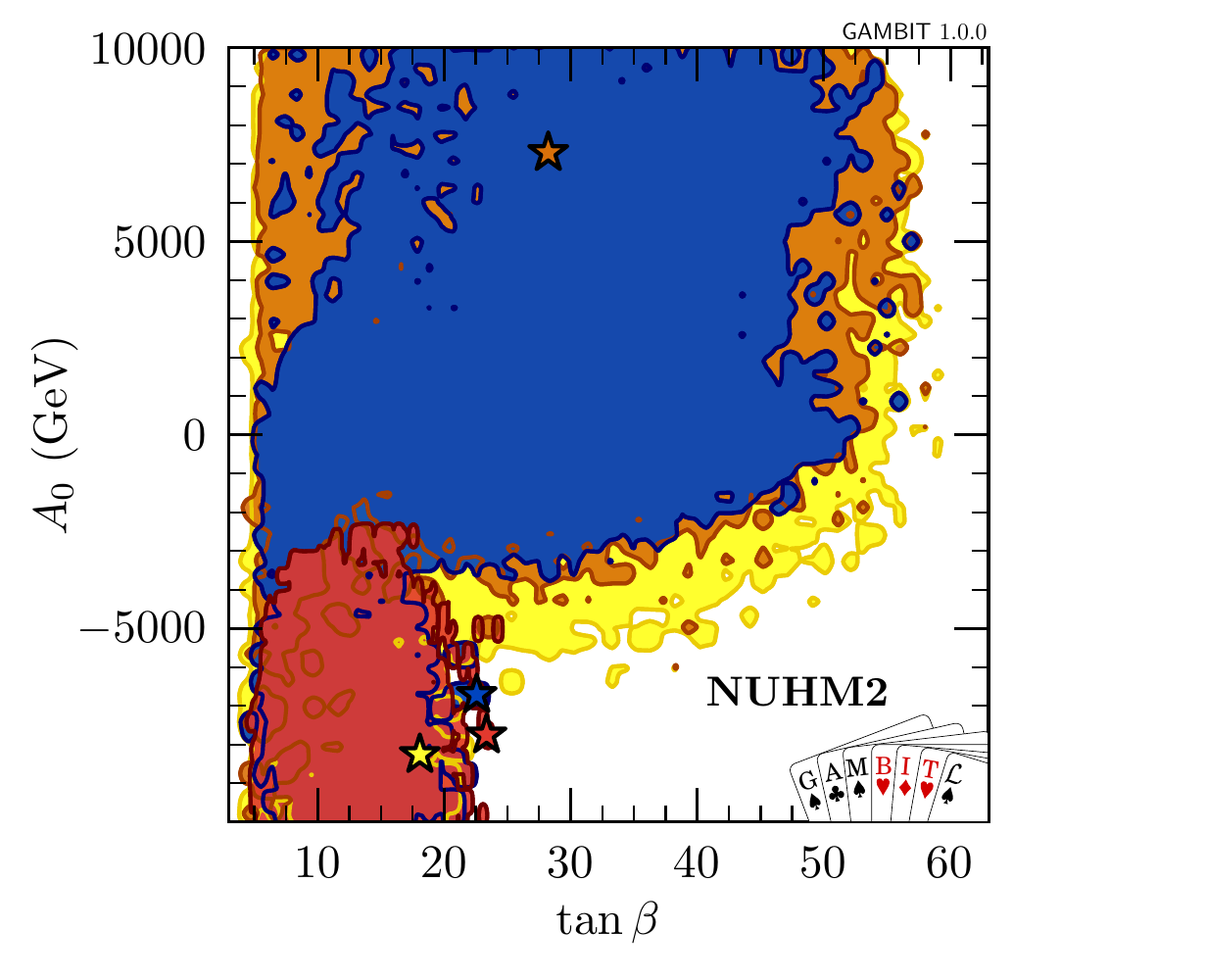}\\
  \includegraphics[height=4mm]{figures/rdcolours4.pdf}
  \caption{\textit{Left:} Profile likelihoods in the NUHM2, in terms of the $m_0-m_{1/2}$ and $A_0-\tan\beta$ planes.  \textit{Right:} corresponding mechanisms to avoid exceeding the observed relic density of DM.}
  \label{fig:2d_parameter_plots_nuhm2}
\end{figure*}

\begin{figure*}[t]
  \centering
  \includegraphics[width=0.49\textwidth]{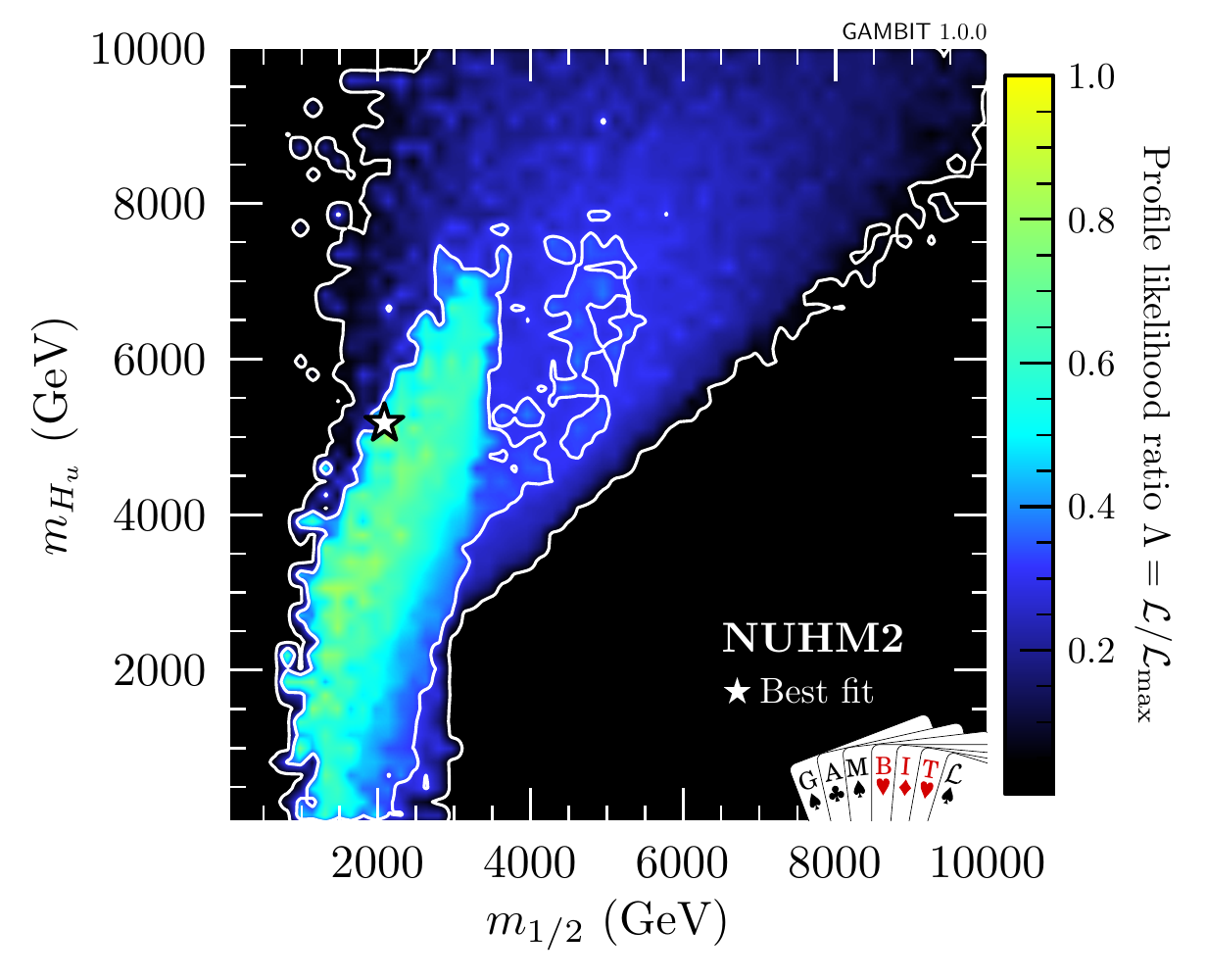}
  \includegraphics[width=0.49\textwidth]{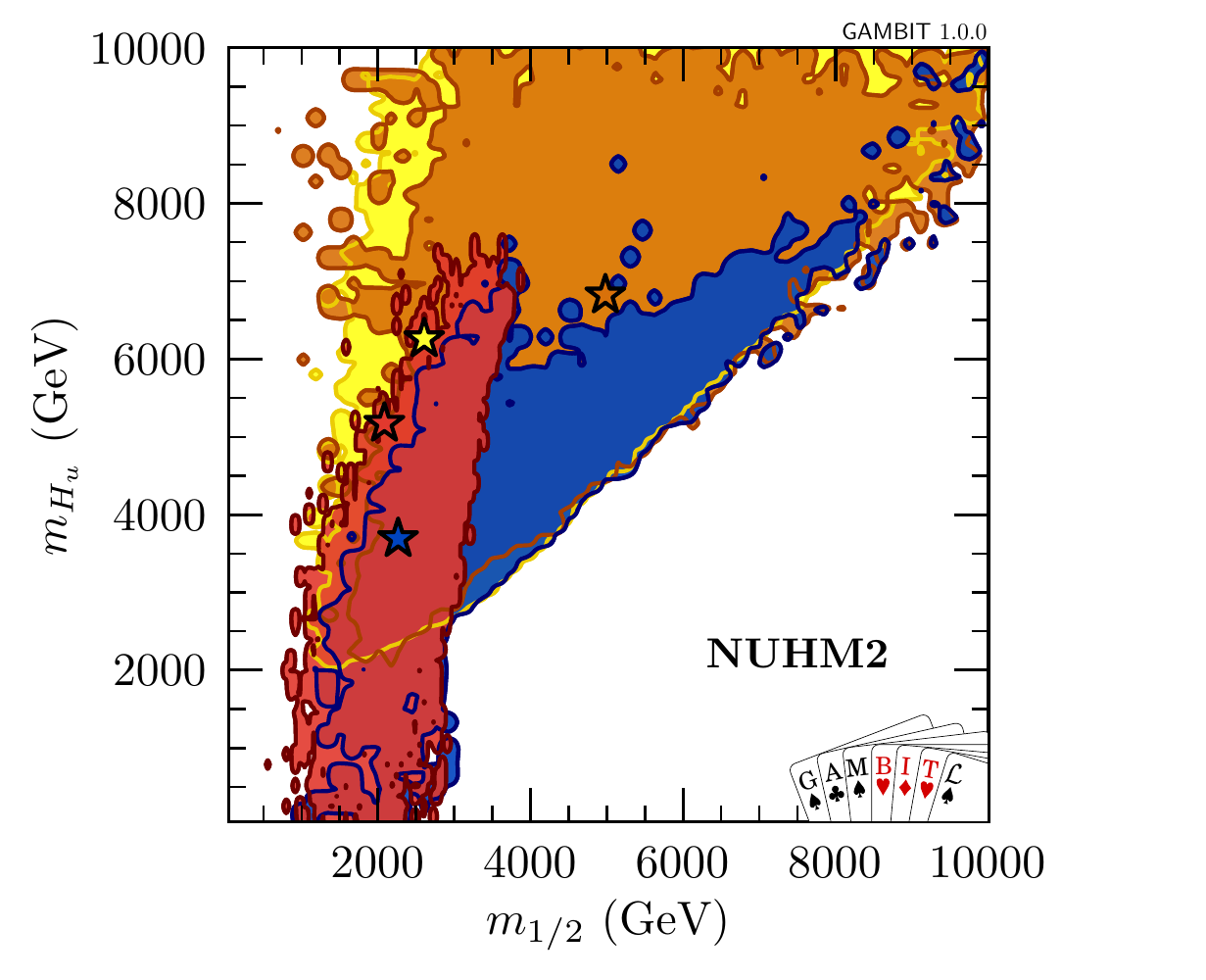}\\
  \includegraphics[width=0.49\textwidth]{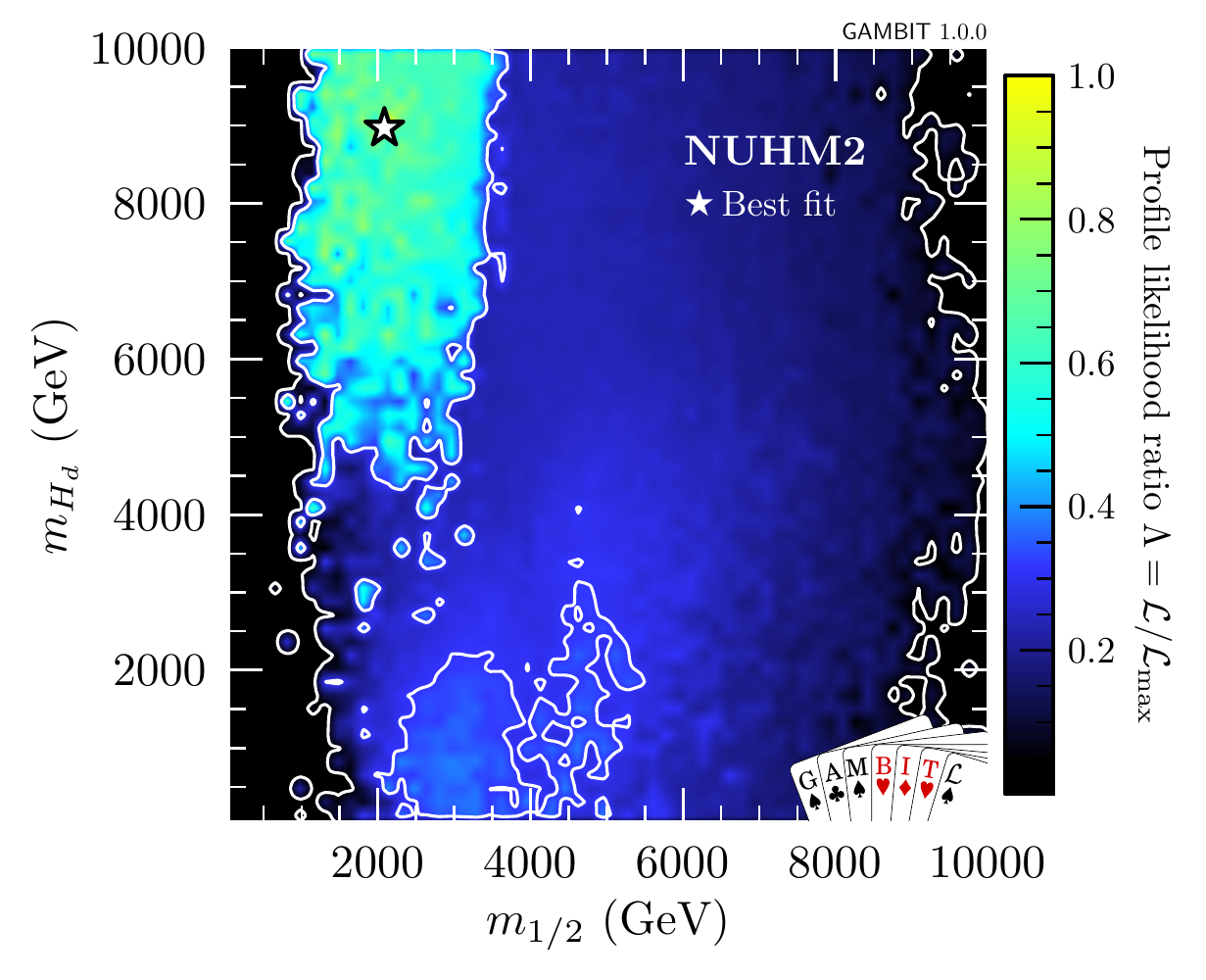}
  \includegraphics[width=0.49\textwidth]{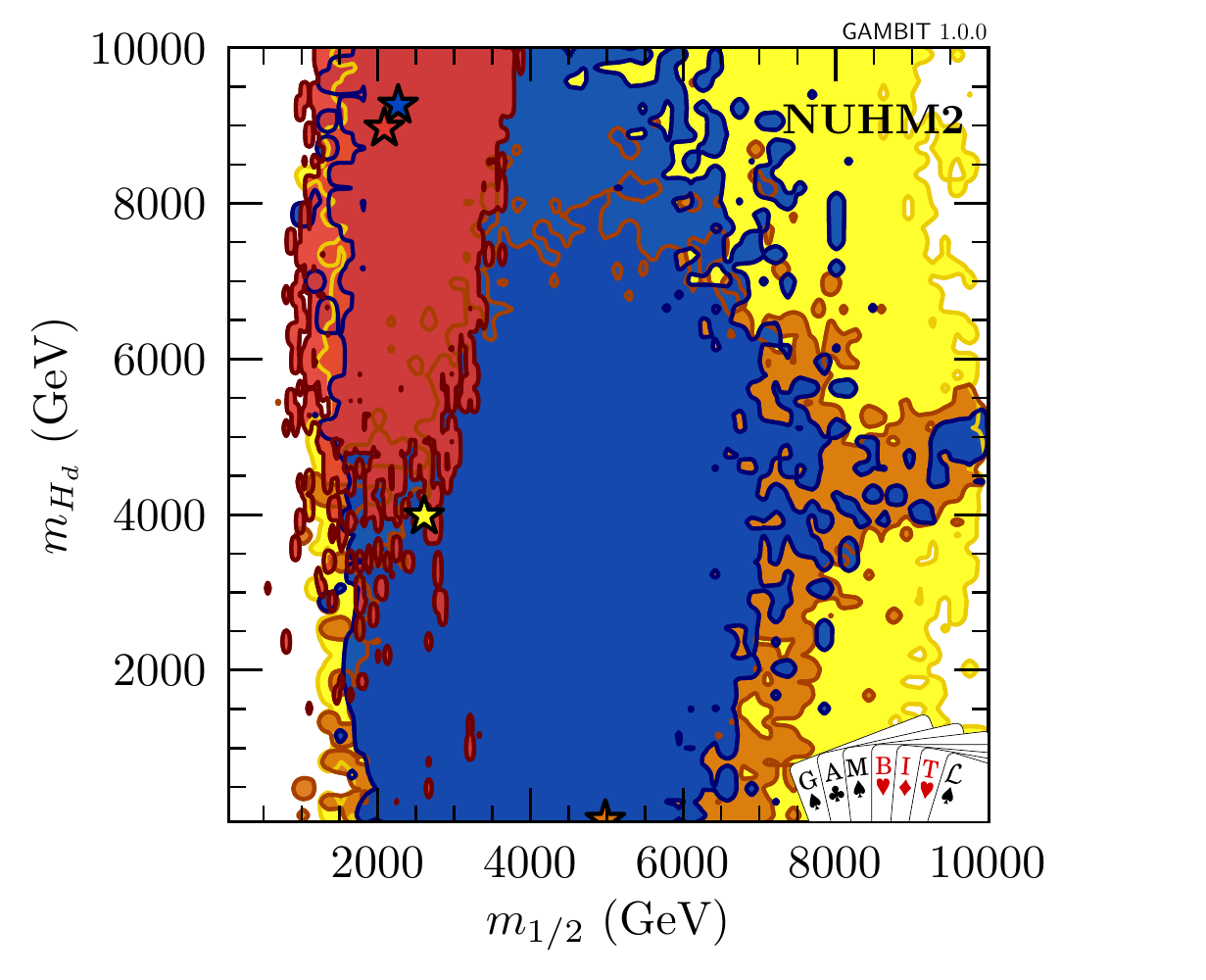}\\
  \includegraphics[height=4mm]{figures/rdcolours3.pdf}
  \caption{As per Fig.\ \ref{fig:2d_parameter_plots_nuhm2}, but for the $m_{H_u}$--$m_{1/2}$ (top) and $m_{H_d}$--$m_{1/2}$ (bottom) planes. }
  \label{fig:2d_parameter_plots_nuhm2_set2}
\end{figure*}

\begin{figure}[t]
\centering
\includegraphics[width=\columnwidth]{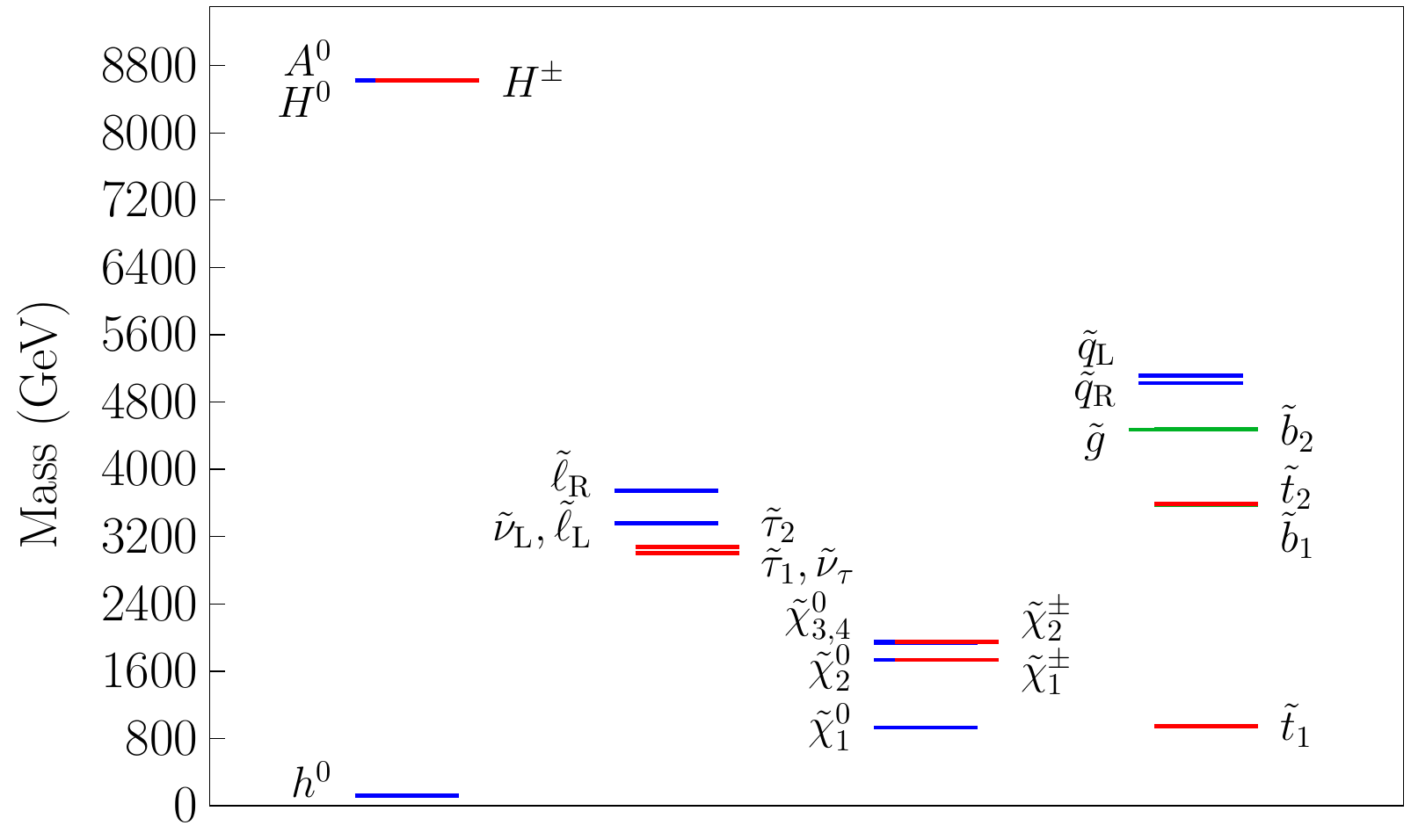}
\caption{Sparticle mass spectrum of the NUHM2 best-fit point.}
\label{fig:nuhm2-bf-spectrum}
\end{figure}

\begin{figure}[t]
\centering
\includegraphics[width=0.49\textwidth]{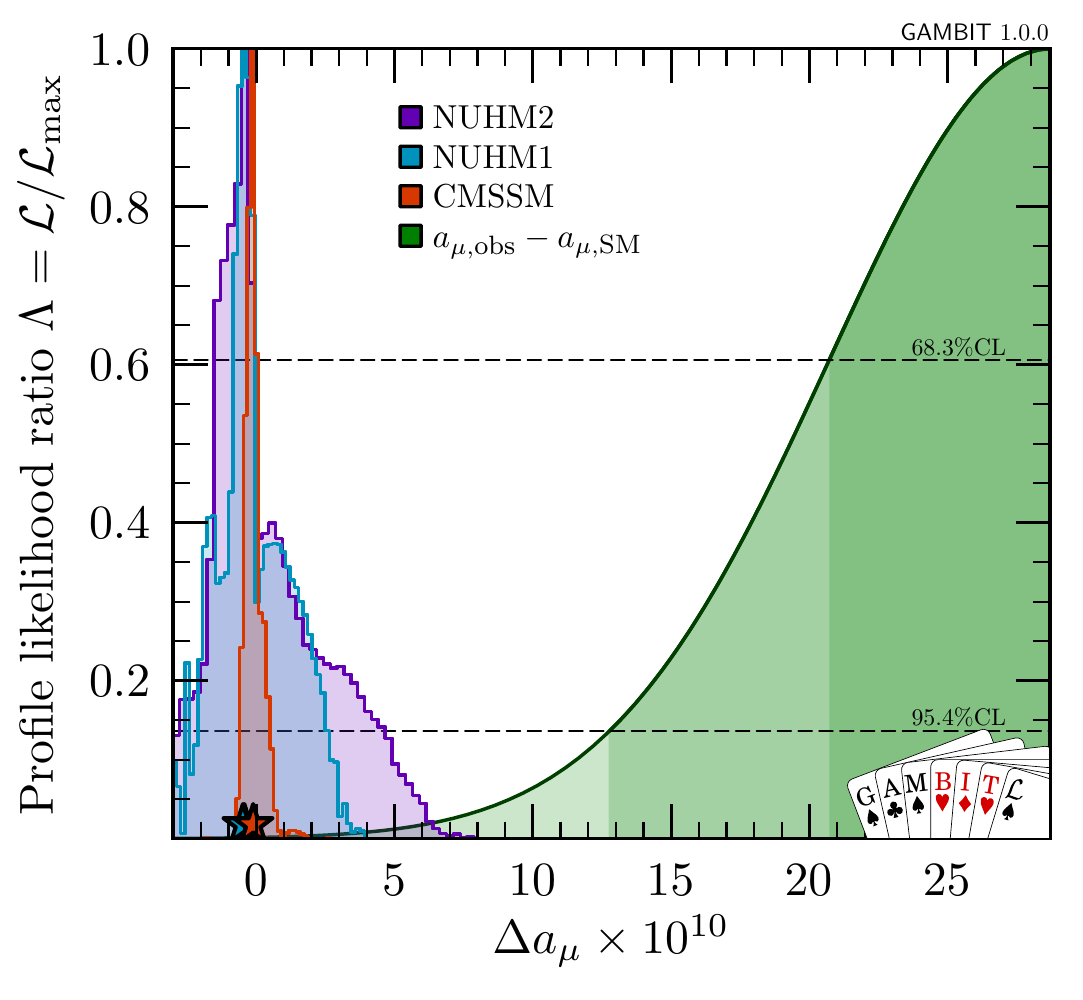}
\caption{1D profile likelihood ratio for $\Delta a_{\mu}$ in the CMSSM (red), NUHM1 (blue) and NUHM2 (purple). For comparison we show a Gaussian likelihood for the observed discrepancy $a_{\mu,\text{obs}} - a_{\mu,\text{SM}} = (28.7 \pm 8.0)\times10^{-10}$ (green), adding the experimental and theoretical uncertainties in quadrature.}
\label{fig:gm2}
\end{figure}

\subsection{NUHM2}
\label{sec:NUHM2}
The NUHM2 results are shown in Figs.~\ref{fig:2d_parameter_plots_nuhm2} and~\ref{fig:2d_parameter_plots_nuhm2_set2}. Figure \ref{fig:2d_parameter_plots_nuhm2}
shows results in the $m_0$--$m_{1/2}$ and $\tan\beta$--$A_{0}$ planes, with
plots of the profile likelihood ratio on the left and the annihilation
mechanism, defined at the start of section \ref{sec:CMSSM}, on the right.

In comparison to the NUHM1 results, the stau co-annihilation region is
significantly extended, covering higher values of $m_{1/2}$, and lower
$A_0$ and $\tan \beta$. This is due to modification of the RG flow for
the soft scalar stau masses that occurs when the soft Higgs masses are
split at the GUT scale. A similar effect has been observed and
discussed for this model in Ref.\ \cite{Buchmueller:2014yva}. As in
the NUHM1 despite the low values of $m_0$ our models within the
2$\sigma$ contours do not generate first and second generation squark
masses below $\sim$2\,TeV.

In Fig.\ \ref{fig:2d_parameter_plots_nuhm2_set2} we show the structure
of the $m_{H_u}$--$m_{1/2}$ and $m_{H_d}$--$m_{1/2}$ planes in the
same format as the $m_0$--$m_{1/2}$ and $\tan\beta$--$A_{0}$ planes.
The $m_{H_u}$--$m_{1/2}$ plot is quite similar to the $m_H$--$m_{1/2}$
plot (Fig.\ \ref{fig:2d_parameter_plots_nuhm1_set2}) for the NUHM1
model discussed in section \ref{sec:NUHM1}, while in contrast there is
not much structure in the $m_{H_d}$--$m_{1/2}$ plane.\footnote{For
brevity we omit plots showing the $m_{H_u}$--$m_0$ and $m_{H_d}$--$m_0$
planes, which exhibit the same behaviour.}

This could be anticipated from the NUHM1 results
(again Fig.\ \ref{fig:2d_parameter_plots_nuhm1_set2}), as the structure was
caused by the fact that smaller $m_{H_u}$ at the GUT scale leads to
$\mu \gg M_1$ at the SUSY scale, making bino DM the only
possibility. As in the NUHM1 case, the $A/H$-funnel mechanism for the bino
again does not work because $\mu$ is too large to allow $m_A \approx 2
m_{\tilde\chi_1^0} \approx 2 M_1$.  However, the extra freedom in the Higgs
sector from splitting $m_{H_u}$ and $m_{H_d}$ at the GUT scale does
allow a better LHC Higgs likelihood at smaller $m_0$ and $m_{1/2}$, so that
the stau and stop co-annihilation regions can be found when
$m_{H_u} \lesssim m_0, m_{1/2}$.

As with the NUHM1, we checked the three charge- and colour-breaking conditions mentioned in Sec.\ \ref{sec:CMSSM}.  The results were as in the NUHM1: some individual parameter combinations are affected, but the overall inference is not.

We give a table of best-fit NUHM2 points in Table~\ref{tab:nuhm2-bf-1}, with the mass spectrum for the overall best fit shown in Figure~\ref{fig:nuhm2-bf-spectrum}. Once again, the overall best fit is obtained for a point that satisfies the relic density bound through stop co-annihilation. The $\tilde{t}_1$ mass is 950\,GeV, but the mass difference with the $\tilde{\chi}_1^0$ is now less than 20\,GeV, making this very difficult to resolve at the LHC. The neutralino-chargino sector features a pure bino $\tilde{\chi}_1^0$, wino-dominated $\tilde{\chi}_2^0$ and $\tilde{\chi}_1^\pm$, and Higgsino-dominated $\tilde{\chi}_3^0$,  $\tilde{\chi}_4^0$ and $\tilde{\chi}_2^\pm$. The large mass of the $\tilde{\chi}_2^0$ adds to the problem of the $\tilde{t}_1-\tilde{\chi}_1^0$ mass difference, making this a particularly challenging scenario for collider searches.

For the NUHM2, $\Delta\ln\mathcal{L} = 36.362$, indicating a better fit than either the CMSSM or NUHM1.  This is expected to some extent, as the NUHM2 has one more free parameter than the NUHM1. Indeed, accounting for the extra freedom in the fit (i.e. adopting a canonical degree of freedom of 38 instead of 39 or 40 -- see previous subsections), and computing the implied $p$-value, the result is just $5.9\times10^{-4}$.  This actually disfavours the NUHM2 compared to the NUHM1 ($p = 7.1\times 10^{-4}$ for 39 dof) and CMSSM ($p = 9.4\times 10^{-4}$ for 40 dof), because its additional parameter does not provide a sufficiently large improvement to the overall fit.

We have not commented so far on the ability of any of the models to
explain the large discrepancy between the measured value of
$a_{\mu} \equiv (g-2)_\mu/2$ and that predicted by the SM \cite{1010.4180,Hagiwara:2011af}.  This is because, with the heavy spectra
found, a sizeable supersymmetric contribution to $\Delta a_{\mu}$ is not expected.
However, in Ref.\ \cite{Buchmueller:2014yva} (see right panel of Fig.\
12 in that paper), it was found that although the best fit for the NUHM2 predicts
a very small $a_{\mu}$, there are points within the $2 \sigma$ contours
that predict significantly larger values of around $2\times 10^{-9}$,
which may give some grounds for optimism. In contrast, the MSSM contribution to $a_\mu$
within the $2 \sigma$ confidence regions of our CMSSM, NUHM1 and NUHM2 fits is below $5\times
10^{-10}$. We show this visually in Fig.\ \ref{fig:gm2}.
Therefore, with the latest data and using \gmtwocalc to
obtain the most precise calculation available of the supersymmetric contributions to the anomalous
magnetic moment of the muon, we find that none of
the GUT-scale models that we consider can make a significant
contribution to resolving this discrepancy.

A similar point can be made about the flavour anomalies associated with the angular observables in $B^0\to K^{*0}\mu^+\mu^-$ decays and the ratios $R_D$ and $R_{D*}$; none of the best-fits or 95\% CL regions of our scans indicate any ability for these data to be explained within the CMSSM, NUHM1 or NUHM2.

\begin{figure*}[tp]
  \centering
  \includegraphics[width=0.49\textwidth]{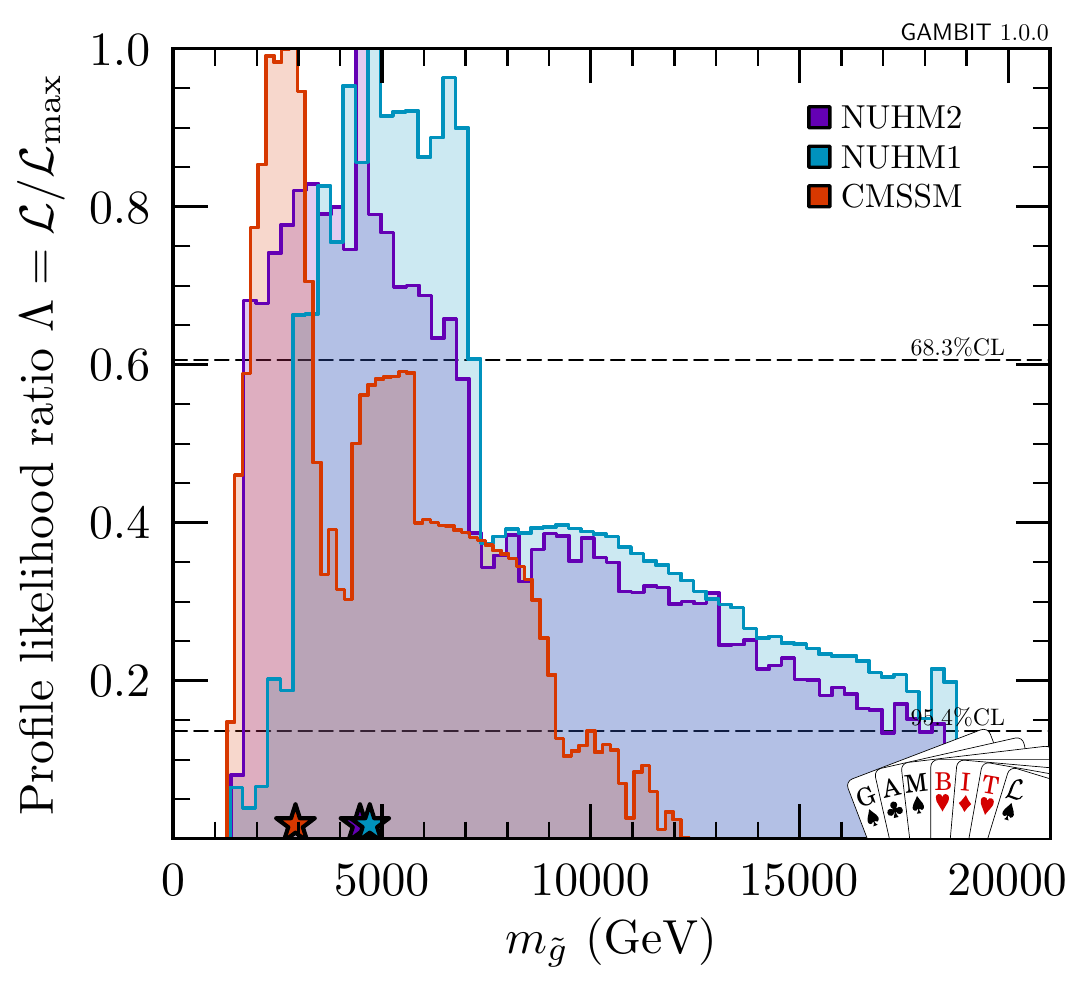}
  \includegraphics[width=0.49\textwidth]{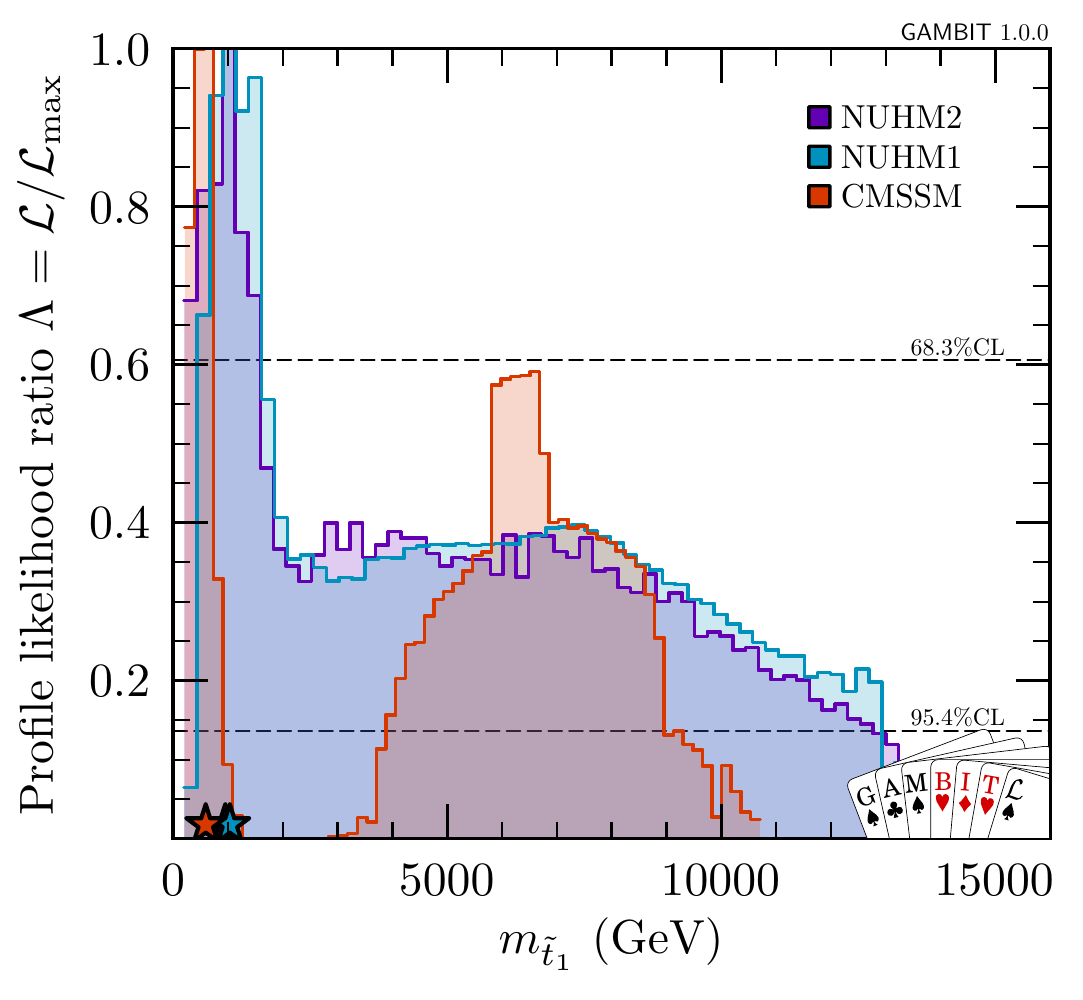}\\
  \includegraphics[width=0.49\textwidth]{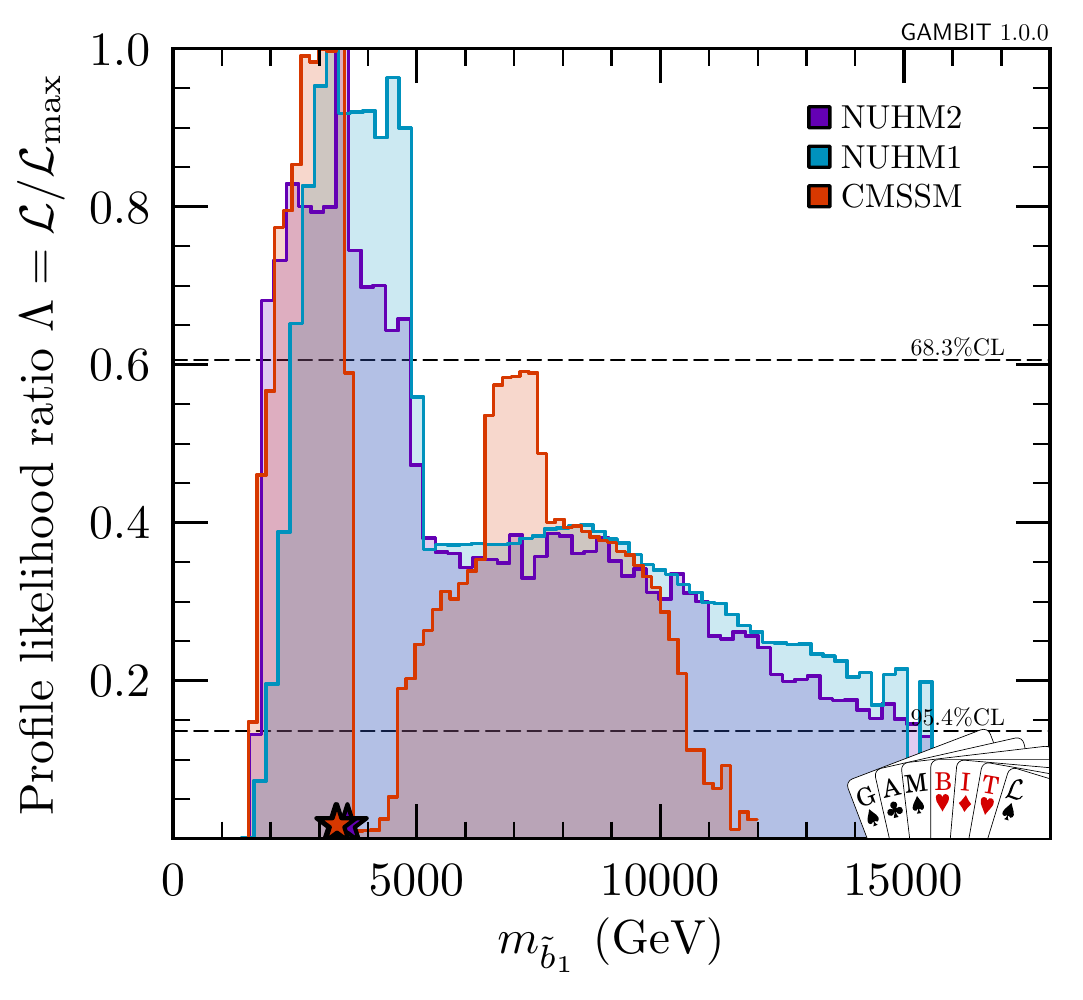}
  \includegraphics[width=0.49\textwidth]{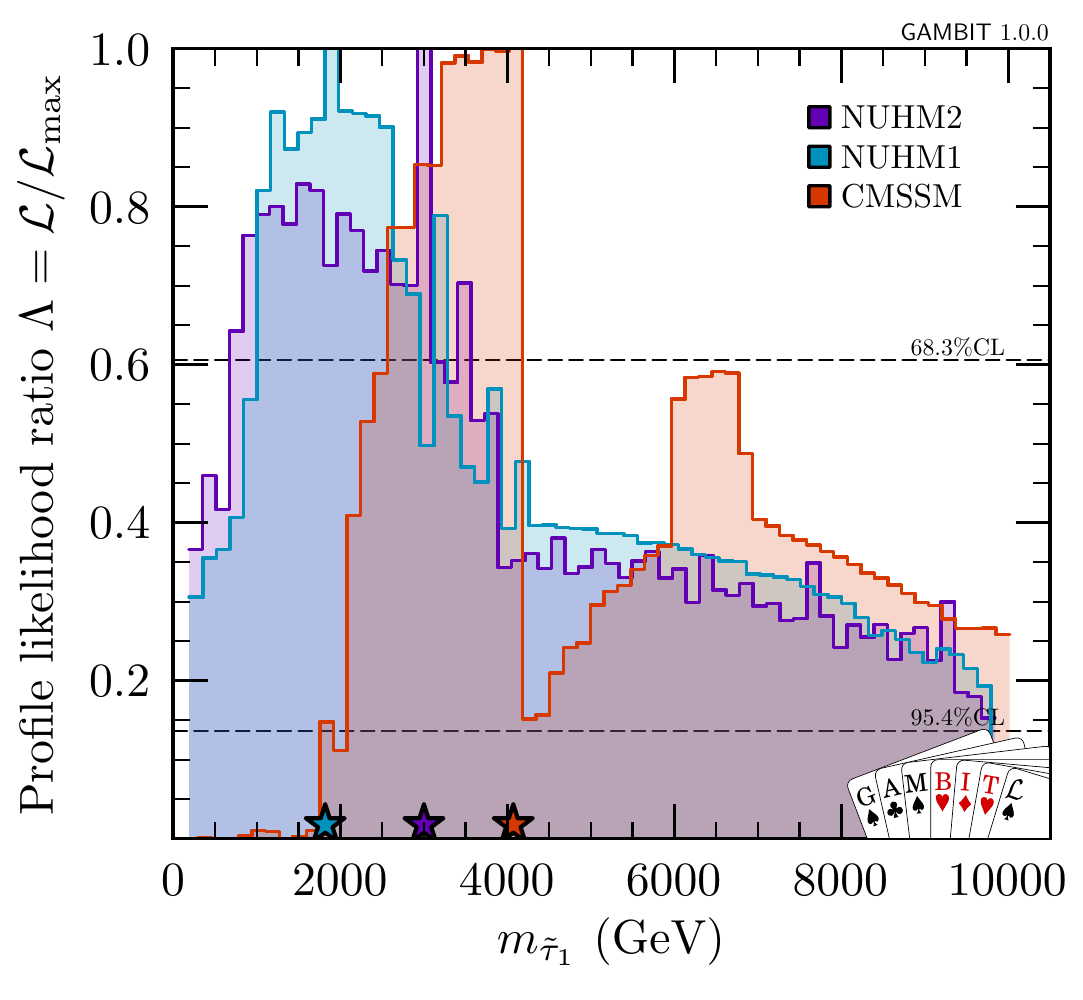}\\
  \includegraphics[width=0.49\textwidth]{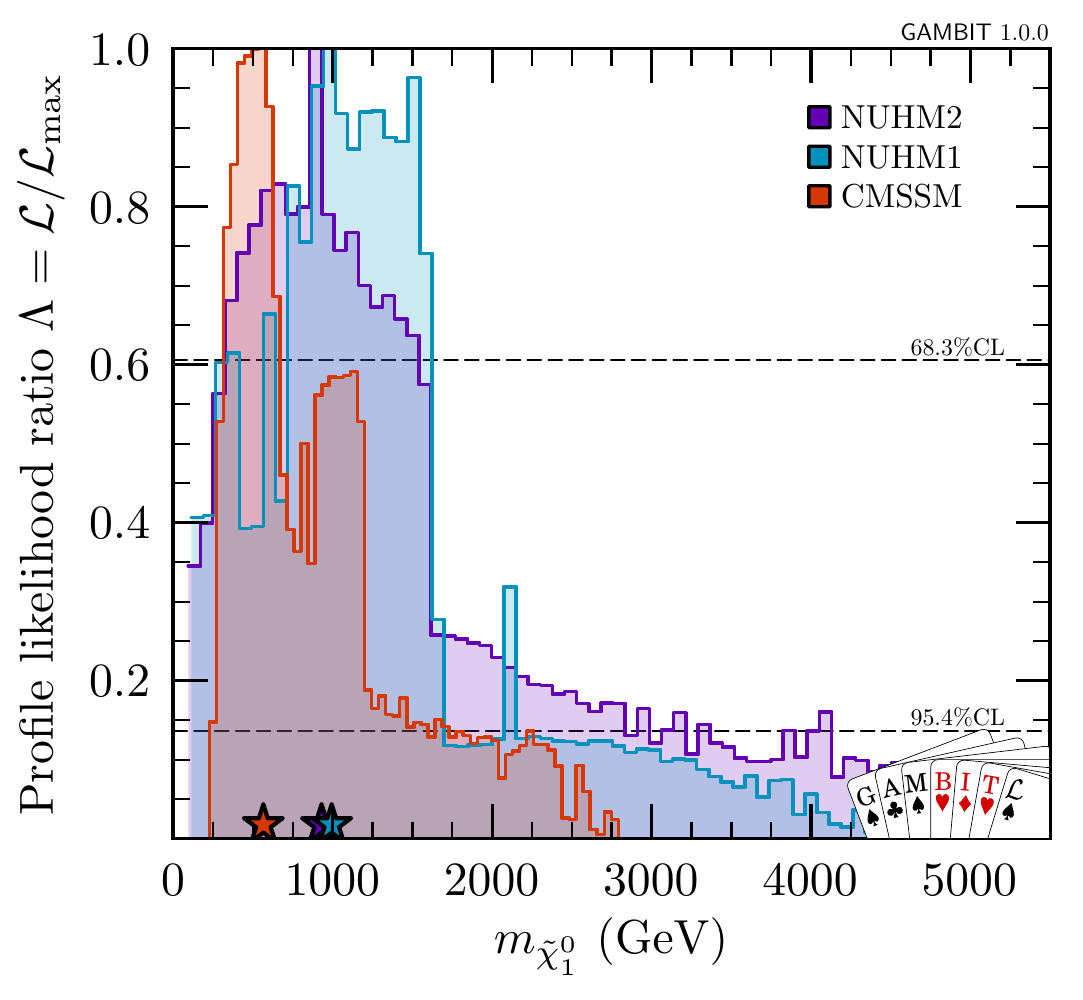}
  \includegraphics[width=0.49\textwidth]{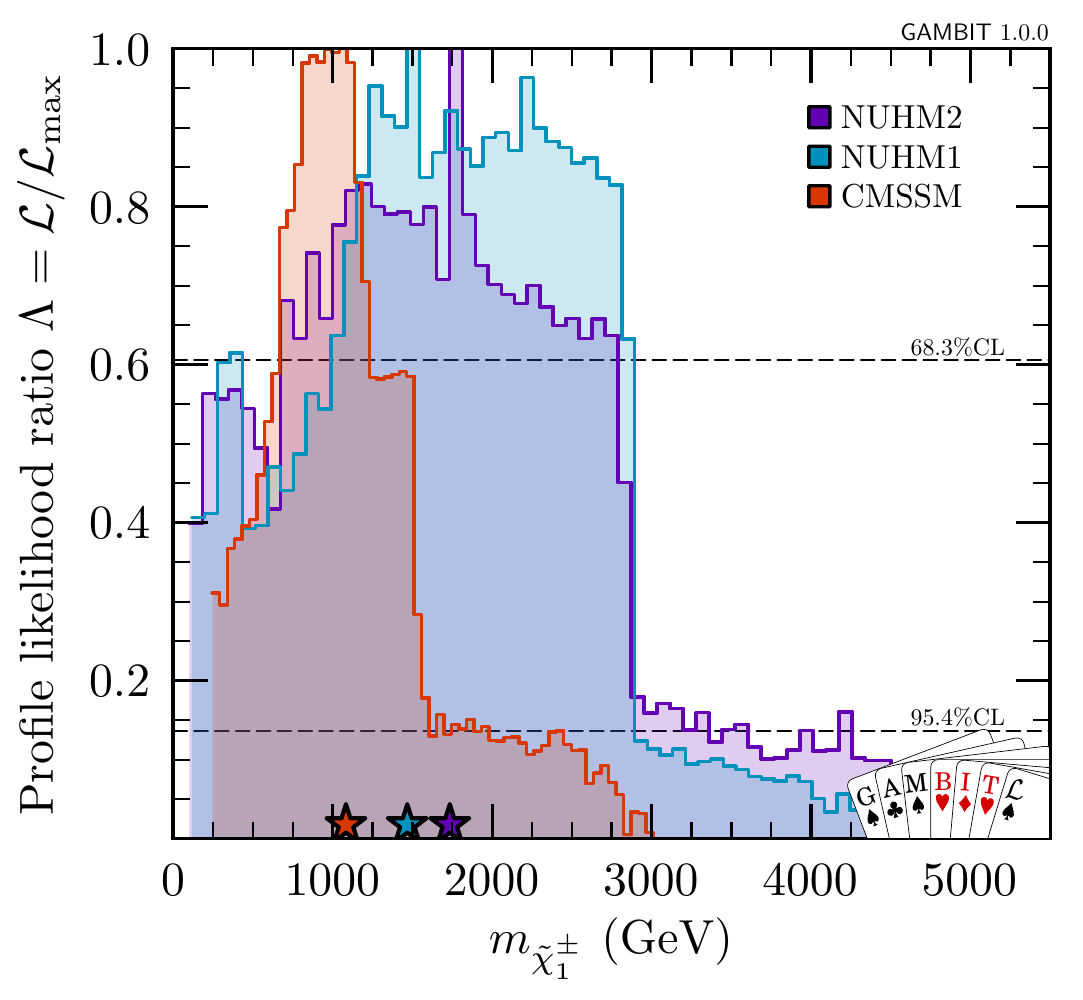}
  \caption{1D profile likelihoods for the masses of $\tilde{g}$, $\tilde{t}_1$, $\tilde{b}_1$, $\tilde{\tau}_1$, $\tilde{\chi}_1^0$ and $\tilde{\chi}_1^\pm$ in the CMSSM (red), NUHM1 (blue) and NUHM2 (purple).}
  \label{fig:1d_sparticle_masses}
\end{figure*}

\begin{figure*}[tp]
  \centering
  \includegraphics[width=0.49\textwidth]{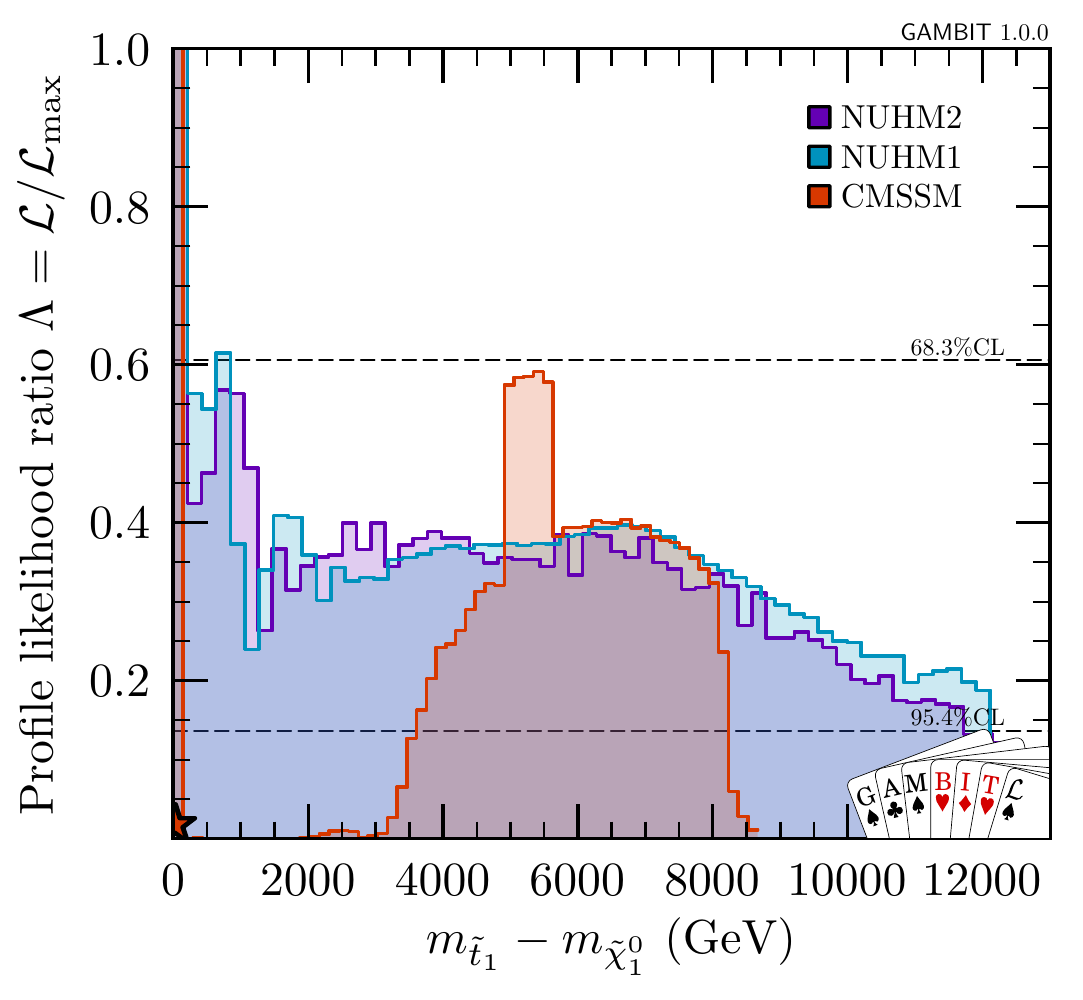}
  \includegraphics[width=0.49\textwidth]{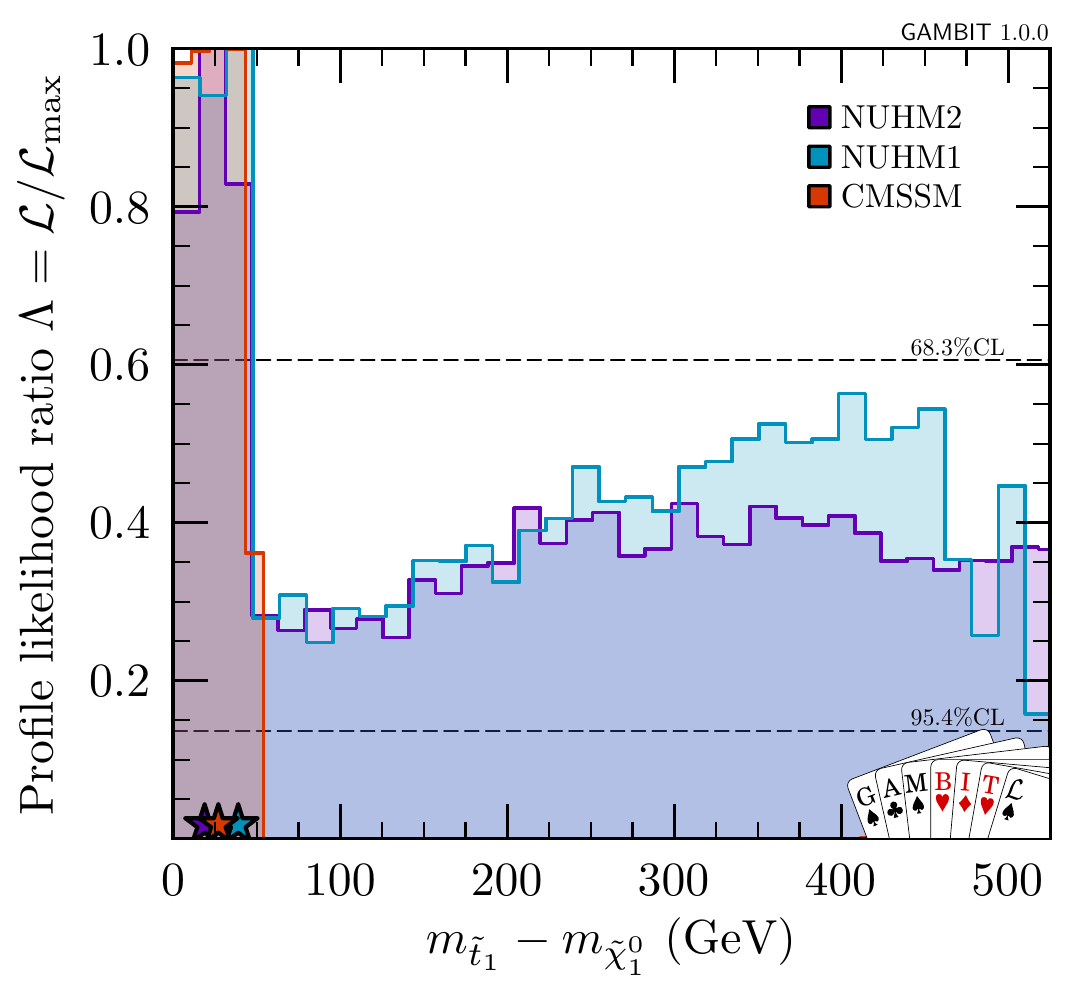}\\
  \includegraphics[width=0.49\textwidth]{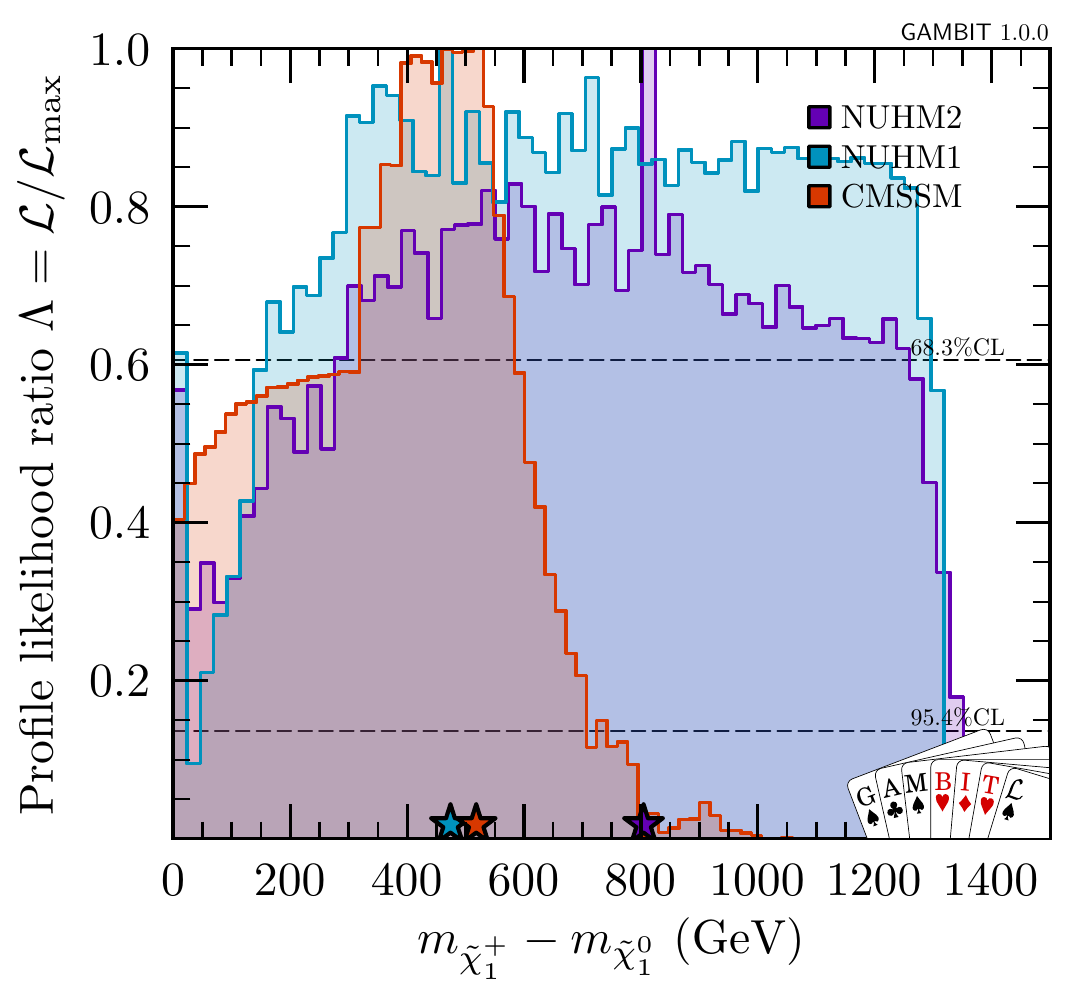}
  \includegraphics[width=0.49\textwidth]{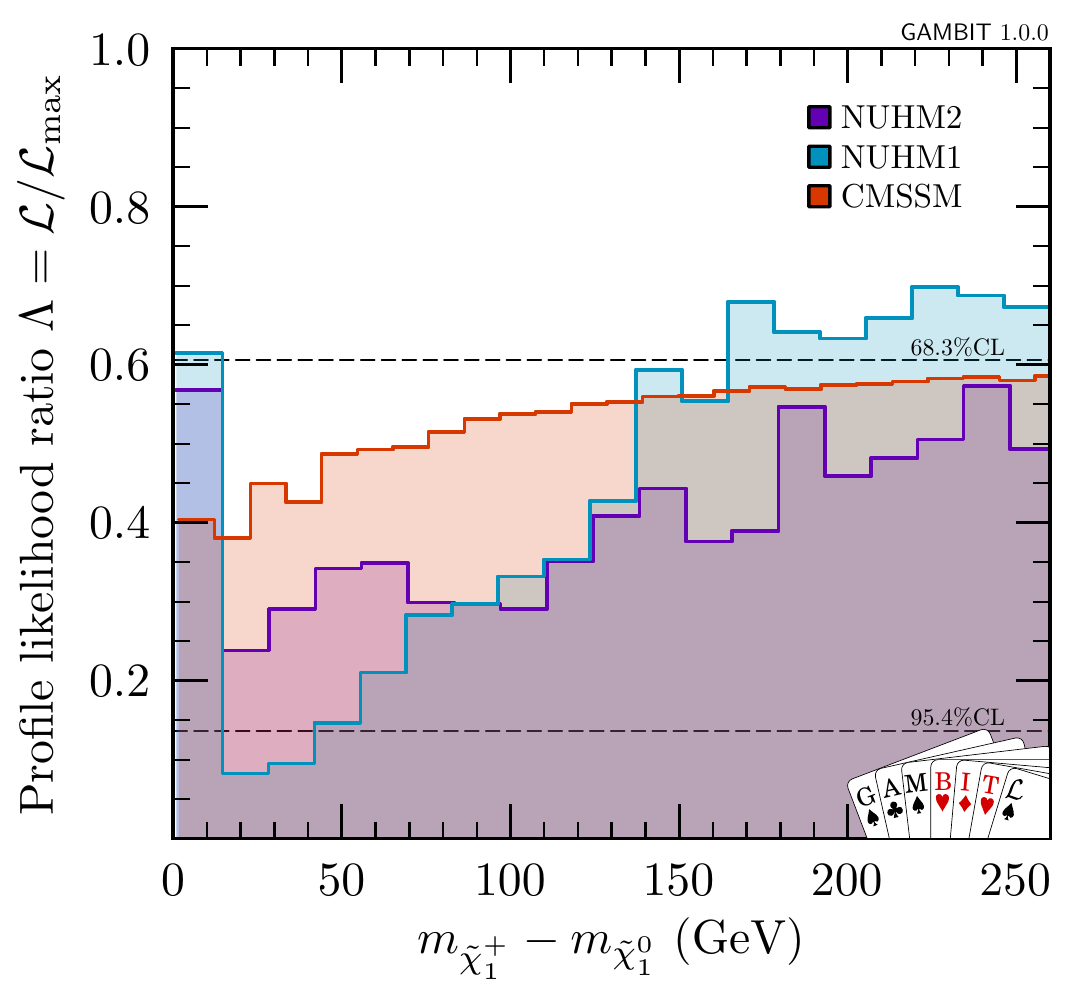}
  \caption{\textit{Left:} 1D profile likelihoods for the mass differences $m_{\tilde{t}_1}-m_{\tilde{\chi}^0_1}$ (top) and $m_{\tilde{\chi}^+_1}-m_{\tilde{\chi}^0_1}$ (bottom), in the CMSSM (red), NUHM1 (blue) and NUHM2 (purple). \textit{Right:} as per left-hand plots, but zoomed to focus on the smallest mass differences.}
  \label{fig:1d_sparticle_mass_differences}
\end{figure*}

\begin{figure*}[pt]
 \centering
 \includegraphics[width=0.49\textwidth]{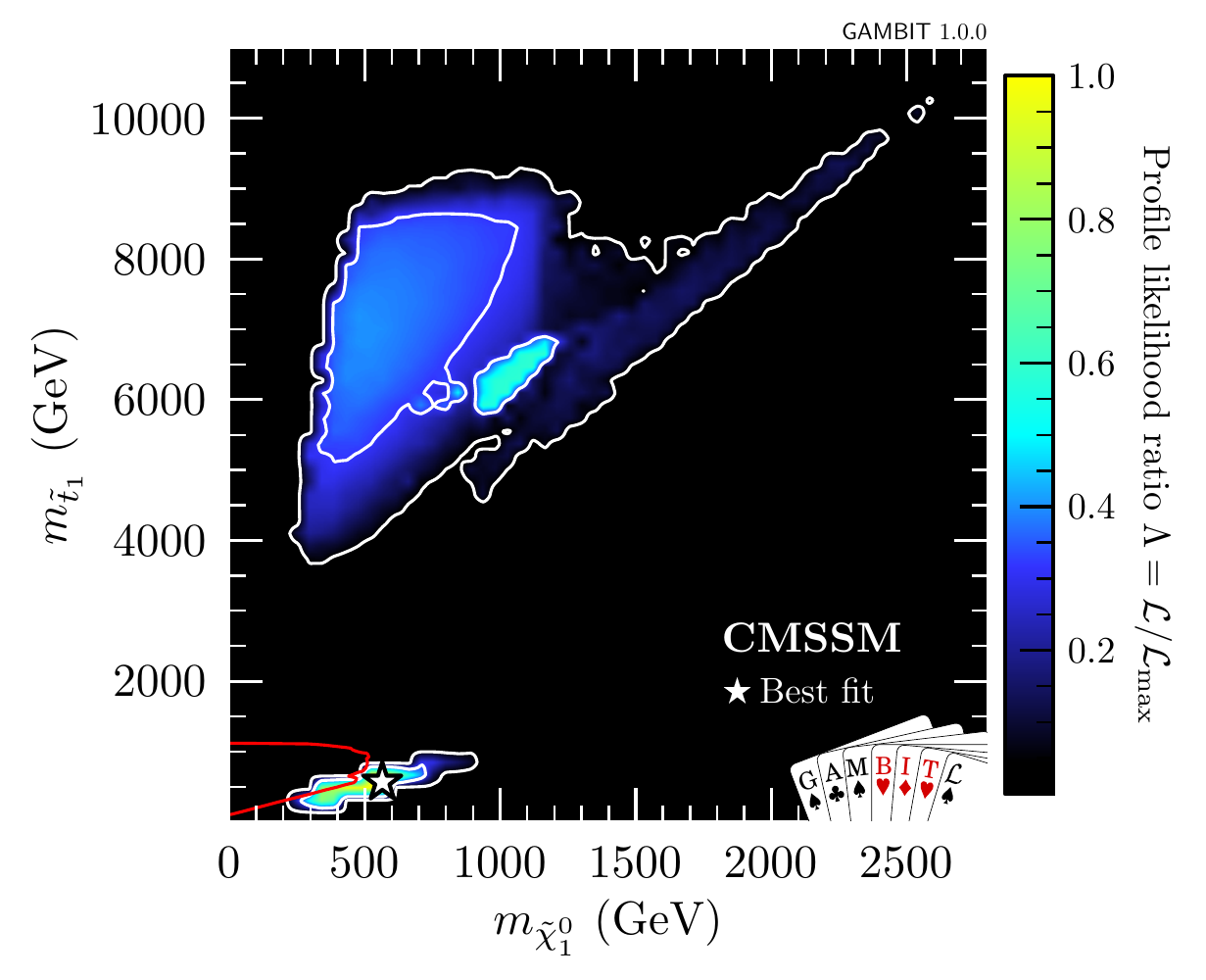}
 \includegraphics[width=0.49\textwidth]{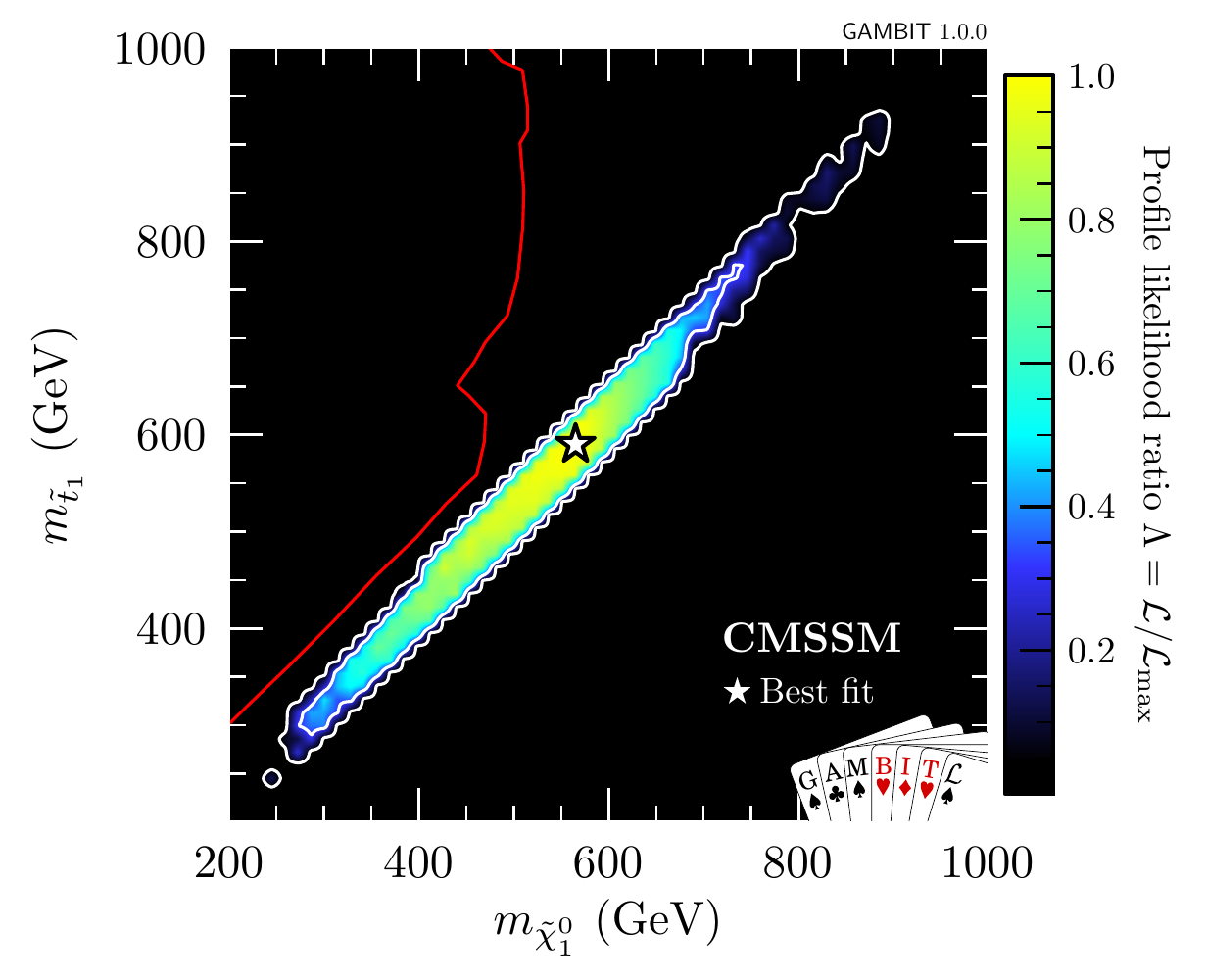}
 \caption{2D profile likelihoods for the CMSSM, plotted in the  $\tilde{t}_1-\tilde{\chi}^0_1$ mass plane. \textit{Left:} the full range of neutralino masses present in the combined sample.  \textit{Right:} as per the lefthand panel, but zoomed in to focus on the low-mass region. Superimposed in red is the latest CMS Run II simplified model limit for  $\tilde{t}_1$ pair production, followed by decay to $t$ quarks and the lightest neutralino~\cite{CMSStopSummary}. This limit should be interpreted with caution (for details see main text).}
 \label{fig:chi10-stop}
\end{figure*}

\begin{figure*}[pt]
 \centering
 \includegraphics[width=0.49\textwidth]{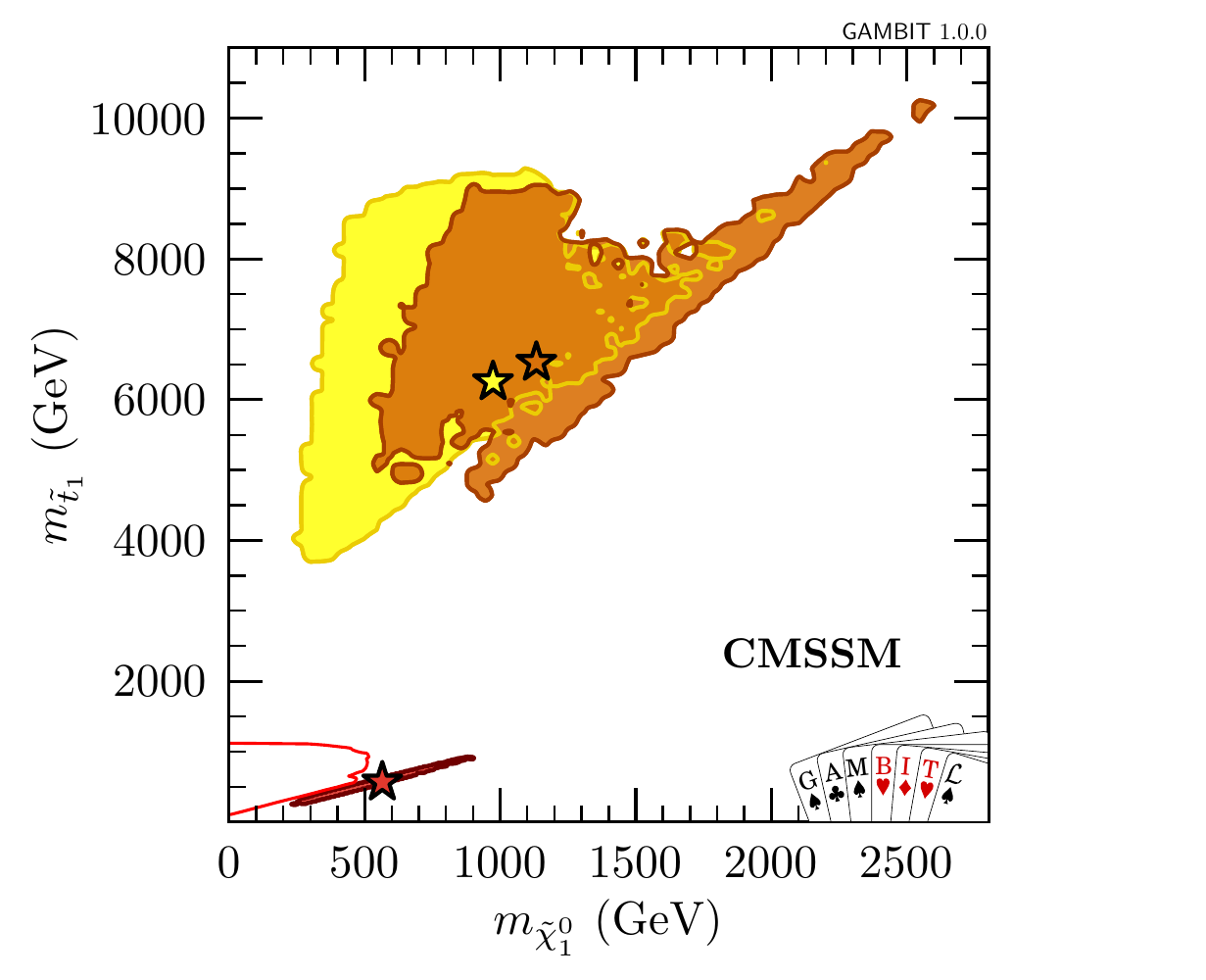}
 \includegraphics[width=0.49\textwidth]{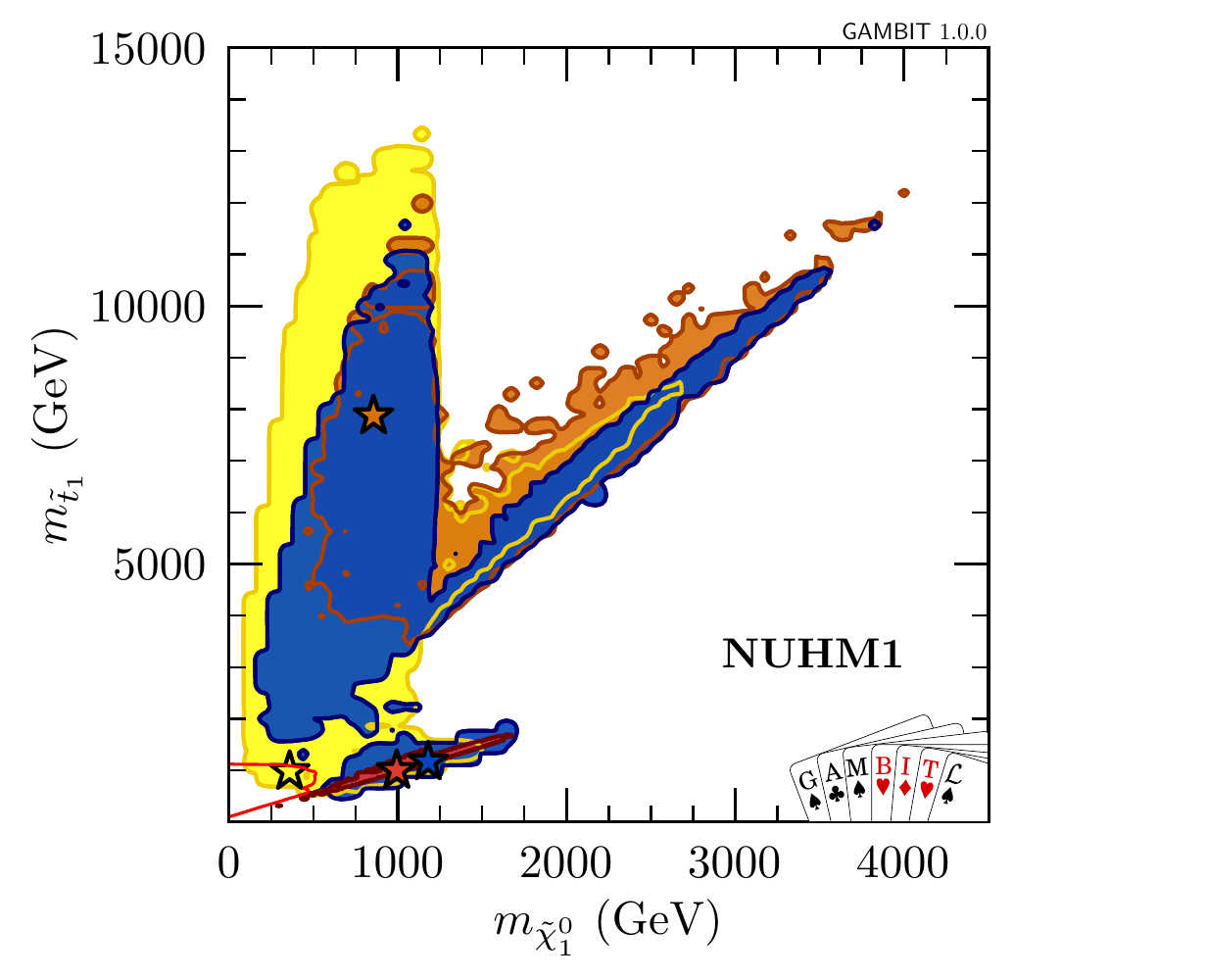}\\
 \includegraphics[height=4mm]{figures/rdcolours4.pdf}
 \caption{95\% CL 2D profile likelihoods in the $\tilde{t}_1-\tilde{\chi}^0_1$ mass plane, coloured according to the mechanism(s) active in depleting the relic density.  \textit{Left:} the CMSSM. \textit{Right:} the NUHM1.  Superimposed in red is the latest CMS Run II simplified model limit for $\tilde{t}_1$ pair production and subsequent decay to $t$ quarks and the lightest neutralino~\cite{CMSStopSummary}. This limit should be interpreted with caution (for details see main text).}
 \label{fig:colour-stop}
\end{figure*}

\begin{figure*}[pt]
 \centering
 \includegraphics[width=0.49\textwidth]{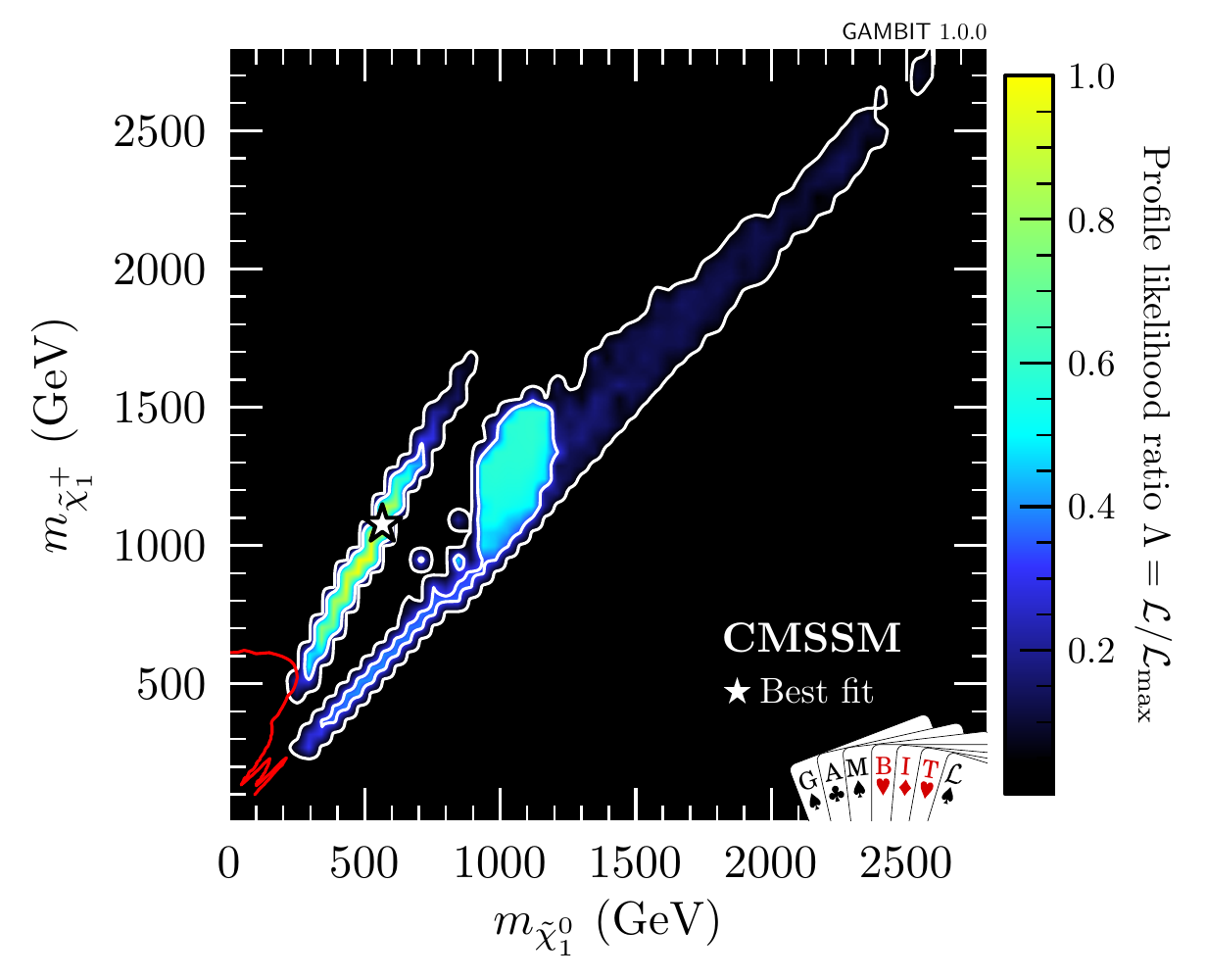}
 \includegraphics[width=0.49\textwidth]{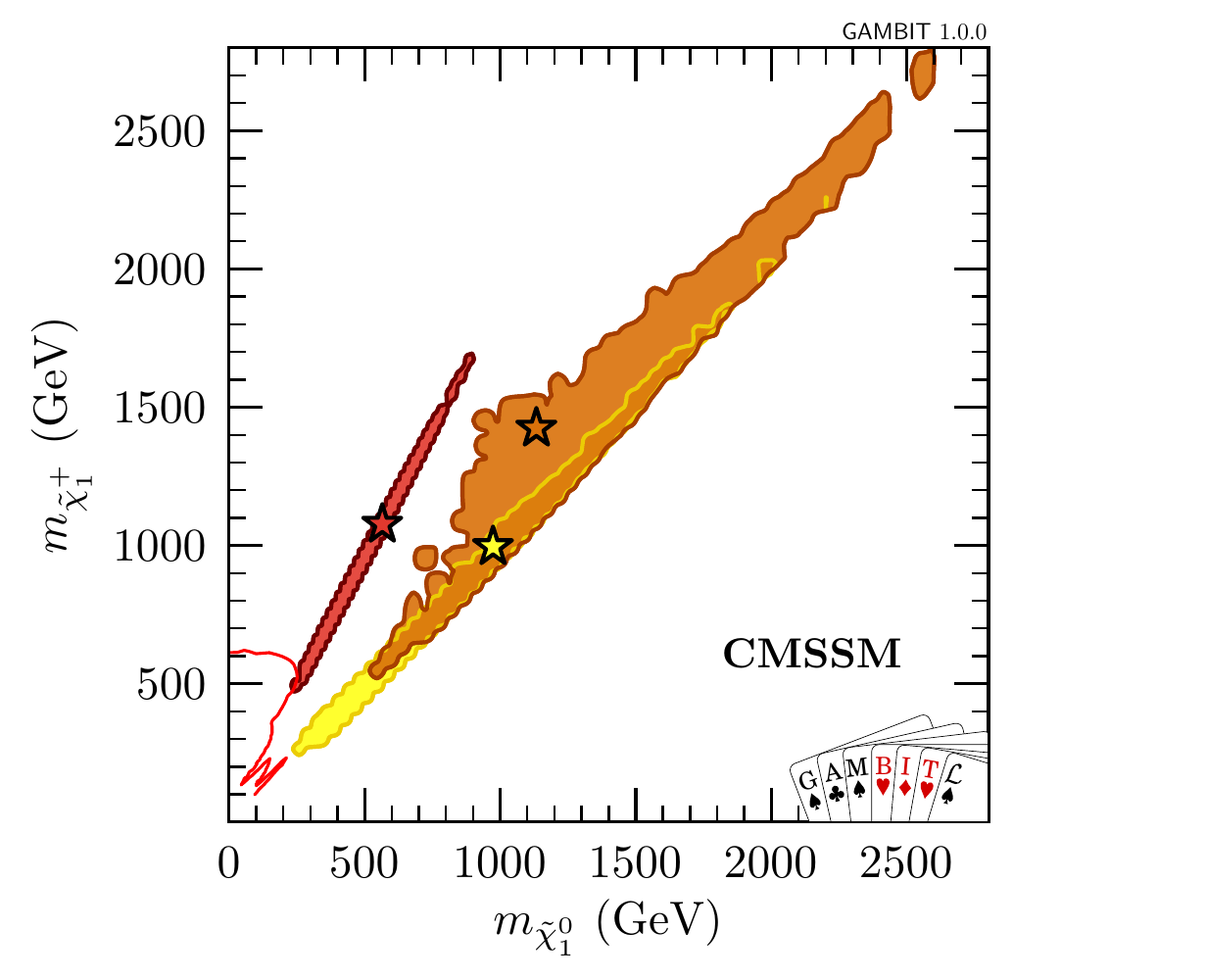}\\
 \includegraphics[height=4mm]{figures/rdcolours3.pdf}
 \caption{\textit{Left:} Profile likelihood for the CMSSM in the  $\tilde{\chi}_1^\pm-\tilde{\chi}^0_1$ mass plane. \textit{Right:} Colour-coding shows the mechanism(s) that allow models within the 95\% CL region to avoid exceeding the observed relic density of DM. Superimposed in red is the latest CMS Run II simplified model limit for $\tilde{\chi}_1^\pm \tilde{\chi}_1^0$ pair production and decay with decoupled sleptons~\cite{CMSEWSummary}. This limit should be interpreted with caution (for details see main text).}
  \label{fig:cmssm-charge-chi}
\end{figure*}

\subsection{Discovery Prospects}
In the following we discuss the discovery prospects of the CMSSM, NUHM1 and NUHM2, given current constraints.  We first address prospects at the LHC, followed by direct and indirect detection of DM.

\begin{table*}
\small
\center
\sisetup{round-mode = places, round-precision = 3,round-integer-to-decimal}
\begin{tabular}{l S[table-format=+3.3] S[table-format=+3] S[table-format=+3] S[table-format=+3] S[table-format=+3] S[table-format=+3]}
Likelihood term & {\ \ Ideal} & {$A/H$-funnel} & {\ \ \ $\tilde{\tau}$ co-ann.} & {\ \ \ \ $\tilde{t}$ co-ann.} & {\ $\tilde{\chi}_1^{\pm}$ co-ann.} & {\ \ \ $\Delta\ln\mathcal{L}_\mathrm{BF}$} \\
\hline
LHC sparticle searches & 0 &  0.000  & 0.000 & 0.000 & 0.000  &  0 \\
LHC Higgs & -37.7339 &  -38.563  &  -37.928  &  -37.980  &  -38.484  &  .2461  \\
LEP Higgs & 0 &  0.000  &  0.000  &  0.000  &  0.000  &  0 \\
ALEPH selectron & 0 &  0.000  & 0.000  & 0.000  & 0.000  &  0 \\
ALEPH smuon & 0 &  0.000  &  0.000  &  0.000  &  0.000  &  0  \\
ALEPH stau & 0 &  0.000  &  0.000  &  0.000  &  0.000  &  0 \\
L3 selectron & 0 &  0.000  &  0.000  &  0.000  &  0.000  &  0  \\
L3 smuon & 0 & 0.000  &  0.000  &  0.000  &  0.000  &  0 \\
L3 stau & 0 & 0.000  &  0.000  &  0.000  &  0.000  &  0 \\
L3 neutralino leptonic & 0 & 0.000  & 0.000  & 0.000  & 0.000  &  0  \\
L3 chargino leptonic & 0 &  0.000  &  0.000  &  0.000  &  0.000  &  0 \\
OPAL chargino hadronic & 0 &  0.000  &  0.000  &  0.000  &  0.000  &  0 \\
OPAL chargino semi-leptonic & 0 &  0.000  &  0.000  &  0.000  &  0.000  &  0 \\
OPAL chargino leptonic & 0 &  0.000  &  0.000  &  0.000  &  0.000  &  0 \\
OPAL neutralino hadronic & 0 &  0.000  &  0.000  &  0.000  &  0.000  &  0  \\
$B_{(s)}\to \mu^+\mu^-$ & 0 &  -1.972  &  -2.037  &  -2.033  &  -2.030  &  2.033  \\
Tree-level $B$ and $D$ decays & 0 &  -15.553  &  -15.283  &  -15.283  &  -15.290  &  15.283 \\
$B^0\to K^{*0}\mu^+\mu^-$ & -184.260 &  -195.596  &  -195.475  &  -195.043  &  -194.415  &  10.783  \\
$B\to X_s\gamma$ & 9.799 &  8.865  &  8.797  &  8.550  &  8.077  &  1.249  \\
$a_\mu$ & 20.266 &  14.086  &  13.756  &  13.842  &  13.876  &  6.424  \\
$W$ mass & 3.281 &  3.060  & 3.078  & 3.074  & 3.097  &  .207  \\
Relic density & 5.989 &  5.989  &  5.989  &  5.989  &  5.989  &  0 \\
PICO-2L & -1 &  -1.000  &  -1.000  &  -1.000  &  -1.000  &  0 \\
PICO-60 F & 0 &  0.000  &  0.000  &  0.000  &  0.000  &  0  \\
SIMPLE 2014 & -2.972 &  -2.972  &  -2.972  &  -2.972  &  -2.972  &  0  \\
LUX 2015  & -0.64 &  -0.657  &  -0.641  &  -0.641  &  -0.671  &  .001  \\
LUX 2016  & -1.467 &   -1.501  &  -1.468  &  -1.470  &  -1.529  &  .003 \\
PandaX 2016  & -1.886 &  -1.909  &  -1.887  &  -1.888  &  -1.929  &  .002  \\
SuperCDMS 2014 & -2.248 &  -2.248  &  -2.248  &  -2.248  &  -2.248  &  0 \\
XENON100 2012 & -1.693 &   -1.685  &  -1.693  &  -1.692  &  -1.678  &  .001  \\
IceCube 79-string & 0.0 &  -0.021  &  0.000  &  0.000  &  -0.108  &  0 \\
$\gamma$ rays (\textit{Fermi}-LAT dwarfs) & -33.244 &  -33.398  &  -33.371  &  -33.369  &  -33.398  &  .125 \\
$\rho_0$ & 1.142 &  1.141  &  1.137  &  1.141 &  1.131  &  .001 \\
$\sigma_s$ and $\sigma_l$ & -6.115 &  -6.115  &  -6.116  &  -6.115  &  -6.116  &  0 \\
$\alpha_s(m_Z)(\MSBar)$ & 6.500 &  6.447  &  6.499  &  6.496  &  6.496  &  .004 \\
Top quark mass & -0.645 &  -0.652  &  -0.661  &  -0.646  &  -0.645  &  .001 \\
\hline
Total & -226.927 & -264.255 & -263.524 & -263.289 & -263.855 & 36.362 \\
\hline
&  &  &  &  & \\
Quantity &  & {$A/H$-funnel} & {\ \ \ $\tilde{\tau}$ co-ann.} & {\ \ \ \ $\tilde{t}$ co-ann.} & {\ $\tilde{\chi}_1^{\pm}$ co-ann.} &  \\
\cmidrule{1-6}
$A_0$ &  &  7337.758  &  -6666.073  &  -7706.626  &  -8213.109 \\
$m_0$ &  &  4945.237  &  1582.304  &  3294.531  &  2697.314 \\
$m_{1/2}$ &  &  4981.246  &  2265.444  &  2085.463  &  2607.561 \\
$m_{H_u}$ &  &  6845.748  &  3714.036  &  5196.468  &  6282.001 \\
$m_{H_d}$ &  &  93.459  &  9285.571  &  8990.311  &  4005.580 \\
$\tan\beta$ &  &  28.221  &  22.567  &  23.345  &  18.075 \\
$\mathrm{sgn}(\mu)$ &  &  \multicolumn{1}{r}{$+$}  &  \multicolumn{1}{r}{$-$}  &  \multicolumn{1}{r}{$-$}  &  \multicolumn{1}{r}{$-$} \\
$m_t$ & &  173.246  &  173.479  &  173.388  &  173.328 \\
$\alpha_s(m_Z)(\MSBar)$ & &  0.119  &  0.119  &  0.119  &  0.119 \\
$\rho_0$ & &  0.396  &  0.388  &  0.405  &  0.381 \\
$\sigma_s$ &  &  43.162  &  42.562  &  43.121  &  43.323 \\
$\sigma_l$ &  &  57.980  &  58.022  &  57.890  &  57.764 \\
\cmidrule{1-6}
$M_1$ & &  2277.442  &  1004.143  &  925.176  &  1157.614 \\
$\mu$ & &  537.021  &  -2480.773  &  -1928.496  &  -382.757 \\
$m_{\tilde{t}_1}$ & &  7589.989  &  1030.595  &  948.763  &  1217.299 \\
$m_{\tilde{\tau}_1}$ & &  4633.573  &  1083.376  &  3001.595  &  2261.195 \\
$m_{A}$ & &  1176.568  &  9151.605  &  8624.785  &  3808.674 \\
$m_{h}$ & &  125.377  &  124.398  &  125.173  &  125.414 \\
$m_{\tilde{\chi}_1^0}$ & &  553.377  &  1004.076  &  930.008  &  391.009 \\
$($\%bino, \%Higgsino$)$ & & \multicolumn{1}{r}{$(0,100)$} & \multicolumn{1}{r}{$(100,0)$} & \multicolumn{1}{r}{$(100,0)$} & \multicolumn{1}{r}{$(0,100)$} \\
$m_{\tilde{\chi}_2^0}$ & &  -555.848  &  1868.405  &  1734.260  &  -396.274 \\
$($\%bino, \%Higgsino$)$ & & \multicolumn{1}{r}{$(0,100)$} & \multicolumn{1}{r}{$(0,1)$} & \multicolumn{1}{r}{$(0,6)$} & \multicolumn{1}{r}{$(0,100)$} \\
$m_{\tilde{\chi}_1^\pm}$ & &  554.943  &  1868.573  &  1734.450  &  394.095 \\
$($\%wino, \%Higgsino$)$ & & \multicolumn{1}{r}{$(0,100)$} & \multicolumn{1}{r}{$(99,1)$} & \multicolumn{1}{r}{$(94,6)$} & \multicolumn{1}{r}{$(0,100)$} \\
$m_{\tilde{g}}$ & &  9979.887  &  4715.895  &  4471.116  &  5436.877 \\
$\Omega h^2$ & &  {$3.06\times10^{-2}$} &  {$6.76\times10^{-2}$} &  {$4.49\times10^{-2}$} & {$1.81\times10^{-2}$} \\
\cmidrule{1-6}
\end{tabular}
\caption{\label{tab:nuhm2-bf-1} Best-fit points in the NUHM2, for each of the regions characterised by a specific mechanism for suppressing the relic density of dark matter. Here we show the likelihood contributions,  parameter values at each point, and some quantities relevant for the interpretation of mass spectra at the different best fits.  We also give likelihood components for a canonical `ideal' likelihood (see text), along with its offset from the global best-fit point. SLHA1 and SLHA2 files corresponding to the best-fit point in each region can be found in the online data associated with this paper \cite{the_gambit_collaboration_2017_801642}.}
\end{table*}

\subsubsection{LHC}
\label{sec:lhc}
In Fig.~\ref{fig:1d_sparticle_masses}, we show the 1D profile likelihood ratio for the masses of the gluino, lightest (third generation) squarks, lightest stau, lightest chargino and lightest neutralino in the CMSSM, NUHM1 and NUHM2. The likelihood is generally low for coloured sparticles light enough to be in reach of LHC Run II, but there is an interesting peak of high likelihood at low stop masses for all three models, centred on the best-fit masses of 592, 1030 and 950\,GeV for the CMSSM, NUHM1 and NUHM2 respectively.  At least naively, this appears worthy of further investigation for each model, in terms of the potential for discovery at the LHC.

Concentrating first on the profile likelihood for $m_{\tilde{t}_1}$ in the CMSSM, the first consideration is the mass difference $m_{\tilde{t}_1}-m_{\tilde{\chi}^0_1}$ for models with a low stop mass, as experimental prospects generally deteriorate rapidly for more compressed spectra. The CMSSM 1D profile likelihood ratio for the mass difference $m_{\tilde{t}_1}-m_{\tilde{\chi}^0_1}$ is shown in the top panels of Figure~\ref{fig:1d_sparticle_mass_differences} in red, while Figure~\ref{fig:chi10-stop} shows the 2D profile likelihood in the $\tilde{t}_1-\tilde{\chi}^0_1$ mass plane. The low-mass stop solutions all satisfy the relic density constraint through stop co-annihilation, giving stop--neutralino mass differences below $\sim 50$~\GeV. For very small mass differences, below the mass of the $b$ quark, these points could be probed by long-lived particle searches at the LHC.  We defer a detailed study of this to future work.

If the stop decays promptly, however, this region can in principle be probed by LHC compressed spectra searches, particularly in the recent Run II updates that were not included in our initial scan. Although we plan a detailed analysis of the full range of recent LHC results in a forthcoming paper, some insight can be gained by examining the recent 36 fb$^{-1}$ simplified model limits presented by the CMS experiment \cite{Sirunyan:2017cwe,CMS-PAS-SUS-16-036,CMS-PAS-SUS-16-049,CMS-PAS-SUS-16-051,CMS-PAS-SUS-17-001} at 13\,TeV. They carried out stop searches in a variety of final states, and interpreted them in terms a model in which stop pair production is immediately followed by decay to a (possibly off-shell) top quark and the lightest neutralino. Although this is not necessarily the case for our models, the simplified model limit acts as a guide to the strongest possible exclusion potential of these Run II searches.  We show this limit in Fig.~\ref{fig:chi10-stop} as a red line. The low-mass part of our 2$\sigma$ best-fit region remains out of reach of the latest CMS search. We have also checked that the models in this region emerge almost unscathed when compared to recent ATLAS limits on compressed stop scenarios~\cite{Aad:2015pfx,Aaboud:2016tnv,ATLASStopSummary},\footnote{We note that the ATLAS limit assumes a 100\% branching fraction for the process $\tilde{t}_1\rightarrow c \tilde{\chi}^0_1$. We have checked that this agrees closely with the branching fractions returned by \decaybit and \susyhit for our best-fit stop co-annihilation point.} but there is some hope that at least the lower parts of this region will be probed in the near future. Completely excluding the stop co-annihilation region in the CMSSM would require probing compressed spectra in lightest stop decays up to a stop mass of approximately 900\,GeV.  Although finding such models is challenging at the LHC, stop pair-production is within the kinematic reach of a multi-TeV linear collider for the whole region, and dedicated analysis, similar to searches for Higgsino-dominated neutralinos, should be effective in constraining such models.

This picture changes in the NUHM1, which is most easily seen by examining which mechanism for obeying the relic density constraint is active in each region of the $\tilde{t}_1-\tilde{\chi}^0_1$ mass plane. Figure~\ref{fig:colour-stop} shows that, whereas the entire CMSSM 95\% CL region at low stop masses arises from stop co-annihilation, the extra freedom in the NUHM1 model allows the existence of points with low stop mass that generate the required relic density through either the stau co-annihilation or chargino co-annihilation mechanisms (or indeed some combination thereof).
There is hence a region with stop masses below 1\,TeV that would exhibit larger $\tilde{t}_1-\tilde{\chi}^0_1$ mass differences, making future discovery at the LHC an easier prospect. Indeed, comparison with the most recent CMS simplified model limits demonstrates that part of this chargino co-annihilation region may already have been probed~\cite{CMSStopSummary}.
Still, the region of highest likelihood is the stop co-annihilation region, with $m_{\tilde{t}_1}-m_{\tilde{\chi}^0_1} \lesssim 50$~\GeV. This can be seen in the top panels of Figure~\ref{fig:1d_sparticle_mass_differences}  in blue. Excluding the stop co-annihilation mechanism entirely in the NUHM1 is more difficult than in the CMSSM, requiring the ability to probe compressed spectra for $\tilde{t}_1$ masses up to approximately 1700\,GeV, as seen in Figure~\ref{fig:colour-stop}. The situation in the NUHM2 model (not shown) is qualitatively similar.

\begin{figure*}[pt]
  \centering
 \includegraphics[width=0.49\textwidth]{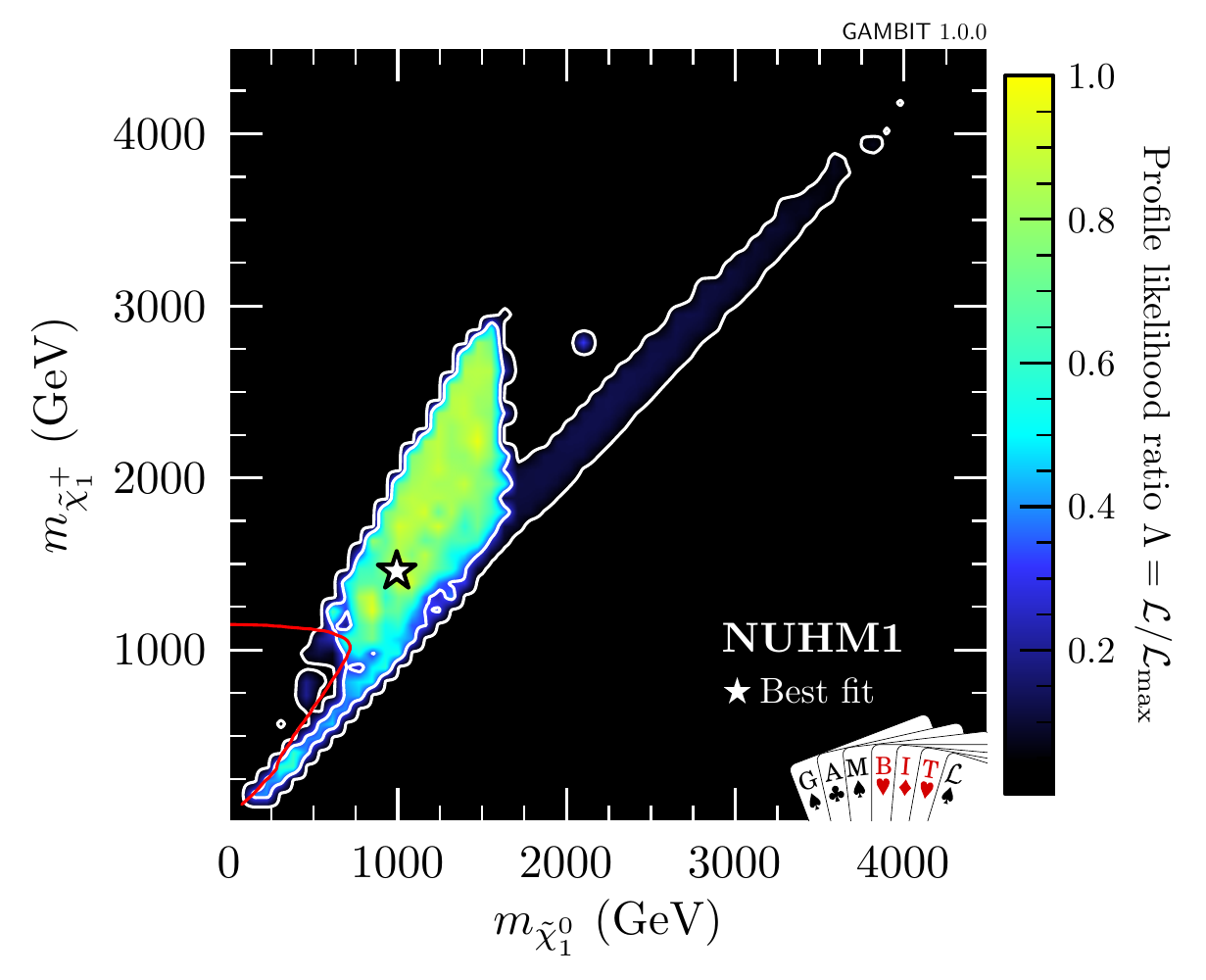}
 \includegraphics[width=0.49\textwidth]{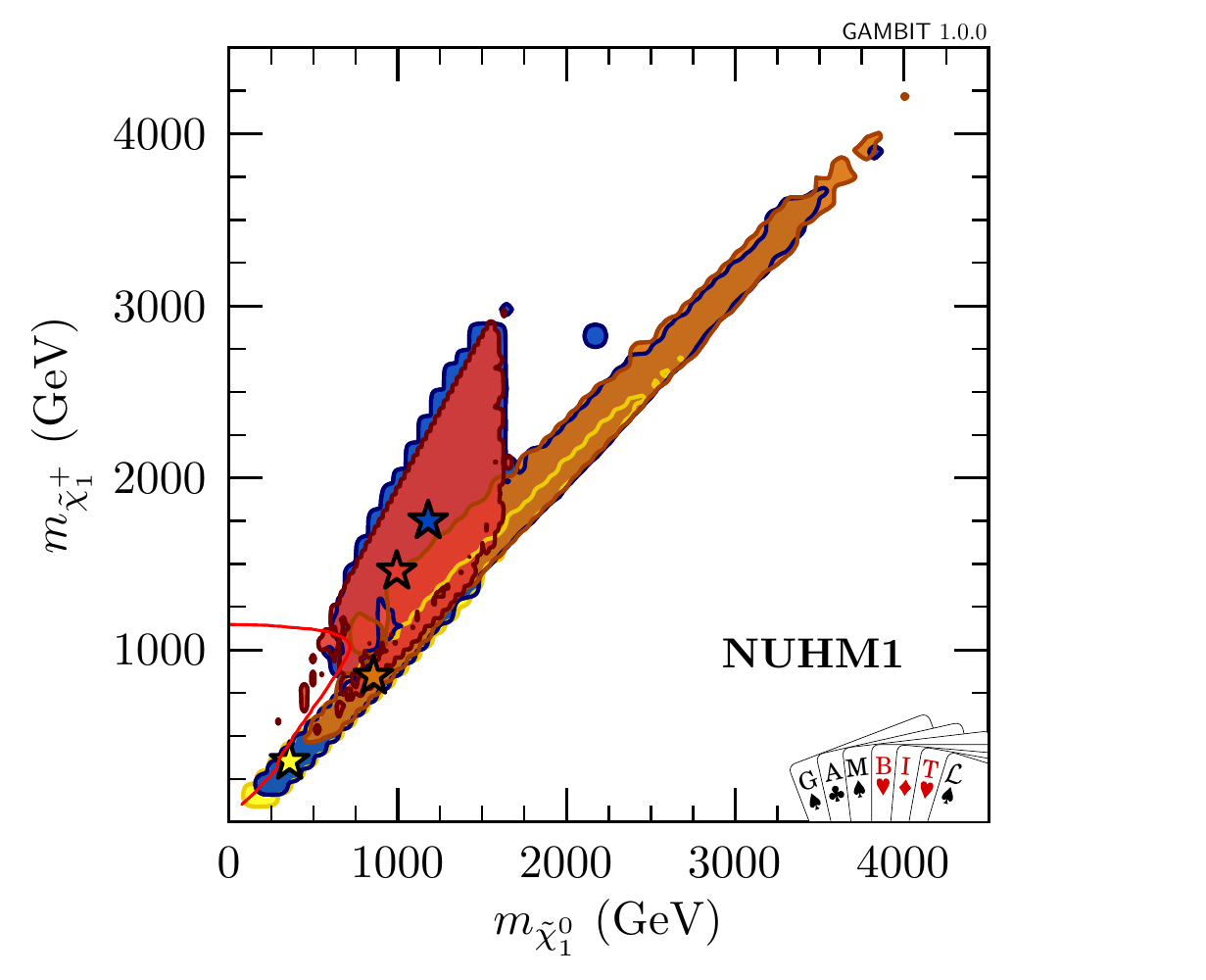}\\
 \includegraphics[height=4mm]{figures/rdcolours4.pdf}
 \caption{\textit{Left:} Profile likelihood for the NUHM1 in the  $\tilde{\chi}_1^\pm-\tilde{\chi}^0_1$ mass plane. \textit{Right:} Colour-coding shows the mechanism(s) that allow models within the 95\% CL region to avoid exceeding the observed relic density of DM. Superimposed in red is the latest CMS Run II simplified model limit for $\tilde{\chi}_1^\pm \tilde{\chi}_1^0$ pair production and decay via sleptons~\cite{CMSEWSleptonSummary}. This limit should be interpreted with caution (for details see main text).}
  \label{fig:nuhm1-charge-chi}
\end{figure*}

\begin{figure*}[pt]
  \centering
 \includegraphics[width=0.49\textwidth]{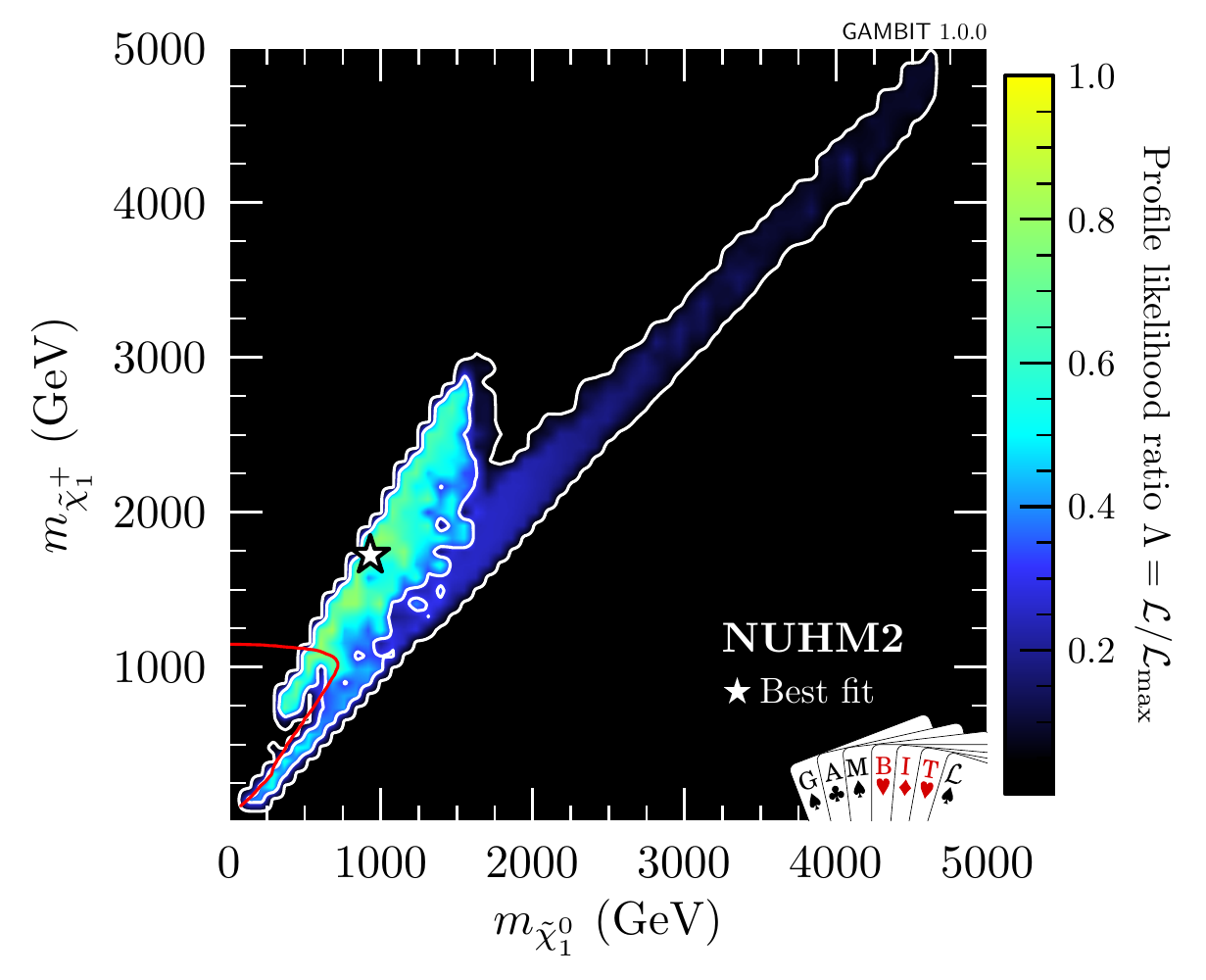}
 \includegraphics[width=0.49\textwidth]{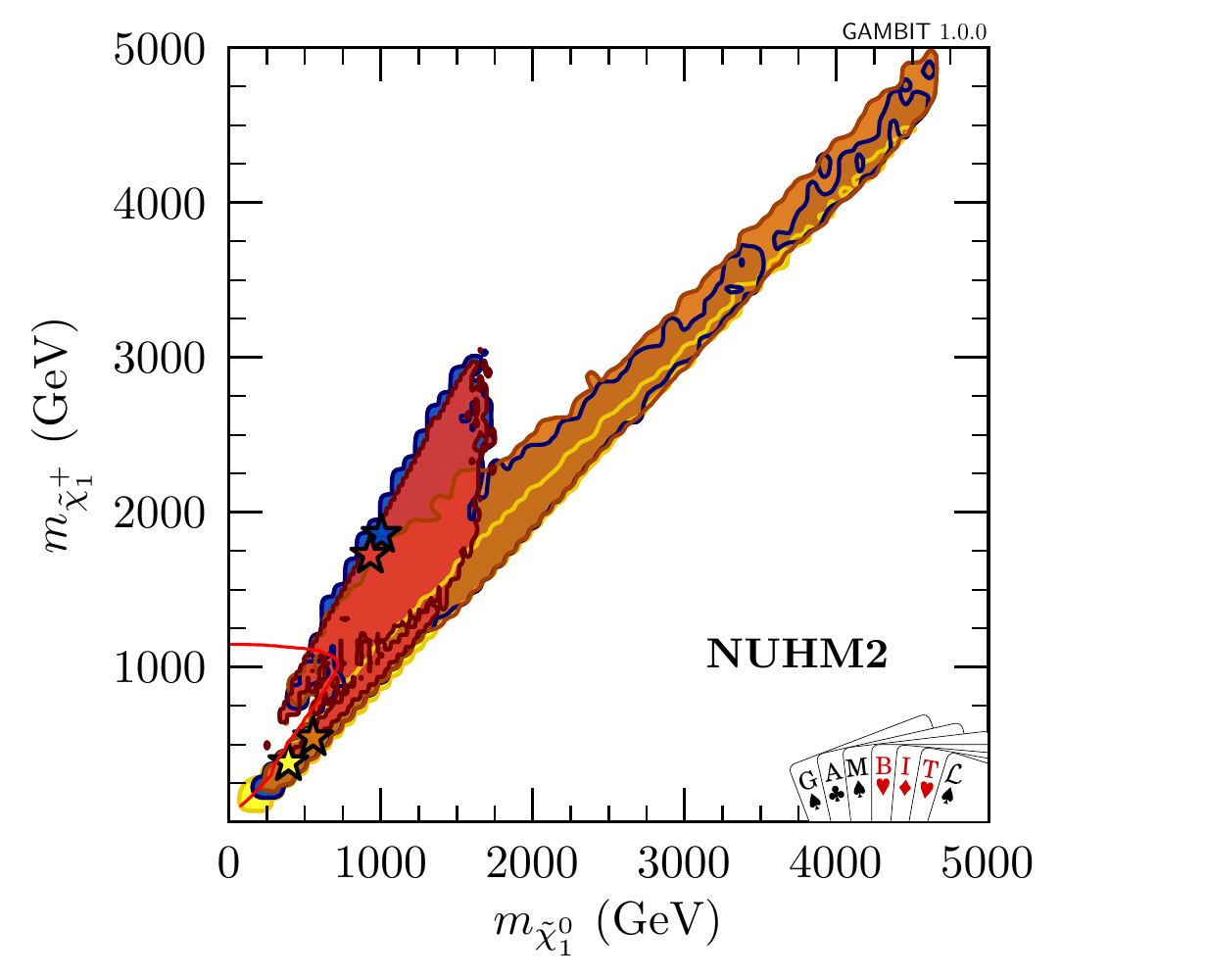}\\
 \includegraphics[height=4mm]{figures/rdcolours4.pdf}
 \caption{\textit{Left:} Profile likelihood for the NUHM2 in the  $\tilde{\chi}_1^\pm-\tilde{\chi}^0_1$ mass plane. \textit{Right:} Colour-coding shows mechanism(s) that allow models within the 95\% CL region to avoid exceeding the observed relic density of DM. Superimposed in red is the latest CMS Run II simplified model limit for $\tilde{\chi}_1^\pm \tilde{\chi}_1^0$ pair production and decay via sleptons~\cite{CMSEWSleptonSummary}. This limit should be interpreted with caution (for details see main text).}
  \label{fig:nuhm2-charge-chi}
\end{figure*}

Figure~\ref{fig:1d_sparticle_masses} also shows the presence of a region with relatively small $\tilde{\tau}_1$ masses, particularly in the NUHM1 and NUHM2. These masses are, however, already too large to lead to substantial stau production, given the small direct production cross-section~\cite{Fuks:2013lya} at 13~TeV.

Finally, we consider the prospects for discovery of charginos and neutralinos in the CMSSM, NUHM1 and NUHM2, as there are high-likelihood model points with relatively low $\tilde{\chi}^0_1$ and $\tilde{\chi}^\pm_1$ masses in all three models. In Figure~\ref{fig:cmssm-charge-chi}, we show the profile likelihood ratio in the $\tilde{\chi}_1^\pm-\tilde{\chi}^0_1$ mass plane for the CMSSM, as well as a version colour-coded by the mechanism for depleting the relic density. For low masses, there is always a strict correlation between the $\tilde{\chi}_1^\pm$ and $\tilde{\chi}^0_1$ masses in the CMSSM. The $m_{\tilde{\chi}_1^\pm} \sim m_{\tilde{\chi}^0_1}$ correlation is a consequence of a Higgsino-dominated lightest neutralino, which is always accompanied by a Higgsino-dominated chargino with a similar mass and leads to chargino co-annihilation. In the stop co-annihilation region, the neutralino is dominantly bino, and the chargino is mostly wino, with a mass about twice that of the neutralino. This comes from the approximate $2:1$ ratio between the low-scale wino and bino mass parameters, produced by RGE running from the common GUT-scale input value $m_{1/2}$. We also show the envelope of the latest CMS simplified model interpretations for $\tilde{\chi}_1^\pm\tilde{\chi}^0_1$ production and decay with decoupled sleptons~\cite{CMS-PAS-SUS-16-039,CMS-PAS-SUS-16-043,CMS-PAS-SUS-16-034,CMS-PAS-SUS-16-048}. This should only be used as an indicator of the optimum CMS exclusion power, as we have not performed a detailed examination of our model points to check that the EW gaugino mixing matrices and decay branching ratios match the CMS assumptions. Only the low-mass tip of the stop co-annihilation part of our 2$\sigma$ region has been probed by the most recent CMS analyses, and the best fit point is far beyond the current LHC reach.

\begin{figure*}[tp]
  \centering
  \includegraphics[width=0.49\textwidth]{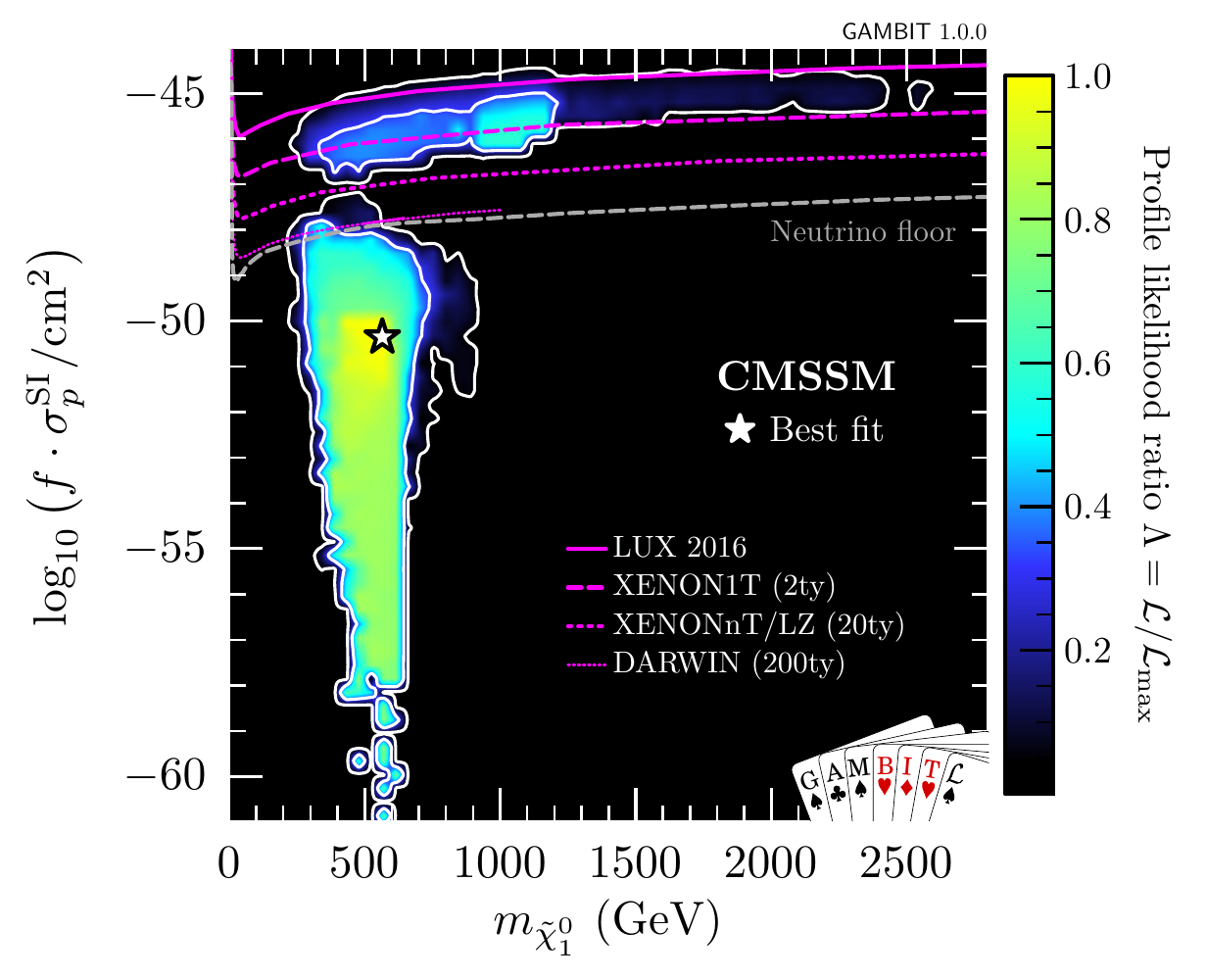}
  \includegraphics[width=0.49\textwidth]{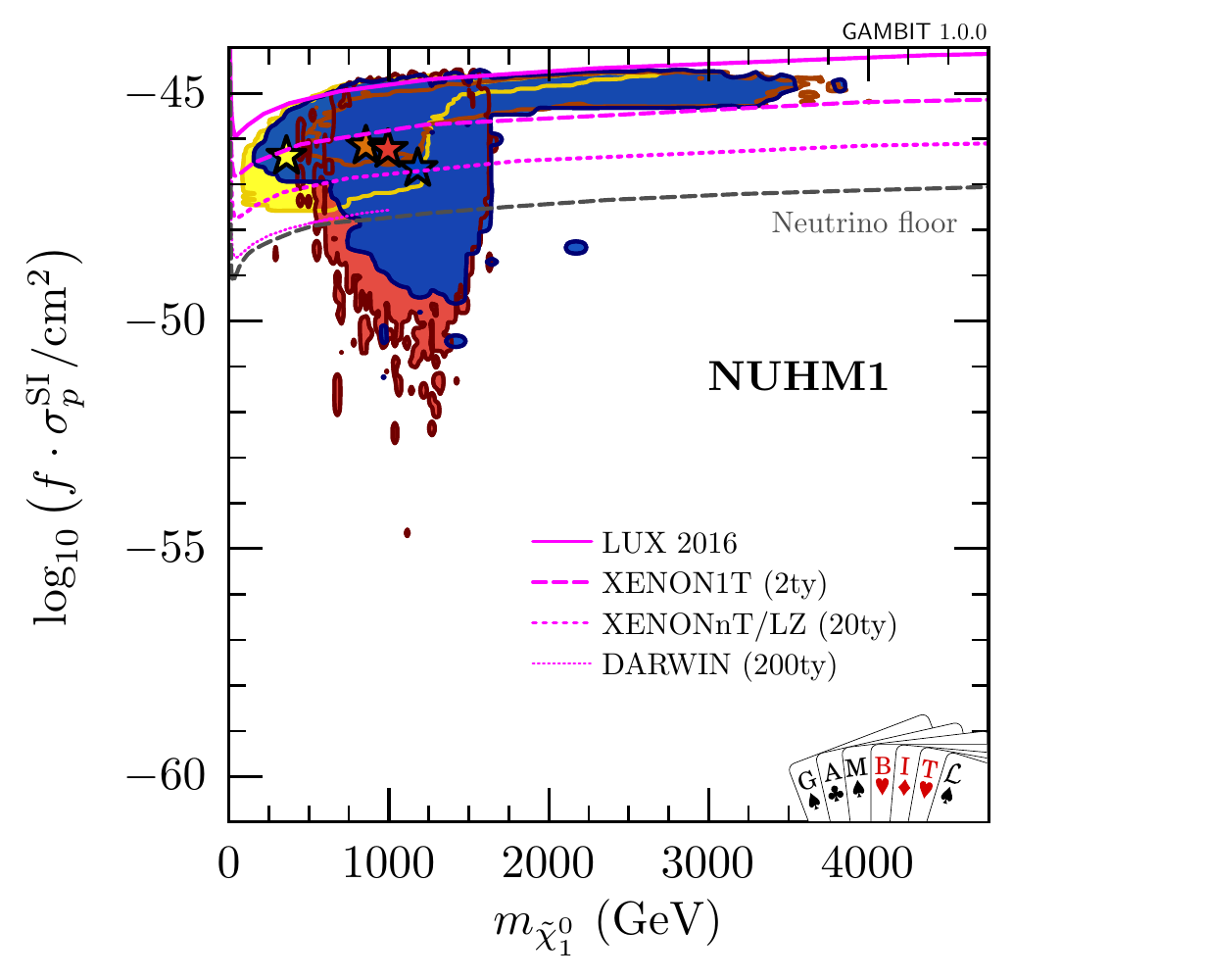}\\
  \includegraphics[width=0.49\textwidth]{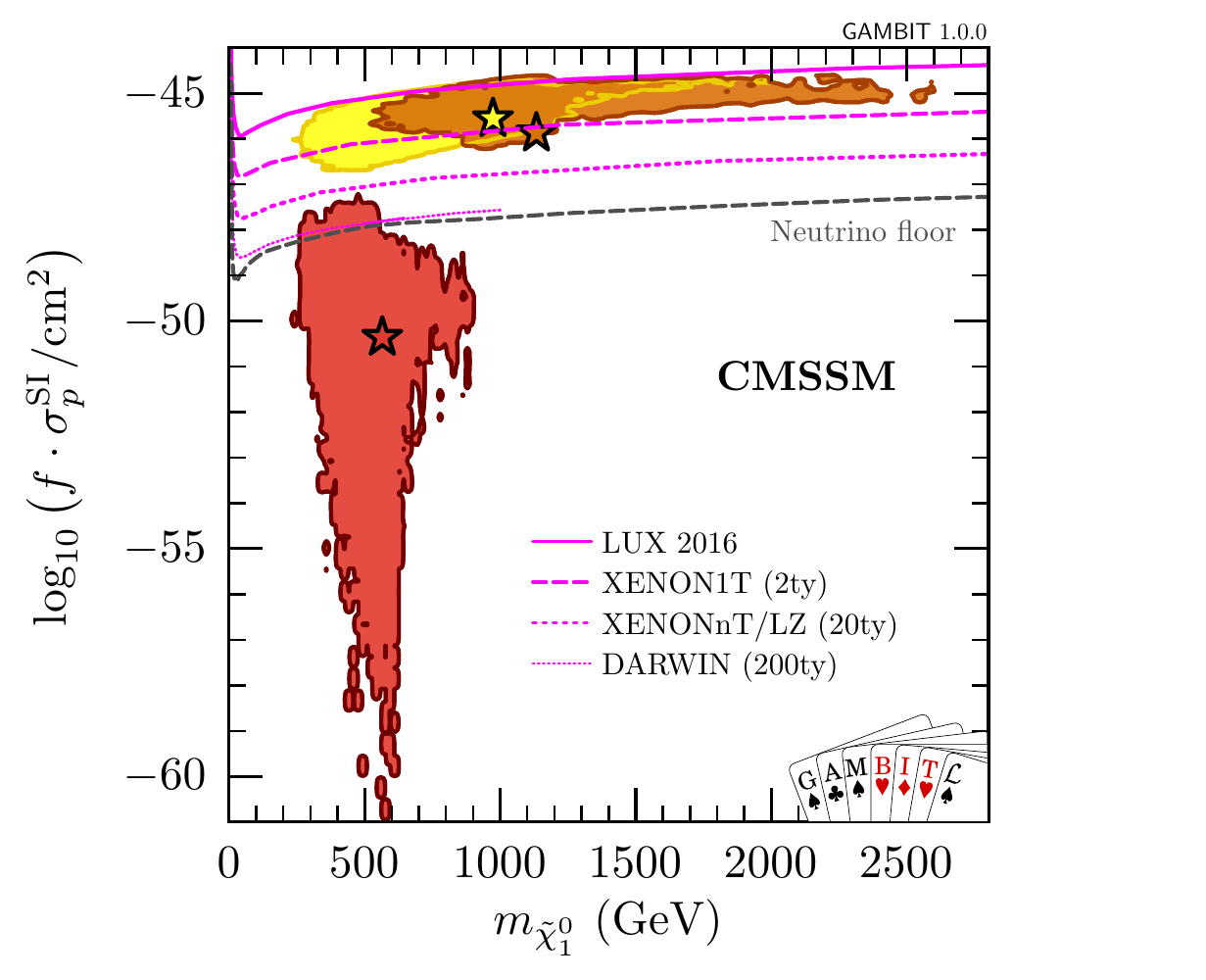}
  \includegraphics[width=0.49\textwidth]{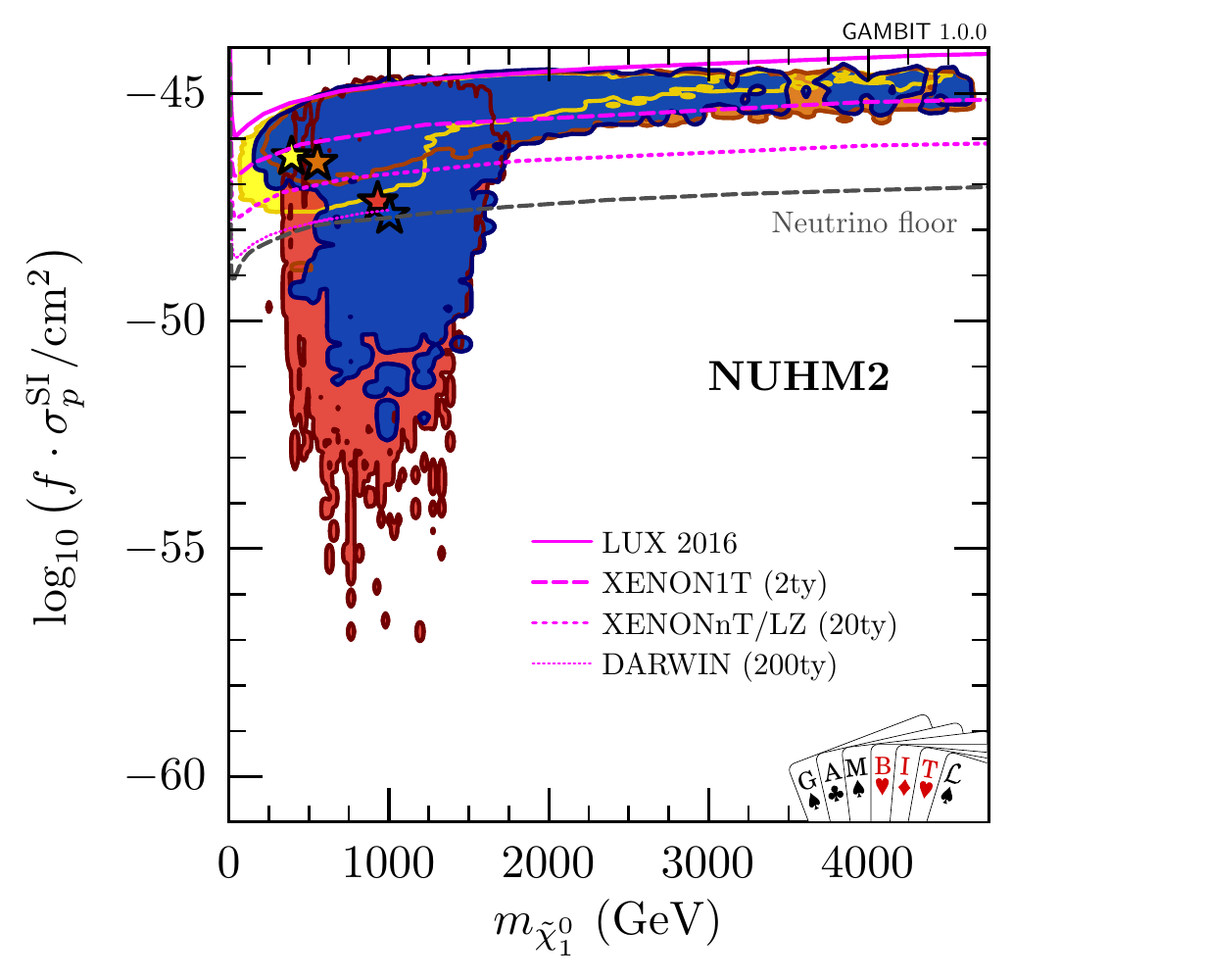}\\
  \includegraphics[height=4mm]{figures/rdcolours4.pdf}
  \caption{The spin-independent neutralino-proton cross-section. \textit{Upper Left:} Profile likelihood in the CMSSM.  \textit{Lower Left:} Colour-coding shows the active mechanism(s) by which CMSSM models avoid exceeding the observed relic density of DM, through either chargino co-annihilation, the $A/H$ funnel, or stop co-annihilation. \textit{Top Right:} Colour-coded regions in the NUHM1, now also featuring stau co-annihilation (blue).  \textit{Bottom Right:} Colour-coded regions of the NUHM2. 90\% CL exclusion limits are overlaid from the complete LUX exposure~\cite{LUXrun2}, the projected reach of XENON1T with two years of exposure, the projected reach of XENONnT/LZ with 20 tonne-years of exposure~\cite{XENONnTLZ} (around 1--3 years of data), and the projected reach of DARWIN with 200 tonne-years of exposure~\cite{DARWIN} (around 3--4 years of data).
  The ``neutrino floor'', where the coherent neutrino background starts to limit the experimental sensitivity, is indicated by the dashed grey line \cite{Billard:2013qya}.  The exact position of this limit is subject to several caveats; see \cite{Billard:2013qya} for further details.
  \label{fig:2d_direct_search_SI}
}
\end{figure*}

\begin{figure*}[tp]
  \centering
  \includegraphics[width=0.49\textwidth]{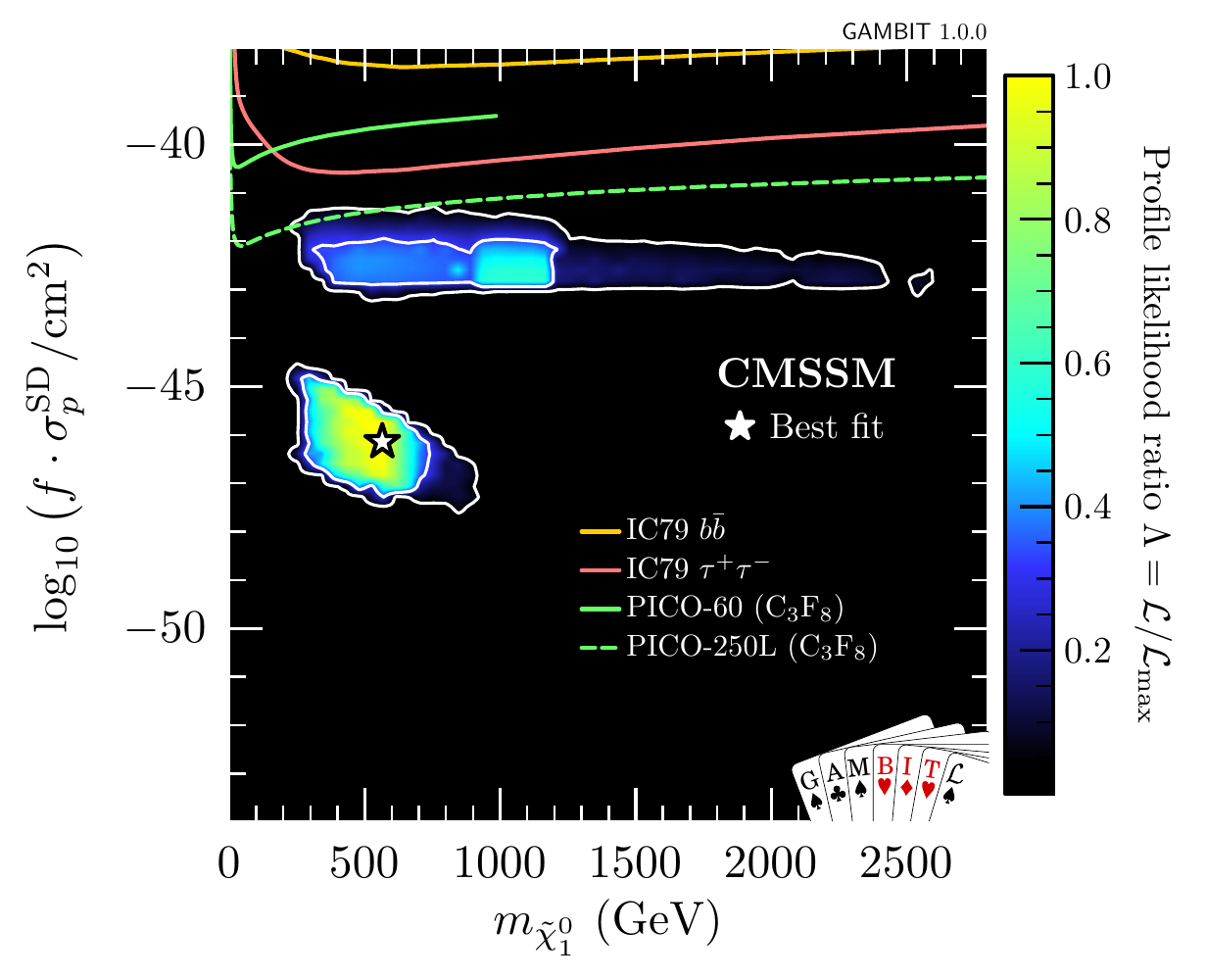}
  \includegraphics[width=0.49\textwidth]{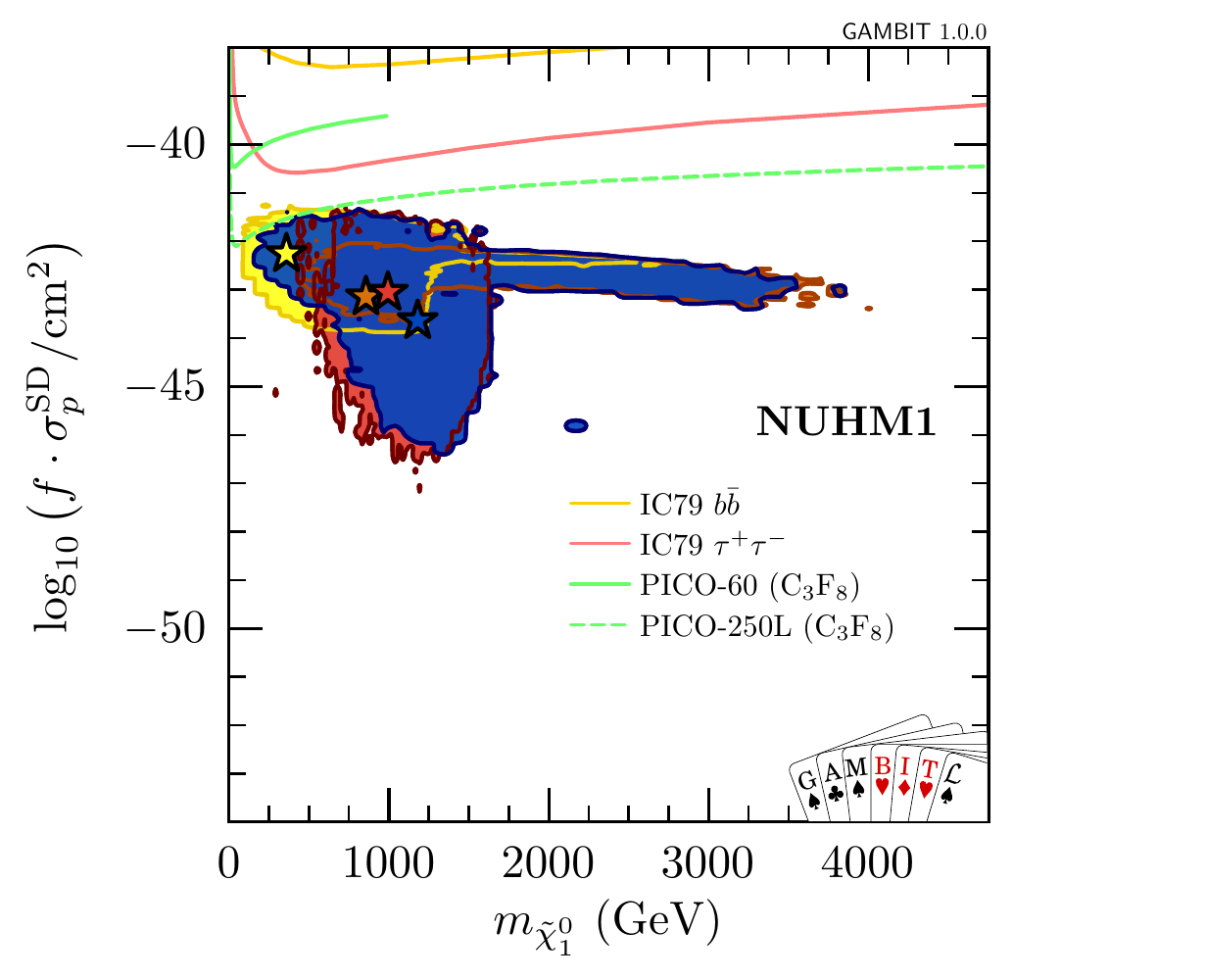}\\
  \includegraphics[width=0.49\textwidth]{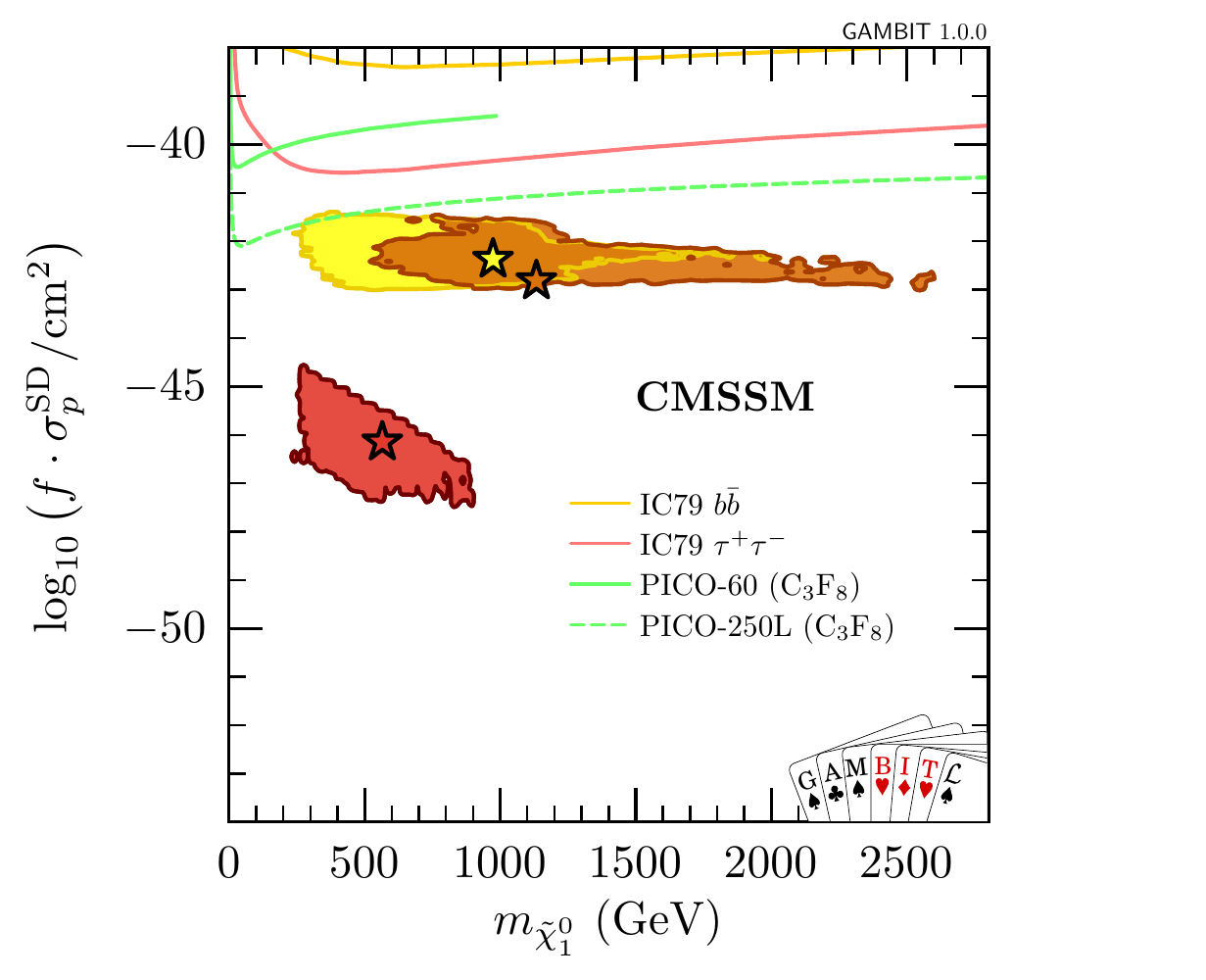}
  \includegraphics[width=0.49\textwidth]{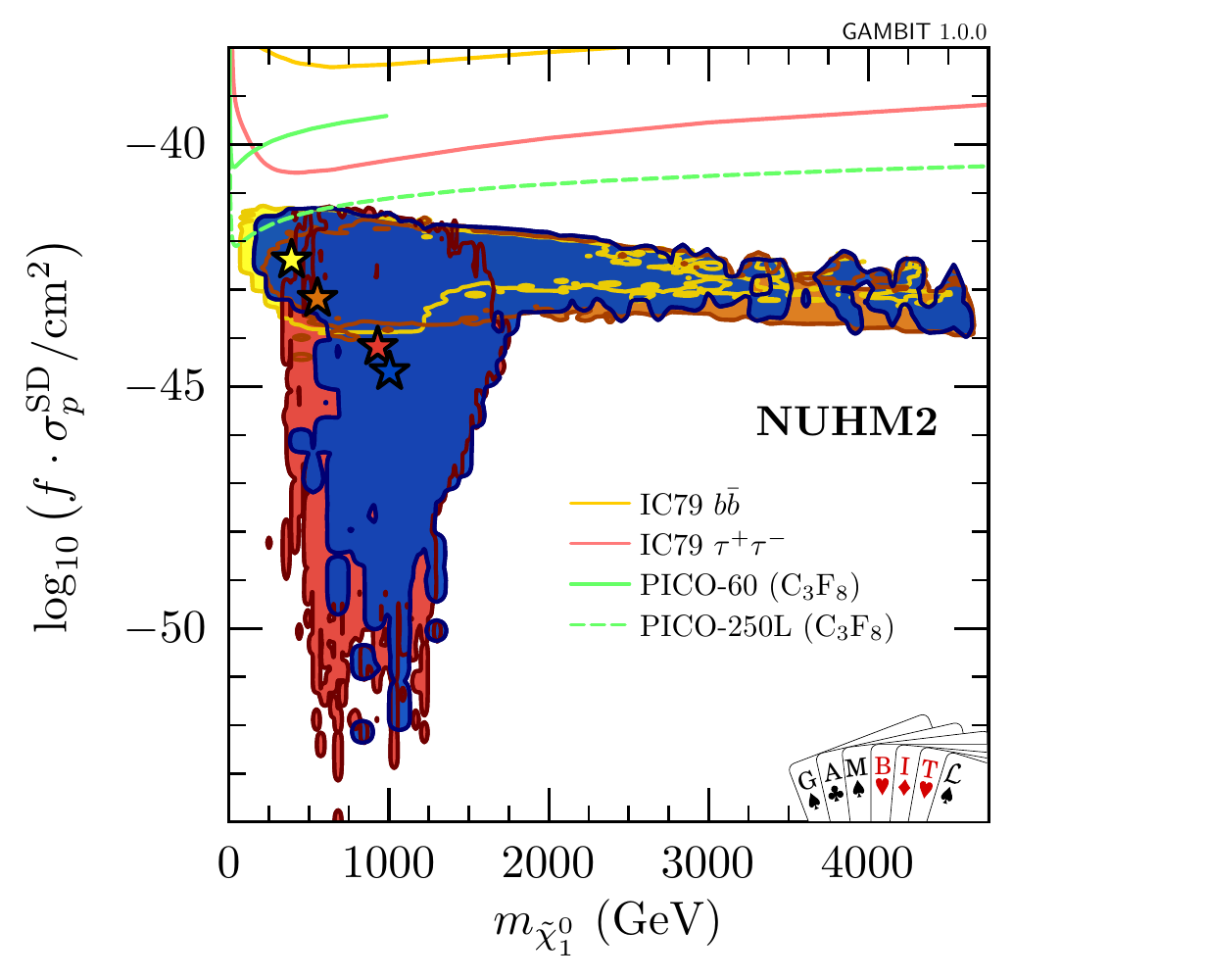}\\
  \includegraphics[height=4mm]{figures/rdcolours4.pdf}
  \caption{The spin-dependent neutralino-proton cross-section. \textit{Upper Left:} Profile likelihood in the CMSSM.  \textit{Bottom Left:} Colour-coding shows the active mechanism(s) by which CMSSM models avoid exceeding the observed relic density of DM, through either chargino co-annihilation, the $A/H$ funnel, or stop co-annihilation. \textit{Top Right:} Colour-coded regions in the NUHM1, now also featuring stau co-annihilation (blue).  \textit{Bottom Right:} Colour-coded regions of the NUHM2. 90\% CL exclusion limits are overlaid from the 79-string IceCube search for DM~\cite{IC79,IC79_SUSY}, assuming dark matter annihilation in the Sun to $\bar b b$ (yellow solid) and $\tau^+\tau^-$ (red solid) final states, from PICO-60~\cite{PICO60_2} (green solid), and projected limits from PICO-250 \cite{PICO250} (green dashes).
 }
  \label{fig:2d_direct_search_SD}
\end{figure*}

Figure~\ref{fig:nuhm1-charge-chi} shows the profile likelihood ratio in the $\tilde{\chi}_1^\pm-\tilde{\chi}^0_1$ mass plane for the NUHM1. The low-mass region now sees a contribution from stau co-annihilation in addition to chargino co-annihilation. This is interesting for LHC searches, as the assumption of decoupled sleptons clearly no longer applies. The recent CMS simplified model interpretations include a model where the sleptons are not decoupled, but the interpretation is even more fraught than that of the previous simplified models we have considered. The slepton masses are fixed in these scenarios, and one can generically expect the strength of the exclusions to decrease as one departs both from the mass assumptions, and from the branching ratio assumptions.
Nevertheless, we show this limit in Figure~\ref{fig:nuhm1-charge-chi} in order to demonstrate the most optimistic possible exclusion, compared to our 2D profile likelihood.
Almost the entire region with compressed spectra remains unprobed. As the bottom right plot of Figure~\ref{fig:1d_sparticle_mass_differences} shows, the highest likelihood in the degenerate region is obtained for chargino--neutralino mass differences less than $15$~\GeV, which is small enough to escape the CMS searches.
However, the part of our low-mass 95\% CL region without degenerate $\tilde{\chi}_1^\pm-\tilde{\chi}^0_1$ masses may be within current LHC reach.
Furthermore, there is hope that the LHC would prove capable of exploring part of the 68\% CL region in the near future. The situation in the NUHM2 is shown in Figure~\ref{fig:nuhm2-charge-chi}.
The main difference with the NUHM1 is that the low-mass region potentially explored by the recent CMS analysis updates falls within our 68\% CL preferred region.

\subsubsection{Direct detection}
Now we turn to a discussion of the discovery prospects at future direct DM detection experiments.  In Figs.\ \ref{fig:2d_direct_search_SI}
and \ref{fig:2d_direct_search_SD}, we show the spin-independent (SI) and spin-dependent (SD) nuclear scattering cross-sections of the lightest neutralino as a function of its mass, scaled for the fraction of the local density of DM in neutralinos. We give the full profile likelihood for the CMSSM only. The other panels show $2\sigma$ confidence regions for each model, with colour-coded mechanisms for reducing the relic density to or below the observed value.

In the CMSSM, the chargino co-annihilation and $A/H$-funnel regions largely overlap, predicting SI cross-sections in the range $10^{-46}$--$10^{-44}$ cm$^2$ and neutralino masses from 200 to 2500\,GeV. This region will be almost fully probed by XENON1T after two years of data-taking (long dashed curve), and could be completely excluded by XENONnT or LZ with 1--3 years of data (short dashed curve). The stop region has significantly lower SI cross-sections, with an LSP that can be almost pure bino in nature. Most of this region is outside of the projected reach of future multi-tonne detectors, although the proposed $\sim$50-tonne DARWIN experiment may probe it slightly (dotted curve).  As discussed in Sec.\ \ref{sec:lhc}, this region is also difficult to see at the LHC, but may be within the reach of a future linear collider.

The NUHM1 and NUHM2 display similar properties, with large parts of the chargino co-annihilation and $A/H$-funnel regions able to be tested in the near future, including models with very heavy LSPs (in the NUHM2 case, up to 4.5\,TeV).  Some of the chargino co-annihilation region at relatively low LSP masses ($< 1250$\,GeV) will remain untested by future direct detection experiments until DARWIN.  Parts of both the stop and stau co-annihilation regions will escape all direct detection, even DARWIN, although other parts of this region at higher masses will be easily detected or excluded by XENON1T.

In Fig.~\ref{fig:2d_direct_search_SD} we show the rescaled spin-dependent neutralino-proton scattering cross-sections for the CMSSM, NUHM1 and NUHM2.  Here we overplot current PICO limits not included in our scan~\cite{PICO60_2}, and sensitivity estimates for PIC0-250, a scaled-up version of PICO-60~\cite{PICO250}. We also show the IceCube 79-string limits from Ref.~\cite{IC79_SUSY}, for two different annihilation final states.  We can see that the preferred regions are relatively far from the current limits, so future direct detection experiments are unlikely to probe them further.  However, the proposed neutrino telescopes IceCube-PINGU \cite{Aartsen:2014oha} and KM3NeT-ORCA \cite{Adrian-Martinez:2016fdl} may have sufficient additional sensitivity to test the models with largest SD cross-sections. Although estimates of the expected sensitivity of these experiments to $\sigma_\mathrm{SD}$ exist, those estimates do not (yet) extend above DM masses of 100\,GeV, so at present they can tell us little about prospects for discovery of the CMSSM, NUHM1 or NUHM2.

Fig.\ \ref{fig:2d_direct_search_SI} shows the $2\sigma$ allowed region for the SI cross-section extending to substantially lower values in the CMSSM than the NUHM1 or NUHM2.  This seems surprising, as the CMSSM is a subspace of the NUHM1 and NUHM2, so all viable CMSSM models are indeed also viable NUHM1 and NUHM2 models.  The improvement in the best-fit likelihood in the NUHM1 compared to the CMSSM is not sufficient to explain this effect.  The smallest scattering cross-sections are caused by cancellations in the tree-level matrix elements, which can be tuned to essentially arbitrary accuracy.  A consequence of this is that models become steadily more fine-tuned as the cross-section asymptotically approaches zero, and therefore steadily more difficult to find for sampling algorithms.  What we see here is evidence of the additional numerical difficulty of finding such points in the NUHM1 and NUHM2, due to the additional challenge of dealing with more dimensions, and a more diverse set of viable regions of parameter space.  However, in models where the mass parameters unify at a high scale, loop corrections \cite{Djouadi:2001kba,Djouadi:2000ck} are expected to spoil such carefully-tuned cancellations anyway, holding cross-sections well above the lowest values that we see in the CMSSM \cite{Mandic:2000jz}.  The fact that we have found scattering cross-sections as low as $10^{-60}$\,cm$^2$ in the CMSSM, but not at quite such low values in the NUHM1 and NUHM2, is therefore ultimately of little physical significance.  Even if this isn't physically significant, however, getting as low as $10^{-60}$\,cm$^2$ in the CMSSM is nonetheless quite a remarkable numerical feat, made possible only by our use of \diver. This increases our confidence in the completeness of our sampling in the rest of the parameter space, and in fits of weak-scale MSSM models \cite{MSSM}.

\subsubsection{Indirect detection}
\label{sec:ID}

\begin{figure*}[tp]
  \centering
  \includegraphics[width=0.49\textwidth]{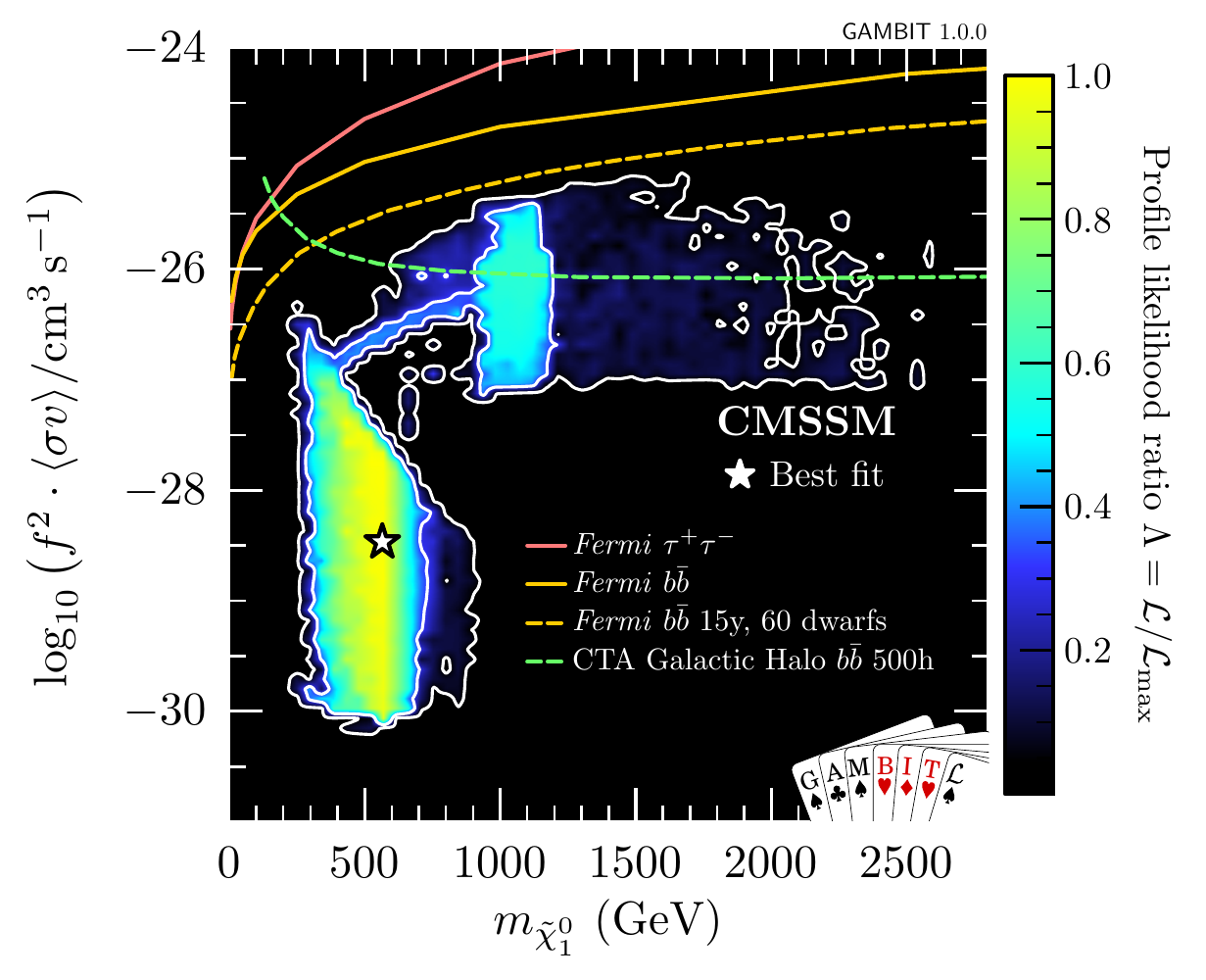}
  \includegraphics[width=0.49\textwidth]{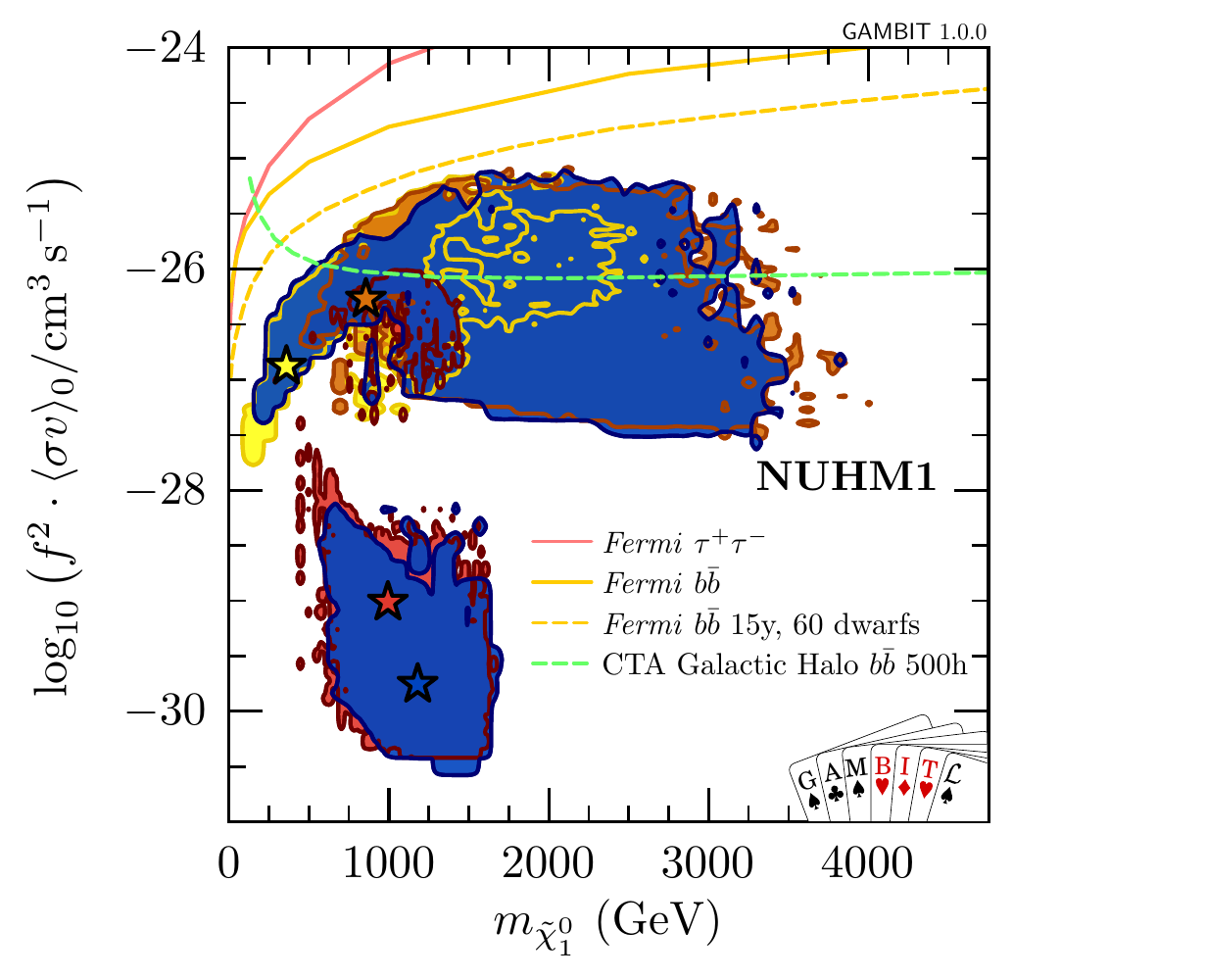}\\
  \includegraphics[width=0.49\textwidth]{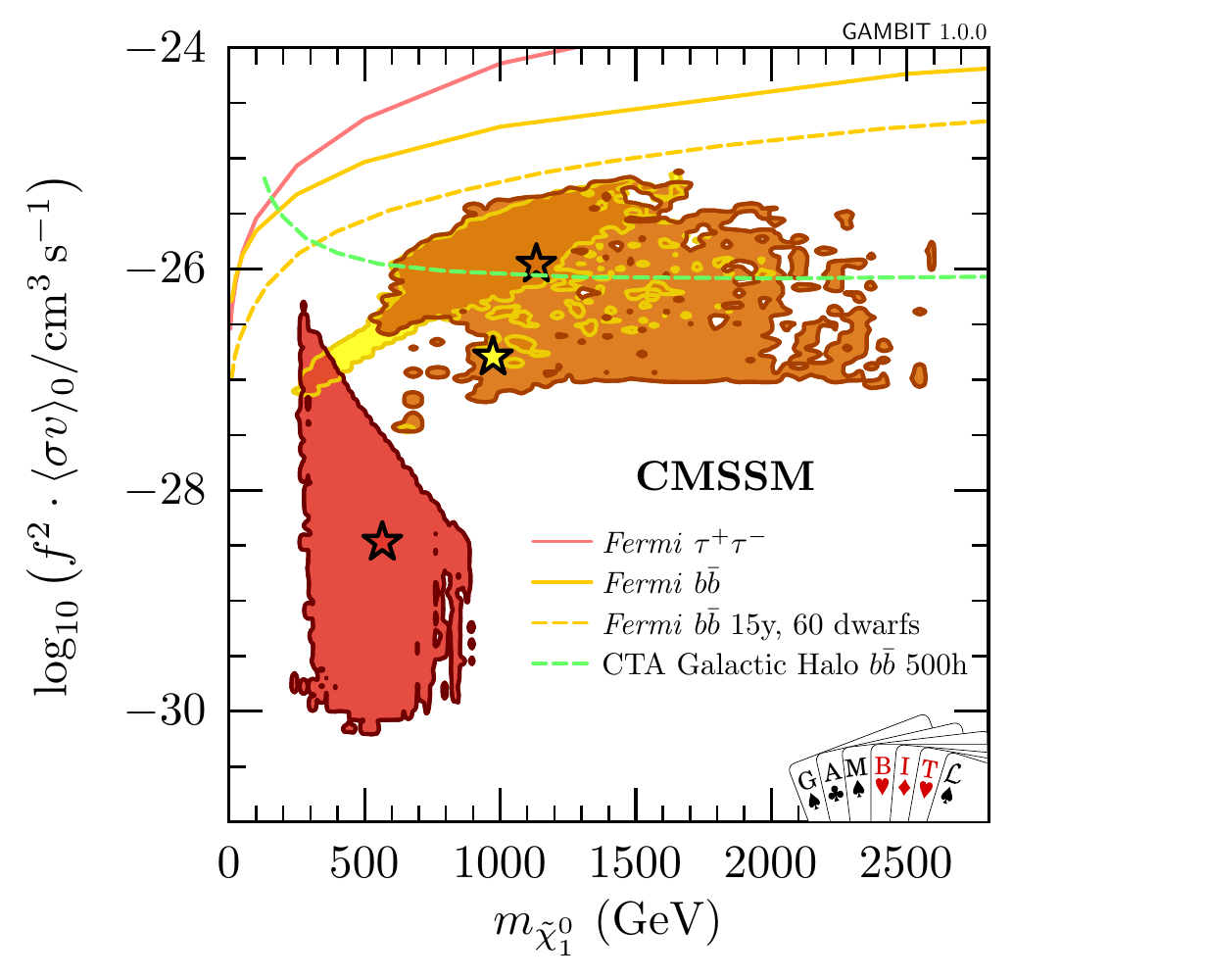}
  \includegraphics[width=0.49\textwidth]{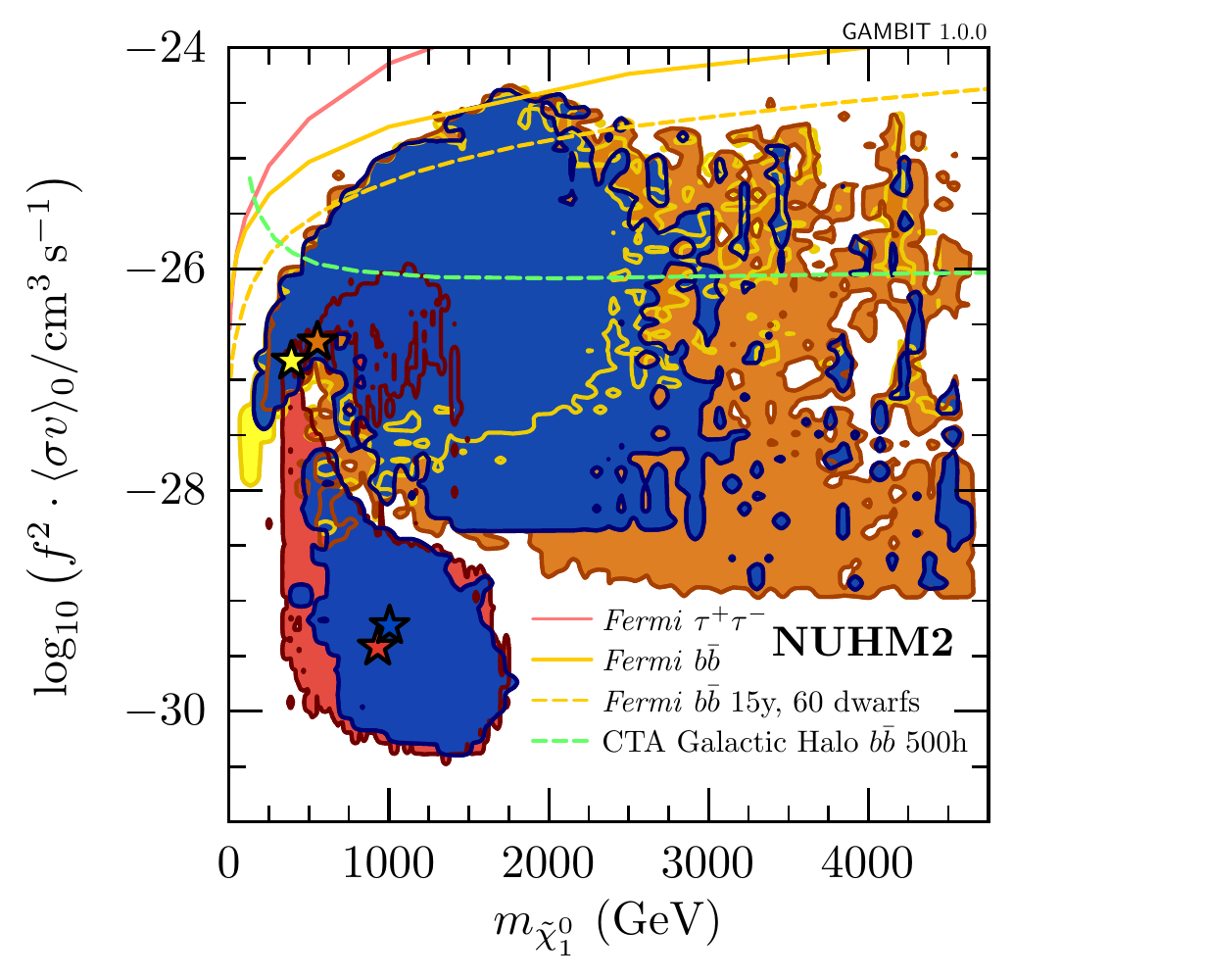}\\
  \includegraphics[height=4mm]{figures/rdcolours4.pdf}
  \caption{The present-day neutralino self-annihilation cross-section. \textit{Upper Left:} Profile likelihood in the CMSSM.  \textit{Bottom Left:} Colour-coding shows the active mechanism(s) by which CMSSM models avoid exceeding the observed relic density of DM, through either chargino co-annihilation, the $A/H$ funnel, or stop co-annihilation. \textit{Top Right:} Colour-coded regions in the NUHM1, now also featuring stau co-annihilation (blue).  \textit{Bottom Right:} Colour-coded regions of the NUHM2. 95\% CL exclusion limits are overlaid from the 6-year \textit{Fermi}-LAT search for DM annihilation in 15 satellite dwarf galaxies \cite{LATdwarfP8}, assuming dark matter annihilation to $\bar b b$ (yellow solid) and $\tau^+\tau^-$ (red solid) final states.  We also show the projected improvement for $b\bar{b}$ final states with 15 years of  LAT data and four times as many dwarfs \cite{Charles:2016pgz} (dashed yellow), and an optimistic projection of the sensitivity to $b\bar{b}$ final states of a Galactic halo search for DM annihilation by the upcoming Cherenkov Telescope Array, assuming 500\,hr of observations and no systematic uncertainties \cite{Carr:2015hta} (green dashes).}
  \label{fig:2d_indirect_search}
\end{figure*}

To assess the discovery prospects for future indirect searches for DM,
in Fig.~\ref{fig:2d_indirect_search} we show the rescaled zero-velocity annihilation cross-section $f^2\cdot\langle\sigma v\rangle_0$, as a function of the mass of the lightest neutralino.  Here $f$ is again the ratio of the neutralino relic density in the model to the observed relic density of DM. Note that we implicitly assume here that if neutralinos are not all of the DM, the other component(s) of DM cluster in the same way as neutralinos, leading to the same $f$ cosmologically, in dwarf galaxies and in the local halo.  Although this needn't be true in general, the general requirements that DM be cold and (almost) non-interacting mean that this should be a reasonably good approximation.

The upper left panel of Fig.~\ref{fig:2d_indirect_search} shows the profile likelihood for the CMSSM, and the remaining panels show the mechanisms by which models in the CMSSM (bottom left), NUHM1 (top right) and NUHM2 (bottom right) avoid producing too much thermal DM.  In the same figure we also indicate, for comparison, current limits
from dwarf galaxy observations by the \textit{Fermi}-LAT \cite{LATdwarfP8}, assuming photon
spectra for DM annihilation to $\bar b b$ and $\tau^+\tau^-$ final states. We also show projected \textit{Fermi} limits for
$\bar b b$ final states \cite{Charles:2016pgz}, assuming 15 years of data on 60 dwarf galaxies (vs 6\,yr and 15 dwarfs in the current limits). Lastly, we show the projected sensitivity of CTA
after 500 hours of observation of the Galactic halo, also assuming $\bar b b$ final states \cite{Carr:2015hta}.  We note that the actual (projected) limits depend on the final state, and hence the lines shown are only indicative for general points in the CMSSM parameter space.  However, as long as the final states are hadronic, the expected variations remain within a factor of about three~\cite{Bringmann:2012ez}.

In general, the largest annihilation cross-sections are expected for the $A/H$ funnel region, where resonant annihilation boosts $\sigma v$.  All models with annihilation cross-sections above the canonical thermal value ($3\times10^{-26}$\,cm$^3$\,s$^{-1}$) exhibit resonant annihilation through the $A$ funnel (note that in the zero-velocity limit, due to the $CP$ properties of the initial state, only the pseudoscalar resonances can contribute).  As one would expect, all regions above this value in Fig.\ \ref{fig:2d_indirect_search} are indeed identified as being part of the $A/H$ funnel region, indicated by the fact that they are shaded orange.  Some parts of these regions are also shaded yellow and/or blue, as some of the parameter points identified as belonging to the $A/H$ funnel region \textit{also} satisfy the necessary mass/composition conditions to be counted as part of the stau and/or chargino co-annihilation regions.  However, in all regions of overlap above $f^2\cdot\langle\sigma v\rangle_0 \sim 3\times10^{-26}$\,cm$^3$\,s$^{-1}$, resonant annihilation via the heavy Higgs bosons is the dominant mechanism in setting the relic density.  Most such models exhibit a relic density below the observed value.\footnote{The fact that $\sigma v$ is set by a resonance in the funnel region means that the present-day annihilation cross-section can be somewhat higher or lower than during freeze-out; models where $\langle\sigma v\rangle_0 > 3\times10^{-26}$\,cm$^3$\,s$^{-1}$ but $\Omega_\mathrm{c}h^2$ matches the observed value exhibit this effect.  This allows the observed relic density to be achieved up to $\langle\sigma v\rangle_0 \sim 7\times10^{-26}$\,cm$^3$\,s$^{-1}$ in the CMSSM, $\langle\sigma v\rangle_0 \sim 10^{-25}$\,cm$^3$\,s$^{-1}$ in the NUHM1, and $\langle\sigma v\rangle_0 \sim 3\times10^{-25}$\,cm$^3$\,s$^{-1}$ in the NUHM2.}

Indeed, most of the high-mass models identified in our scans as having stau and/or chargino co-annihilation also exhibit resonant annihilation through the heavy Higgs funnel.\footnote{We remind the reader that shaded models are all those that exhibit a given relic density mechanism -- not those that \textit{only} exhibit that mechanism.}  Indeed, as discussed at the beginning of Sec.\ \ref{sec:current} in the context of Fig.\ \ref{fig:mchi_oh2}, above DM masses of around a TeV, chargino co-annihilation in the CMSSM is unable to deplete the relic density to the thermal value or below without additional assistance from the $A/H$ resonance -- so in fact, \textit{all} chargino co-annihilation models above about a TeV are hybrid models of some kind.  This point is confirmed by careful study of all other CMSSM, NUHM1 and NUHM2 plots in this section: for $m_\chi\gtrsim1$\,TeV, chargino co-annihilation regions only appear in the presence of the heavy Higgs funnel, indicating that all high-mass chargino co-annihilation models are in fact either hybrids with, or completely dominated by, the heavy Higgs funnel.  The NUHM1 and NUHM2 plots also show that the same situation holds for stau co-annihilation at $m_\chi\gtrsim1.5$\,TeV in the NUHM, which at such masses appears only in combination with the heavy Higgs funnel.

In the CMSSM, \textit{Fermi} will generally only probe the low-likelihood tails of the $A/H$ funnel region and its co-annihilation hybrid, with the exception of $\sim1$\,TeV Higgsinos (see below).  Taking the optimistic predictions of Ref.\ \cite{Carr:2015hta} at face value, CTA will significantly cut into this region.  However, this ignores the impact of detector and background systematics.  Adding a systematic uncertainty of 1\%, the sensitivity would degrade by a factor of $\sim$6 \cite{Silverwood:2014yza}, and hence probe a significantly smaller part of the funnel and/or chargino co-annihilation region. The stop co-annihilation region, on the other hand, will remain largely unconstrained in the CMSSM, even with future indirect detection missions.

Let us stress, however, that indirect detection prospects will generally be
better than indicated by this general discussion. One aspect not taken into account
here is the impact of radiative corrections to the annihilation rate, which are particularly
relevant for large neutralino masses \cite{Bringmann:2007nk,Ciafaloni:2011sa, Bringmann:2017sko}. In the stop
co-annihilation region, for example, the inclusion of Higgs-Strahlung off fermion
final states can increase the total annihilation rate by a factor of a few compared
to what we have implemented \cite{Bringmann:2017sko}. Also antiprotons can be
an efficient complementary probe of compressed mass spectra \cite{Asano:2011ik}.
For heavy neutralinos
with mass-degenerate charginos, another example is a distinct feature from
$W^+W^-\gamma$ final states \cite{Bergstrom:2005ss}, which adds to the already
large monochromatic line signal from such models
\cite{Bergstrom:1997fh,Bern:1997ng,Ullio:1997ke};
such a signal is much more easily distinguished from astrophysical backgrounds than
the spectrum from $\bar b b$ final states assumed for CTA in Fig.~\ref{fig:2d_indirect_search}.  This signal would also appear in observations of the Galactic Centre well
before any dwarf observations shown in Fig.~\ref{fig:2d_indirect_search} or taken into account in our scans.
Last but not least, let us mention the Sommerfeld effect
\cite{Sommerfeld,Hisano:2002fk,Hisano:2004ds}, which leads to
a large enhancement in particular for $\sim$1\,TeV Higgsino DM
\cite{Hisano:2004ds,Hryczuk:2010zi,Catalan:2015cna}, and which we have
not (yet) included in \GB.

In the NUHM1 and NUHM2, \textit{Fermi} appears to be just beginning to constrain $A$-funnel models at masses of around 1.7\,TeV.  The various mechanisms to suppress the relic density are not as well-separated in the $\langle \sigma v\rangle_0$ -- $m_\chi$ plane in NUHM models as in the CMSSM, however. As a consequence, all these mechanisms
can be (partially) tested with CTA. We note that this includes the stau co-annihilation
region.  The sensitivity curves shown here are overly conservative for such models, because
$\tau^+\tau^-\gamma$ final states will be much more constraining than $\bar bb$ spectra \cite{Bringmann12}. Still, even with the most optimistic
assumptions, large parts of the viable parameter spaces of all of the GUT-scale models that we consider here will remain impossible to probe with indirect DM searches.

\section{Conclusions}
\label{sec:conc}

In this paper we have presented state-of-the-art profile likelihood global fits to three constrained versions of the minimal supersymmetric standard model, using \GB.  We have incorporated updated experimental data, additional observables and improved calculations for many quantities compared to previous global fits.  We have also fully explored the parameter space in which the models are not excluded by any experimental measurements, specifically including areas where the neutralino only constitutes a fraction of the dark matter in the Universe.

In the CMSSM, we show that the stau co-annihilation region is finally ruled out at more than 95\% CL.  This comes about due to Run II LHC constraints, difficulty in fitting the Higgs mass in this region, and an overall lifting of the isolikelihood contours defining the boundaries of this region, brought about by our improved sampling in this paper and resulting discovery of what is effectively a better best-fit than in previous works.  The NUHM1 and NUHM2 allow more freedom, permitting lighter staus and a re-appearance of the stau co-annihilation region as a source of equally good fits as other mechanisms for depleting the relic density.  Those include stop co-annihilation, chargino co-annihilation and resonant annihilation through the $A/H$ funnel.  We find that the chargino co-annihilation region also widens substantially in the NUHM1 and NUHM2 compared to the CMSSM, extending to arbitrarily low values of $m_0$.

Current constraints from the LHC push superpartner masses towards the multi-TeV regime, even if one does not demand that the lightest neutralino is the only DM species.  The important exceptions are the lightest neutralinos and charginos, which can still have masses as low as $\sim$100\,GeV without violating any experimental constraints, the lightest stau, which can be as light as $\sim$200\,GeV, and the lightest stop, which can be as light as $\sim$500\,GeV.

Despite very heavy spectra in many parts of the parameter space, future direct detection experiments will fully explore the chargino co-annihilation region, encompassing the so-called `focus point'.

We find a region of good fits at large negative trilinear coupling, where the neutralino relic abundance is set by co-annihilation with the lightest stop.  The trilinear couplings in this region raise questions about colour- and charge-breaking vacua, but our tests indicate that large parts of this region remain unaffected by such considerations.  More detailed investigation would, however, be interesting.  This region has been properly seen only in very recent fits performed contemporaneously with this one \cite{Han:2016gvr}.  Models in this region feature quite light stops ($\sim$500\,GeV), making it very appealing from the point of view of electroweak naturalness.  However, the stop co-annihilation mechanism requires the neutralino-stop mass difference to be quite small, which may constitute a fine-tuning in itself.  A more detailed analysis of naturalness considerations, including a full Bayesian treatment of the fit, would be illuminating.  Models in this region will be challenging to discover at the LHC, and next to impossible at direct detection experiments, but are promising targets for a future linear collider.

We began this study mainly intending to validate the new generic beyond-the-Standard-Model global fitting framework \GB.  In the end however, we have found quite a few genuinely new and interesting results.  This serves to illustrate the utility of a modern and adaptive global fitter such as \GB, where the impacts of different searches on different models can easily be examined and compared whilst retaining a consistent treatment of theoretical assumptions, systematics, nuisances, scanning algorithms, statistical approaches, experimental analyses and external code interfaces.

All input files, samples and best-fit benchmarks produced for this paper are publicly accessible from \textsf{Zenodo} \cite{the_gambit_collaboration_2017_801642}.  The \GB software is available from \href{http://gambit.hepforge.org}{gambit.hepforge.org}.

\begin{acknowledgements}
We thank Andrew Fowlie, Tom\'as Gonzalo, Julia Harz, Sebastian Hoof, Felix Kahlhoefer, James McKay, Roberto Trotta and Sebastian Wild for useful discussions, and Lucien Boland, Sean Crosby and Goncalo Borges of the Australian Centre of Excellence for Particle Physics at the Terascale for computing assistance and resources.  \gambitacknospmare
\end{acknowledgements}

\bibliography{R1}

\end{document}